\documentclass[a4paper,11pt]{article}
\pdfoutput=1
\usepackage[T1]{fontenc}
\usepackage[utf8]{inputenc}
\usepackage{jheppub}
\usepackage{amsmath,amssymb,amsthm,mathtools}
\usepackage{enumerate}
\usepackage{enumitem}
\usepackage{xspace}
\usepackage{subcaption}
\usepackage{lmodern}
\usepackage{cleveref}

\renewcommand{\tabcolsep}{2em}

\allowdisplaybreaks

\makeatletter
\g@addto@macro\bfseries{\boldmath}
\makeatother

\DeclareRobustCommand{\ensuremathrm}[1]{\ensuremath{\mathrm{#1}}\xspace}

\DeclareRobustCommand{\rd}{\ensuremathrm{d}} %

\DeclareRobustCommand{\order}[1]{\ensuremath{\mathcal{O}\left(#1\right)}}

\DeclareRobustCommand{\rs}{\ensuremathrm{s}}
\DeclareRobustCommand{\rem}{\ensuremathrm{em}}
\DeclareRobustCommand{\rT}{\ensuremathrm{T}}
\DeclareRobustCommand{\rR}{\ensuremathrm{R}}
\DeclareRobustCommand{\rF}{\ensuremathrm{F}}

\DeclareRobustCommand{\cut}{\ensuremathrm{cut}\xspace}
\DeclareRobustCommand{\alphas}{\ensuremath{\alpha_\rs}\xspace}
\DeclareRobustCommand{\alphaem}{\ensuremath{\alpha_\rem}\xspace}
\DeclareRobustCommand{\mur}{\ensuremath{\mu_\rR}\xspace}
\DeclareRobustCommand{\muR}{\mur}
\DeclareRobustCommand{\muf}{\ensuremath{\mu_\rF}\xspace}
\DeclareRobustCommand{\muF}{\muf}

\DeclareRobustCommand{\Pg}{{\ensuremath{g}}\xspace}

\DeclareRobustCommand{\Pq}{{\ensuremath{q}}\xspace}
\DeclareRobustCommand{\Paq}{{\ensuremath{\bar{q}}}\xspace}

\DeclareRobustCommand{\GeV}{\ensuremathrm{GeV}\xspace}

\DeclareRobustCommand{\fb}{\ensuremathrm{fb}\xspace}

\DeclareRobustCommand{\nnlojet}{NNLO\scalebox{0.8}{JET}\xspace}
\DeclareRobustCommand{\Matrix}{\textsc{Matrix}\xspace}

\DeclareRobustCommand{\Sherpa}{\textsc{Sherpa}\xspace}
\DeclareRobustCommand{\Diphox}{\textsc{Diphox}\xspace}

\DeclareRobustCommand{\NLO}{\text{NLO}\xspace}

\DeclareRobustCommand{\pt}{\ensuremath{p_\rT}\xspace}

\DeclareRobustCommand{\vecptof}[1]{\ensuremath{\mathbf{p}_{\rT,{#1}}}\xspace}

\DeclareRobustCommand{\vecptg}[1]{\ensuremath{\mathbf{p}_{\rT}^{\gamma_{#1}}}\xspace}

\DeclareRobustCommand{\ptg}[1]{\ensuremath{p_{\rT}^{\gamma_{#1}}}\xspace}
\DeclareRobustCommand{\ptgg}{\ensuremath{p_{\rT}^{\gamma\gamma}}\xspace}
\DeclareRobustCommand{\ptgj}{\ensuremath{p_{\rT}^{\gamma j}}\xspace}
\DeclareRobustCommand{\ptgave}{\ensuremath{\left\langle p_{\rT}^{\gamma} \right\rangle\xspace}}
\DeclareRobustCommand{\ptj}[1]{\ensuremath{p_{\rT}^{j_{#1}}}\xspace}

\DeclareRobustCommand{\Et}[1]{\ensuremath{E_{\rT}^{{#1}}}\xspace}

\DeclareRobustCommand{\Etiso}{\ensuremath{\Et{\text{iso}}}\xspace}
\DeclareRobustCommand{\Etthresh}{\ensuremath{\Et{\text{thr.}}}\xspace}
\DeclareRobustCommand{\epsg}{\ensuremath{\varepsilon_{\gamma}}\xspace}
\DeclareRobustCommand{\Etiso}{\ensuremath{E_{\rT}^{\text{iso}}}\xspace}
\DeclareRobustCommand{\Etisopart}{\ensuremath{E_{\rT}^{\text{iso,part}}}\xspace}

\DeclareRobustCommand{\dR}{\ensuremath{\Delta R}\xspace}

\DeclareRobustCommand{\dygg}{\ensuremath{\Delta y_{\gamma\gamma}}\xspace}
\DeclareRobustCommand{\dRgg}{\ensuremath{\Delta R_{\gamma\gamma}}\xspace}
\DeclareRobustCommand{\dRgj}{\ensuremath{\Delta R_{\gamma j}}\xspace}

\DeclareRobustCommand{\absyg}{\ensuremath{\left| y^{\gamma} \right| }\xspace}

\DeclareRobustCommand{\absdetagg}{\ensuremath{\left| \Delta\eta_{\gamma\gamma} \right| }\xspace}

\DeclareRobustCommand{\thetaetastar}{\ensuremath{{\theta_\eta^\ast}}\xspace}

\DeclareRobustCommand{\Mgg}{\ensuremath{M_{\gamma\gamma}}\xspace}
\DeclareRobustCommand{\MTgg}{\ensuremath{M_{\rT,\gamma\gamma}}\xspace}

\DeclareRobustCommand{\dphigg}{\ensuremath{\Delta\phi_{\gamma\gamma}}\xspace}

\DeclareRobustCommand{\abscosthetaetastar}{\ensuremath{\left|\cos{\thetaetastar}\right|}\xspace}

\preprint{{\raggedleft%
		IPPP/20/42 \\
		ZU-TH 17/20 \\
		CERN-TH-2020-160 \\
}}

\title{Scale and isolation sensitivity of diphoton distributions at the LHC}

\author[a]{Thomas Gehrmann,}
\author[b]{Nigel Glover,}
\author[b,c]{Alexander Huss,}
\author[b]{James Whitehead}
\affiliation[a]{Physik-Institut, Universit\"at Z\"urich, Winterthurerstrasse 190, CH-8057 Z\"urich, Switzerland}
\affiliation[b]{Institute for Particle Physics Phenomenology, Durham University, Durham, DH1 3LE, UK}
\affiliation[c]{Theoretical Physics Department, CERN, 1211 Geneva 23, Switzerland}
\emailAdd{thomas.gehrmann@uzh.ch}
\emailAdd{e.w.n.glover@durham.ac.uk}
\emailAdd{alexander.huss@cern.ch}
\emailAdd{james.c.whitehead@durham.ac.uk}

\abstract{
Precision measurements of diphoton distributions at the LHC display some tension with theory predictions, obtained at
next-to-next-to-leading order (NNLO) in QCD. We revisit the theoretical uncertainties arising from the approximation of
the experimental photon isolation by smooth-cone isolation, and from the choice of functional form for the
renormalisation and factorisation scales.  We find that the resulting variations are substantial overall, and enhanced
in certain regions.  We discuss the infrared sensitivity at the cone boundaries in cone-based isolation in related
distributions.  Finally, we compare predictions made with alternative choices of dynamical scale and isolation
prescriptions to experimental data from ATLAS at 8~TeV, observing improved agreement. This contrasts with previous
results, highlighting that scale choice and isolation prescription are potential sources of theoretical uncertainty
that were previously underestimated.
}


\begin{document}
	\maketitle

	\section{Introduction}
	\label{sec:intro}

	The production of pairs of isolated photons at hadron colliders is important as a test of perturbative QCD, as a
	clean background against which to measure the properties of the Higgs boson \cite{Aaboud:2018xdt,Sirunyan:2020xwk},
	and as a possible channel for the detection of new physics \cite{Aaboud:2017yyg,Sirunyan:2018wnk}.

	These alternatives reflect the different ways pairs of final-state photons can be produced at hadron colliders:
	directly in the partonic hard scattering (`prompt' photons), or as decay products.  Hadrons which may decay to
	photon pairs (e.g. $\eta$  or $\pi^0$ mesons) are produced in huge numbers in the collider environment.  Each such
	decay produces a highly-collimated photon pair, which is typically identified as a single photon accompanied by
	hadronic radiation.  For photonic final-states, such `non-prompt' photons are produced in sufficient abundance to overwhelm
	the prompt photon signal to which they form the background.

	To test our understanding of prompt photon production, it is therefore necessary to impose isolation cuts to
	suppress this overwhelming background.  Schematically, a photon is considered isolated if it is accompanied by relatively low
	levels of hadronic energy.  The standard (`fixed-cone') implementation of this idea is to veto events in which the
	total hadronic transverse energy deposited in a cone of fixed radius about the photon exceeds some threshold.  In practice,
	many additional sophisticated corrections are applied to correct for detector pileup and the fake rate from jets
	misidentified as photons.  These detector effects are unfolded in the experimental analysis to give parton-level
	fiducial cuts, which are then used for the corresponding theory predictions, obtained using numerical Monte Carlo
	calculations.

	This difference between the experimental isolation cuts on the transverse energy detected in calorimeter cells, and
	the corresponding theoretical isolation cuts on the transverse energy of simulated partons, is compounded by the
	theoretical difficulty of computing the fragmentation contribution to the process.  Final-state collinear
	singularities occur wherever a hard parton produced in the short-distance hard scattering undergoes a series of
	splittings, ending with a quark-photon splitting.  These are factorised to all orders into a fragmentation function
	$D_a^\gamma (z; \mu_f)$, encoding the probability that a photon is found in parton $a$ with momentum fraction $z$
	(at fragmentation scale $\mu_f$).  Analogously to parton distribution functions, these obey evolution equations in
	$\mu_f$, with boundary conditions that must be extracted from fits to experimental data.  Uncertainties in the data
	and in the fit propagate to uncertainties in the functions, and hence to predictions made with them.

	The fragmentation contribution could be eliminated entirely by setting the threshold for the isolation criterion to
	zero, but this restriction of the phase space of soft gluon emissions would spoil the necessary cancellation of
	real and virtual divergences in the direct contribution.  Instead, in fixed-order calculations theorists typically
	eliminate it formally, using `smooth-cone', or `Frixione' isolation \cite{Frixione:1998jh}, in which the energy
	threshold for permitted partonic radiation is promoted to a function $\chi(r)$ of the angular separation between
	the photon and the parton.  This function may be chosen freely, subject to the requirement that its limit vanishes
	towards the centre of the cone, with the dependence of the prediction on the unphysical profile function $\chi$ 
	entering as a new source of theoretical uncertainty.

	The finite granularity of the angular resolution of calorimeters makes this condition impossible to implement
	exactly at detectors, though a discretised version has been applied at the level of reconstructed particles at OPAL
	\cite{Abbiendi:2003kf} and investigated for the LHC \cite{Binoth:2010nha}.  Other isolation procedures that can be
	implemented both theoretically and experimentally have recently been proposed, such as `soft-drop isolation'
	\cite{Hall:2018jub}, based on jet substructure techniques and related both to `democratic isolation'
	\cite{Glover:1993xc} and to smooth-cone isolation in specific limits. These however have not yet been commonly adopted.
	As a result, all experimental measurements of final-states containing isolated photons so far performed at the LHC
	use fixed-cone isolation, whilst the majority of next-to-next-to-leading-order (NNLO) QCD predictions
	\cite{Catani:2011qz, Campbell:2016yrh, Chawdhry:2019bji} use smooth-cone isolation.

	In \cite{Siegert:2016bre} a compromise was introduced, called `hybrid-cone' isolation.  Formally a subset of
	smooth-cone isolation, it restricts the profile function $\chi(r)$ to be constant above some inner radius $R_d$,
	resulting in an annulus on which fixed-cone isolation is applied.  If this constant is chosen to match an
	experimental fixed-cone isolation cut, the artificial suppression of the cross-section resulting from the use of
	smooth-cone rather than fixed-cone isolation should be reduced.
	Imposing continuity of the profile function at the boundary $R_d$ leads to `matched-hybrid' isolation, which was 
	used in the photon-isolation study of \cite{Amoroso:2020lgh}.

	Here we apply matched-hybrid isolation to a new NNLO QCD calculation of the production of pairs of isolated prompt
	photons.  The calculation of the NNLO corrections uses antenna subtraction, making this the first such calculation
	with a local subtraction scheme, and avoiding the possible influence of isolation cuts on the power-corrections of
	$q_\rT$- and $N$-jettiness subtraction \cite{Ebert:2019zkb} used in prior NNLO calculations
	\cite{Catani:2011qz,Campbell:2016yrh,Catani:2018krb}.

	We find that relative to the standard smooth-cone parameters used in previous calculations \cite{Catani:2011qz,
	Campbell:2016yrh,Catani:2018krb}, matched-hybrid isolation gives a substantially larger cross-section, though still
	without signs of violating perturbative unitarity at NLO.  The localised effect of the suppression of smooth-cone
	isolation on differential cross-sections is explored and found to be connected kinematically to a similar, but opposing effect,
	resulting from the conventional scale choice $\muR=\muF=\Mgg$.  We explore the effect of
	making an alternative choice, focusing on the average $\pt$ of the identified photons $\ptgave$, and find that the
	resulting prediction accurately describes the 8~TeV ATLAS data \cite{Aaboud:2017vol}.

	\section{Photon isolation}
	\label{sec:isol}

	As outlined above, within fixed-cone isolation a photon is considered isolated if the total hadronic transverse energy
	deposited within a fixed cone of radius $R$ around photon $i$ in the $(\eta, \phi)$-plane, $E_\rT^\text{had} (R)$,
	is smaller than some threshold:
	\begin{align}
		E_{\rT}^\text{had} (R) \leqslant \Etiso (\gamma_i) ,
	\end{align}
	where we allow the threshold to vary between photons and events, typically as an affine function of the transverse
	energy of the photon $\Et{\gamma_i}$, 
	\begin{align}
	\label{eqn:isol_etmax_affine}
		\Etiso (\gamma_i) \coloneqq \Etthresh + \epsg \Et{\gamma_i}.
	\end{align}
	This threshold is set by experiment on a case-by-case basis, differing between studies of different processes.  The
	experimental cut applied to calorimeter cells is unfolded using detector simulations to an approximately equivalent
	fiducial cut on simulated partons. Motivated by experimental studies of diphoton production, we will consider
	$\epsg = 0$.

	Smooth-cone  isolation \cite{Frixione:1998jh} tightens this requirement. Rather than imposing a fixed threshold on
	the total hadronic transverse energy deposited within the cone, it imposes a threshold function on the radial profile of
	hadronic transverse energy deposited within the cone, requiring that
	\begin{align}
	\label{eqn:isol_smooth-cone}
		E_{\rT}^\text{had} (r) &\leqslant \Etiso (\gamma_i) \; \chi(r; R) &\forall r \leqslant R.
	\end{align}
	The function $\chi(r)$ may be chosen freely subject to the requirement that it vetoes exactly-collinear radiation,
	however soft, so that
	\begin{align}
	\label{eqn:isol_chi_r_to_0}
		\lim_{r\to 0} \chi(r; R) = 0.
	\end{align}
	It is typically additionally required to be continuous, monotonic, and such that $\chi(R;R) = 1$ on the boundary of
	the cone.
	We use the original choice introduced in \cite{Frixione:1998jh},
	\begin{align}
	\label{eqn:isol_profile_function}
		\chi(r; R)
		= \left(\frac{1 - \cos r}{1 - \cos R}\right)^n
		\equiv \left(\frac{\sin \frac{1}{2}r}{\sin \frac{1}{2}R}\right)^{2n}.
	\end{align}
	For $R \leqslant \frac{\pi}{2}$, this is approximately equal%
	\footnote{This holds to high precision, since
		\[\left(\frac{1 - \cos r}{1 - \cos R}\right)^n
		= \left(\frac{r}{R}\right)^{2n}
		\left(1 + \frac{n}{12}\left(R^2 - r^2\right) + \order{\left(\frac{R}{2}\right)^4}\right).
		\]
	}
	to the other profile function common in the literature,
	\begin{align}
		\chi(r; R) = \left(\frac{r}{R}\right)^{2n}.
	\end{align}

	Fixed-cone isolation corresponds to the constant profile function $\chi(r) \equiv 1$, which does not satisfy
	\cref{eqn:isol_chi_r_to_0}, and so is not a legitimate smooth-cone choice of $\chi$.  By permitting some amount of
	collinear radiation, fixed-cone isolation leads to a non-zero contribution from the fragmentation process, in
	contrast to smooth-cone profile functions which exclude it.

	For any two isolation schemes with matching $\Etiso$ and $R$ and profile functions $\chi_1 (r)$ and $\chi_2 (r)$, if
	\begin{align}
	\label{eqn:profile_fn_inequality}
		 \chi_1(0) &= \chi_2(0)
		 &\text{and}&&
		 \chi_1 (r) &\leqslant \chi_2 (r) \quad \forall r \leqslant R,
	\end{align}
	it follows that the permitted phase-space for the former is a subset of that for the latter, and so on physical
	grounds we expect that
	\begin{align}
	\label{eqn:dsigma_inequality}
		\rd\sigma_1 \leqslant \rd\sigma_2.
	\end{align}

	Hybrid isolation \cite{Siegert:2016bre} describes a family of profile functions which interpolate between
	smooth-cone isolation with a given profile function, and fixed-cone isolation.  It can be formulated as smooth-cone
	isolation with the profile function
	\begin{align}
	\label{eqn:hybrid_profile_function}
		\chi^\text{hyb}(r; R_d, R) =
		\begin{cases}
		E_1 \; \chi(r; R_d)	&r \in [0, R_d]  \\
		E_2				&r \in (R_d, R].
		\end{cases}
	\end{align}
	As in \cref{eqn:isol_etmax_affine}, $E_1$ and $E_2$ are, in general, affine functions of the photon transverse
	momenta.  For $E_1 \leqslant E_2$, this is equivalent to applying fixed-cone isolation on the cone $r \leqslant R$
	in addition to smooth-cone isolation on an inner cone $r \leqslant R_d$. For $E_1 > E_2$, these two formulations
	differ on the inner annulus  $r \in \left(R_\text{eff}, R_d\right]$ on which $\chi(r; R_d) > E_2/E_1$.  The latter
	formulation is then equivalent to a variant of the former, \cref{eqn:hybrid_profile_function}, with a smaller
	effective radius $R_\text{eff} < R_d$.  In the limit $R_d \to R$, hybrid isolation reduces to smooth-cone isolation
	with the profile function $\chi$, whilst the pointwise limit as $R_d \to 0$ corresponds to the fixed-cone profile
	function, except at $r=0$, where the former is 0 and the latter 1.

	This point is where photonic and partonic radiation are exactly collinear.  Fragmentation in QCD is a strictly
	collinear phenomenon, so these different values of the profile function at $r=0$ correspond to the formal exclusion
	or inclusion of the fragmentation contribution respectively. The quark-to-photon fragmentation function
	$D_q^\gamma (z, \mu_f)$
	contains a divergent and negative NLO
	mass-factorisation term, which compensates for the divergence that would
	otherwise arise from probing the quark-photon collinear limit, and so yields a finite cross-section for fixed-cone
	isolation upon integration.

	From \cref{eqn:dsigma_inequality,eqn:hybrid_profile_function} we can deduce that the hybrid isolation cross-section
	grows as $R_d$ decreases.  This is in accordance with the intuition that additional radiation is permitted within
	the isolation cone.  Because the fragmentation contribution is vetoed by the value of the profile function at
	$r=0$, the $R_d$ parameter acts as the sole regulator of the collinear quark-photon singularity, and we should
	expect the resulting dependence on $R_d$ to be logarithmic.
	It follows that there is some value of the parameter
	$R_d$ for which the hybrid cross-section and the fixed-cone cross-section must coincide, and the divergent
	cross-section
	of vetoed radiation in the inner-cone numerically matches that of the missing fragmentation counterterm.

	\subsection{Matched-hybrid isolation}
	\label{sec:isol_matchedhybrid}
	Throughout we chiefly consider matched-hybrid isolation, where we impose continuity at the boundary between the
	inner-cone and the outer annulus: $E_1 = E_2$. Other choices are discontinuous at $r = R_d$, which is expected to
	lead to instabilities.%
	\footnote{For matched-hybrid isolation, only the derivative $\chi'$ is discontinuous at
	$r=R_d$.  It is possible to define more sophisticated piecewise schemes which are arbitrarily smooth at $R_d$, and
	non-piecewise smooth-cone profile functions with similar properties to hybrid isolation, but we do not consider
	these alternatives further here.}
	In this scheme, when making experimental predictions, once the inner-cone
	profile function $\chi$ is chosen, the parameters $\Etiso$ and $R$ are fixed by the fiducial cuts of the
	experiment.  The only remaining unphysical parameter is then $R_d$, the radius of the inner cone.

	\begin{figure}[tbp]
		\centering
		\includegraphics[width=\textwidth]{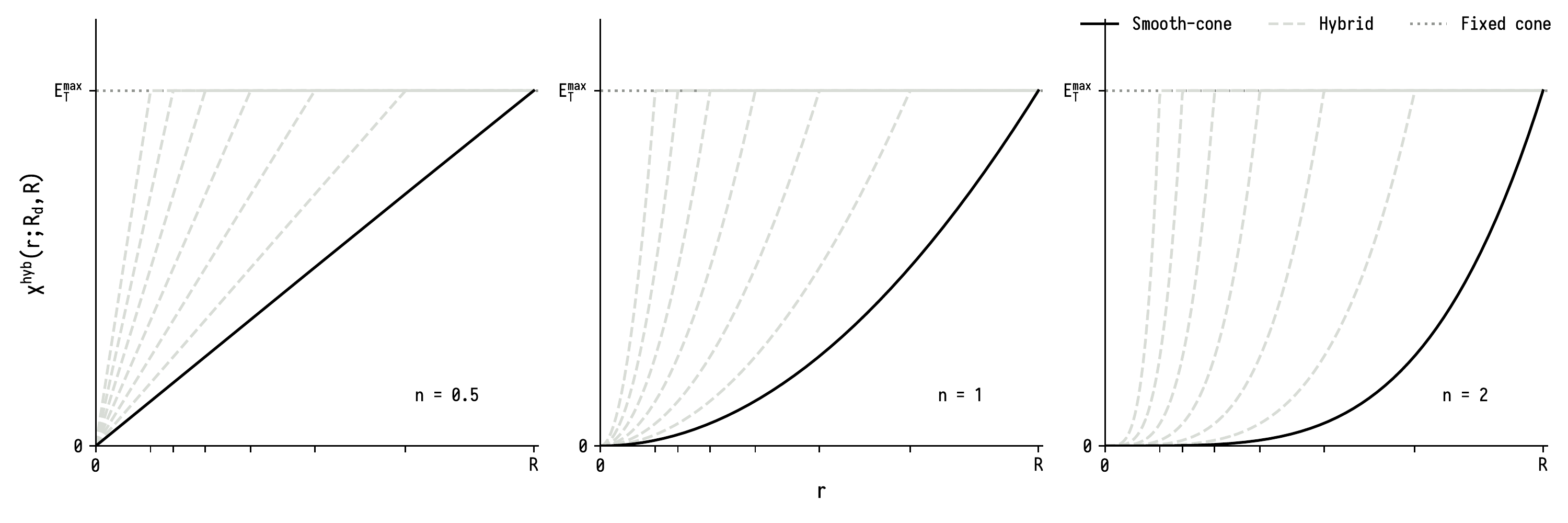}
		\caption{The matched-hybrid isolation profile function $\chi^\text{hyb}$ for $n\in \left\{\frac{1}{2}, 1, 2
		\right\} $ and several choices of the inner cone radius, $R_d$ (dashed).  As $R_d \to 0$ (dotted), the 
		smooth-cone (solid) suppression of
		the collinear singularity is retained, but the numerical deviation from the constant profile function of
		fixed-cone isolation is diminished.  For all values of $R_d$, exactly-collinear radiation is vetoed.}
		\label{fig:isol_chihyb}
	\end{figure}

	Since we are concerned with the physical criterion in \cref{eqn:dsigma_inequality}, we consider the
	hybrid-isolation cross-section relative to the corresponding smooth-cone prediction,
	\begin{align}
		\Delta\sigma \left(R_d\right) = \sigma_\text{hybrid} - \sigma_\text{smooth},
	\end{align}
	using the profile function of \cref{eqn:isol_profile_function}.  We can consider $\Delta\sigma \left(R_d\right)$
	as the physical cross-section resulting from the presence of the generalised isolation measurement function
	\begin{align}
		\Theta\left[\chi^\text{hyb}\left(\left\{p_{i}\right\};\, R_d\right) - E_{\rT}^\text{had} (R)\right]
		- \Theta\left[\chi^\text{smooth}\left(\left\{p_{i}\right\}\right) - E_{\rT}^\text{had} (R)\right].
	\end{align}
	in the integrand.  This is zero for, and hence vetoes, events that are treated commonly by the two isolation
	criteria, and, since $\chi^\text{hyb}(r; R_d, R) \geqslant \chi^\text{smooth}(r; R) $ selects those that are vetoed
	under smooth-cone isolation but permitted under hybrid isolation.  The Heaviside step functions implementing the
	isolation criteria induce discontinuities in the resulting distributions, which will be discussed further in
	\cref{sec:isol_irsensitivity}.

\begin{figure}[htbp]
		\centering
		\includegraphics[width=\textwidth]{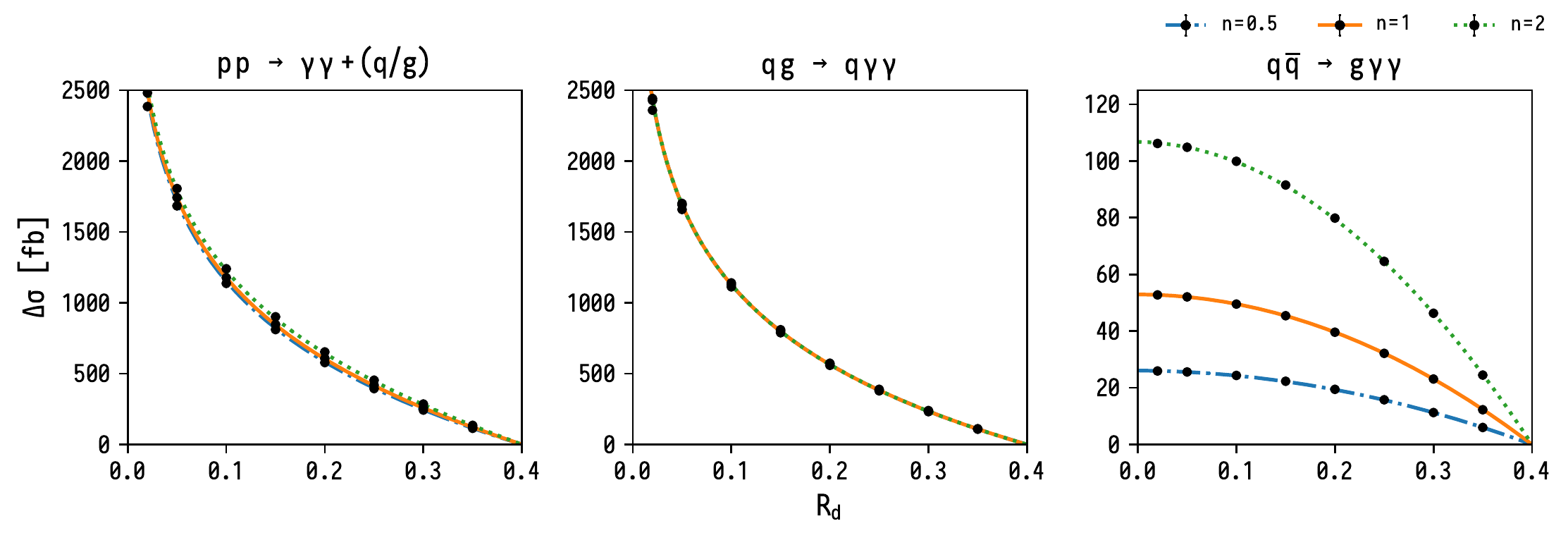}
		\caption{The variation $\Delta\sigma \left(R_d\right) = \sigma_\text{hybrid} - \sigma_\text{smooth}$ at NLO as
		a function of the inner-cone radius $R_d$, for $R=0.4$.  All other parameters are kept constant.  As expected,
		the gluon splitting gives rise to a quadratic dependence, whilst the quark splitting gives rise to a
		logarithmic divergence arising from the integrated collinear singularity.}
		\label{fig:isol_sigma_Rd_NLO}
	\end{figure}

	We begin by summarising the $R_d$-dependence of $\Delta\sigma \left(R_d\right)$, where other parameters are fixed,
	so $R$ and $\Etthresh$ are common to both profile functions.  Where a gluon is emitted inside the cone,
	\begin{align}
	\label{eqn:deltasigma_gluon}
	\Delta\sigma\left(R_d\right) \sim n (R^2 - R_d^2)
	\end{align}
	in accordance with the intuition that the additional cross-section allowed is proportional to the area over which
	the gluon can additionally be emitted, which is the difference in areas between the outer and the inner cone.
	Where a quark is emitted, the collinear behaviour of the splitting function gives
	\begin{align}
	\label{eqn:deltasigma_quark}
	\Delta\sigma\left(R_d\right) \sim \log\frac{R}{R_d}.
	\end{align}
	This behaviour is verified at NLO in \cref{fig:isol_sigma_Rd_NLO}.  The dependence of the inner-smooth-cone
	cross-section on its remaining isolation parameters is unchanged from the detailed description in
	\cite{Catani:2018krb}, whilst the $R$-dependence of the outer cone is that of fixed-cone isolation as described in
	\cite{Catani:2013oma}.

	The $\log {R}/{R_d}$ scaling of \cref{eqn:deltasigma_quark} indicates that the cross-section diverges in the
	small-inner-cone limit, as can be seen in \cref{fig:isol_sigma_Rd_NLO}, and as expected from the above
	discussion.   This is a known feature of narrow-cones in both smooth- and fixed-cone isolation
	\cite{Catani:2002ny}.  It arises because the partition of phase-space into a cone of radius $R$ and its complement
	induces $\log R$ contributions in both, which cancel in their sum.  Any isolation procedure applied only inside the
	photon cone changes the former but not the latter, leading to a miscancellation of logarithms, the remainder of
	which will become large in the small-$R$ limit.

	We must therefore be careful to choose a value of $R_d$ that is large enough to regulate the collinear singularity,
	but small enough to approximate the fixed-cone result better than the smooth-cone value $R_d = R$.
	Ideally, this would be approximately equal to that at which the compensation that was discussed above occurs, to
	reproduce the cross-section given by fixed-cone isolation.

	To determine the value of $R_d$ at which this compensation occurs, in \cite{Amoroso:2020lgh} we compared NLO
	cross-sections and differential distributions obtained at fixed order with hybrid isolation to those obtained using
	\Diphox with fixed-cone isolation.  \Diphox \cite{Binoth:1999qq} is a Monte Carlo event generator implementing the
	NLO QCD corrections to diphoton production, together with the single- and double-fragmentation contributions.  We
	found that for the ATLAS-motivated cuts $\Etthresh = 11~\GeV$, $R = 0.4$ and $R_d = 0.1$ the hybrid isolation
	result was almost fully contained within the \Diphox uncertainty band, except where the fragmentation contribution
	populated regions of phase-space that first enter the fixed-order calculation at the subsequent order of
	perturbation theory.

	At NLO the underlying kinematics restrict the relevance of photon isolation to a relatively minor region of
	phase-space.  The only part of the fixed-order NLO calculation sensitive to the isolation parameters is the real
	emission, and within the real contribution, the final state parton $p_1$ may only enter the isolation cone of the second-hardest
	photon, as they must together balance $\vecptg{1}$.  The collinear invariant being regulated by the isolation
	criterion is therefore
	\begin{align}
		s_{\gamma_2 p_1}
		&\approx \Et{\gamma_2} \, \Et{1} \; \dR_{\gamma_2 p_1}^2
		= \Et{\gamma_2} \, \ptgg \; \dR_{\gamma_2 p_1}^2,
	\end{align}
	where
	\begin{align}
		\ptgg = \left\| \vecptg{1} + \vecptg{2} \right\|
	\end{align}
	is the transverse momentum of the diphoton system, and the last equality is valid only for three-particle
	final-states.

	For any monotonic profile function $\chi$, it follows from \cref{eqn:dsigma_inequality} that the resulting
	isolation criterion is at least as restrictive as fixed-cone isolation with the same boundary condition, so the
	effect of isolation will be confined to $\ptgg = \Et{1} \leqslant \Etiso (\gamma_2)$ purely from kinematic
	constraints.%
	\footnote{As a consequence, for fixed radius $R$ we would expect the constraints imposed by
	unitarity to force a larger choice of $R_d$ for more restrictive isolation thresholds $\Etthresh$, and to permit a
	smaller choice for less strict threshold energies.}
	This implies that any differences between two isolation schemes are only resolved at this order on the strip
	\begin{align}
	\label{eqn:NLO_isolkin}
		\ptg{2} \in
			\left[
				\max\left\{
					\ptg{2,\cut}, \frac{\ptg{1} - \Etthresh}{1+\epsg}
				\right\}, \;
			\ptg{1}
			\right].
	\end{align}
	For asymmetric photon cuts with a $\pt^{\cut}$-gap greater than $\Etthresh$, this would exclude events close to the
	threshold of the photon cuts from isolation dependence entirely, at this order.  For the more conventional case, the
	dependence of the NLO cross-section on the parameters is dominated by events on the threshold of the cuts.

\begin{figure}[tbp]
		\centering
		\includegraphics[width=0.45\textwidth]{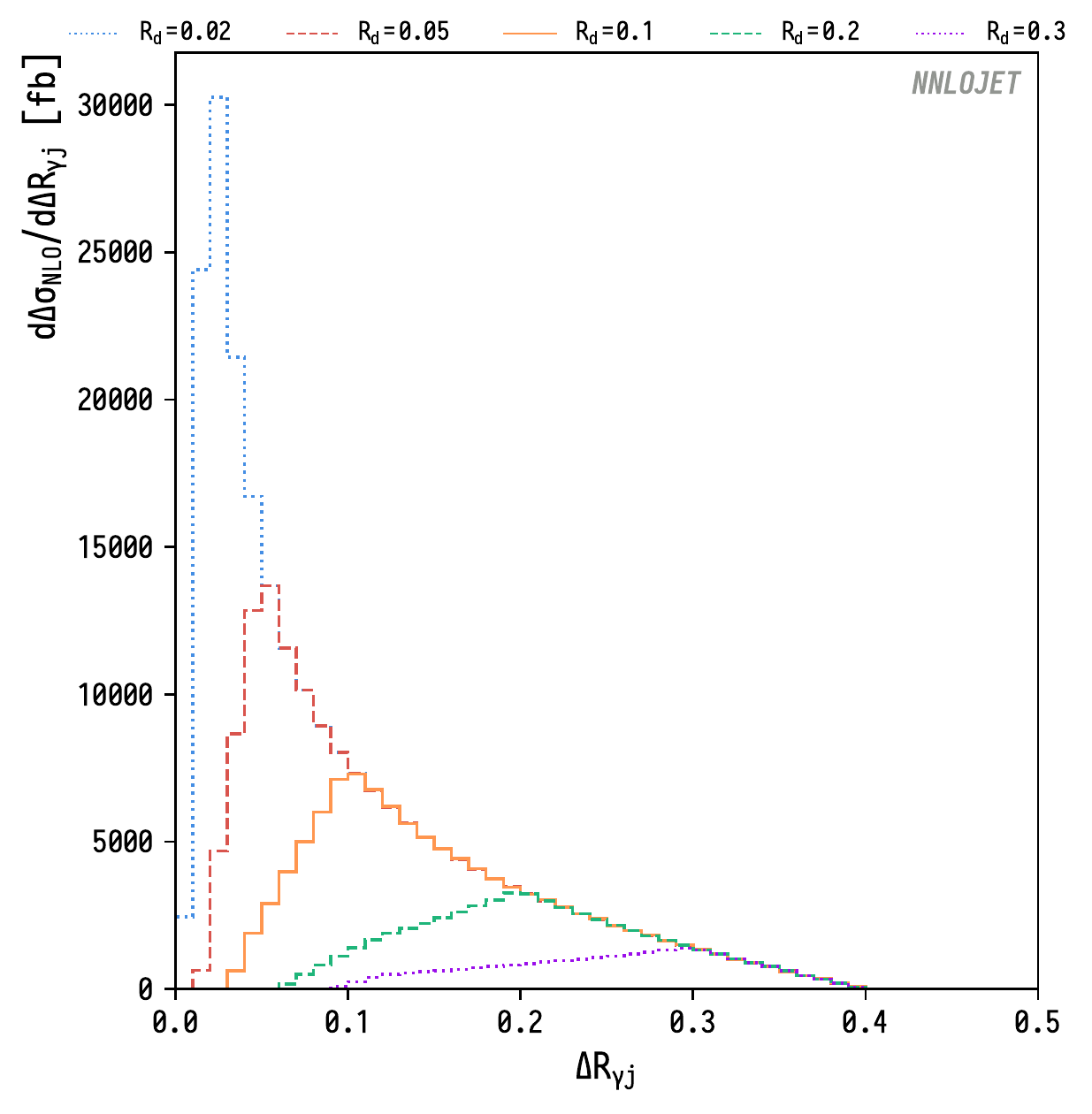}
		\includegraphics[width=0.45\textwidth]{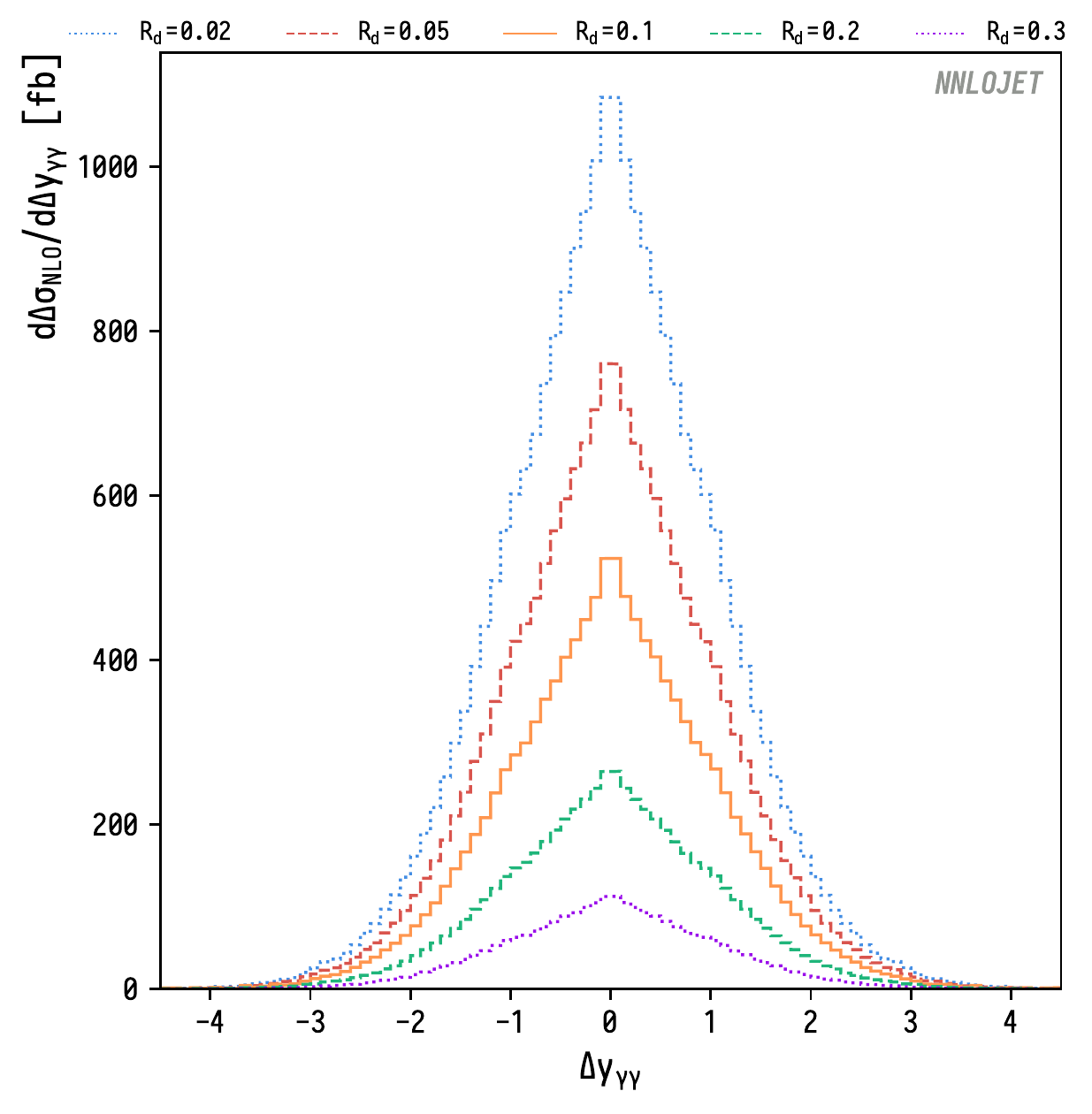}
		\caption{Isolation cone effects at NLO, showing the difference between matched-hybrid and smooth-cone isolation
		$\Delta\sigma$.  The $\rd \Delta\sigma / \rd\Delta R_{\gamma j}$ distribution has regions of highly-local
		sensitivity to $R_d$, whilst the $\rd \Delta\sigma / \rd\Delta y_{\gamma\gamma}$ distribution is sensitive only
		through a small global normalisation.   In the first plot the jet cut is 1 GeV; at this order all jets comprise
		a single parton.  Higher values of the jet cut increase the minimal value of $\Delta R_{\gamma j}$ at which
		partons of that $\pt$ can be emitted and not vetoed, with the minimum approximately  $ R \sqrt{\ptj{} / \Etiso}
		$, leading to steeper slopes to the left of the peak.}
		\label{fig:isol_hybrid_Rd_dists_NLO}
	\end{figure}

\begin{figure}[htbp]
	\centering
		\includegraphics[width=0.48\textwidth]{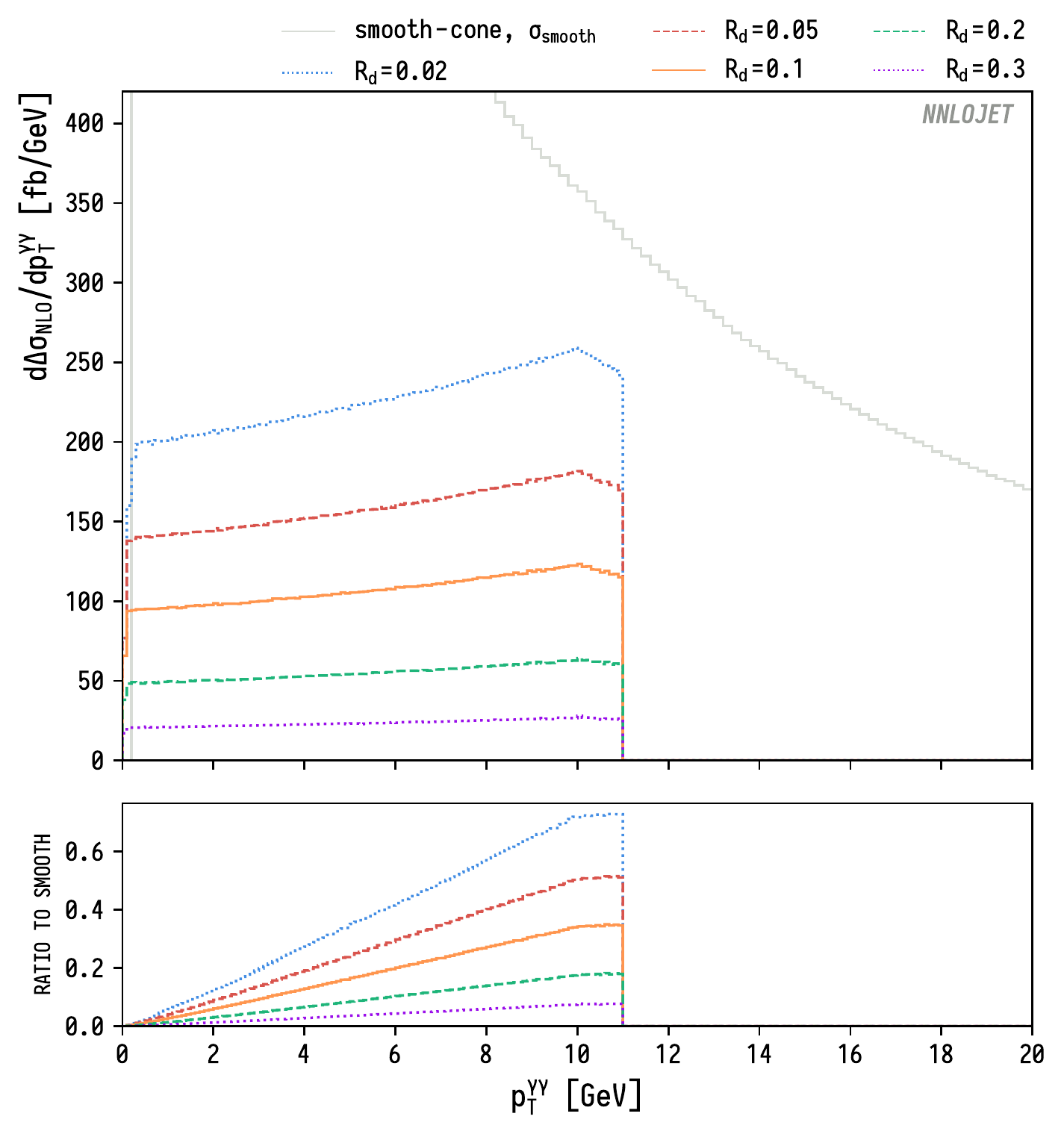}
		\includegraphics[width=0.48\textwidth]{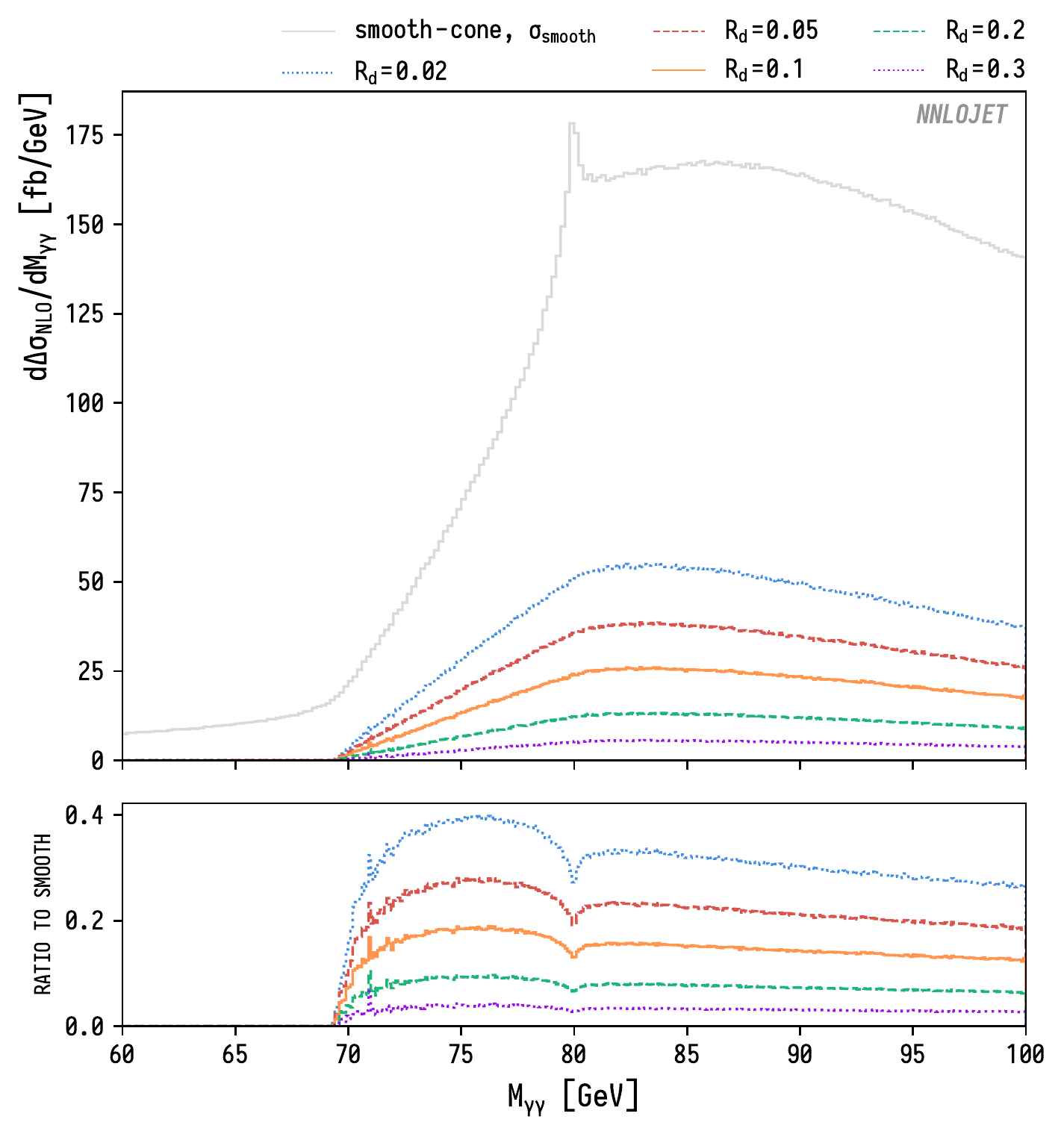}
		\includegraphics[width=0.48\textwidth]{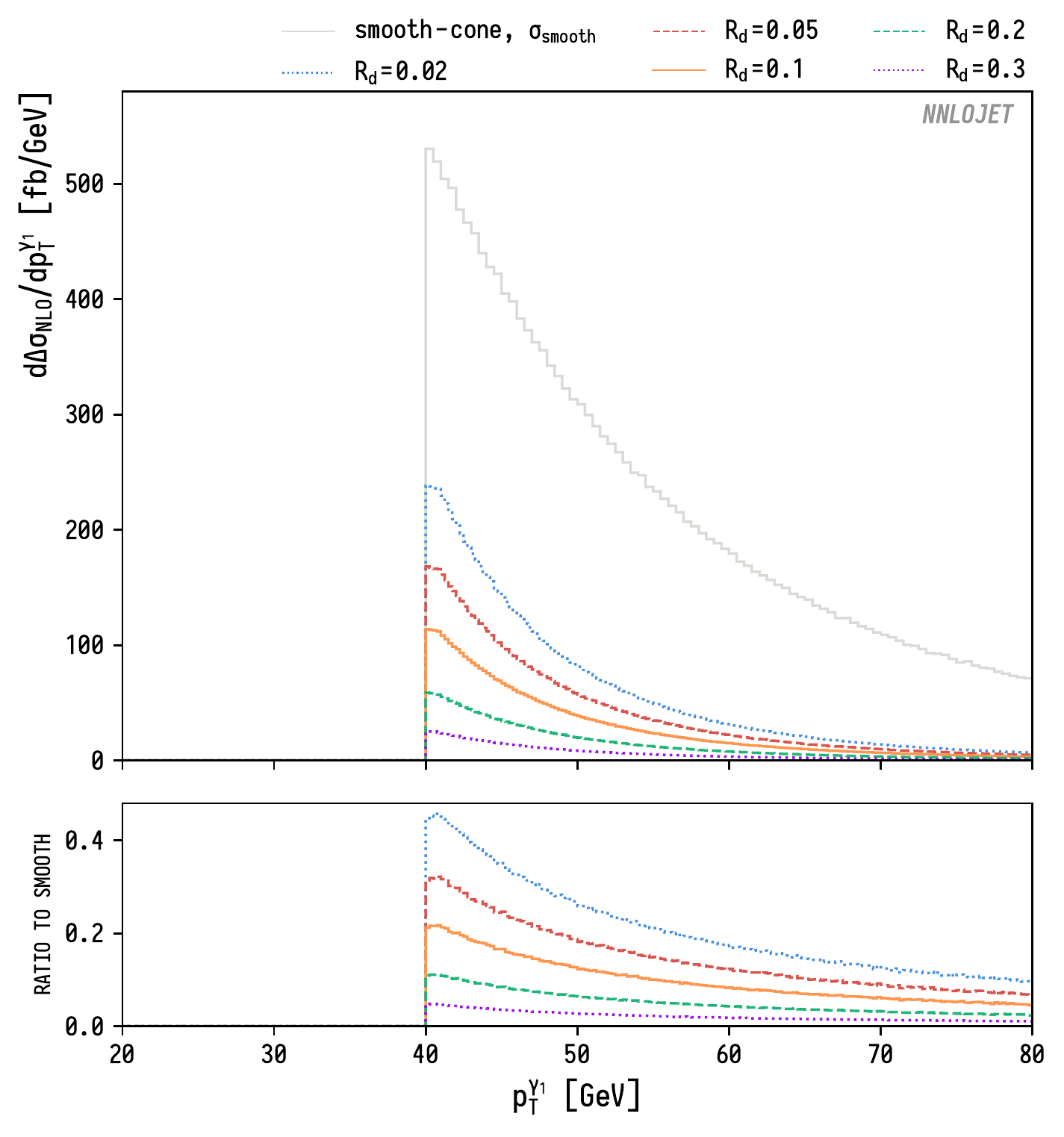}
		\includegraphics[width=0.48\textwidth]{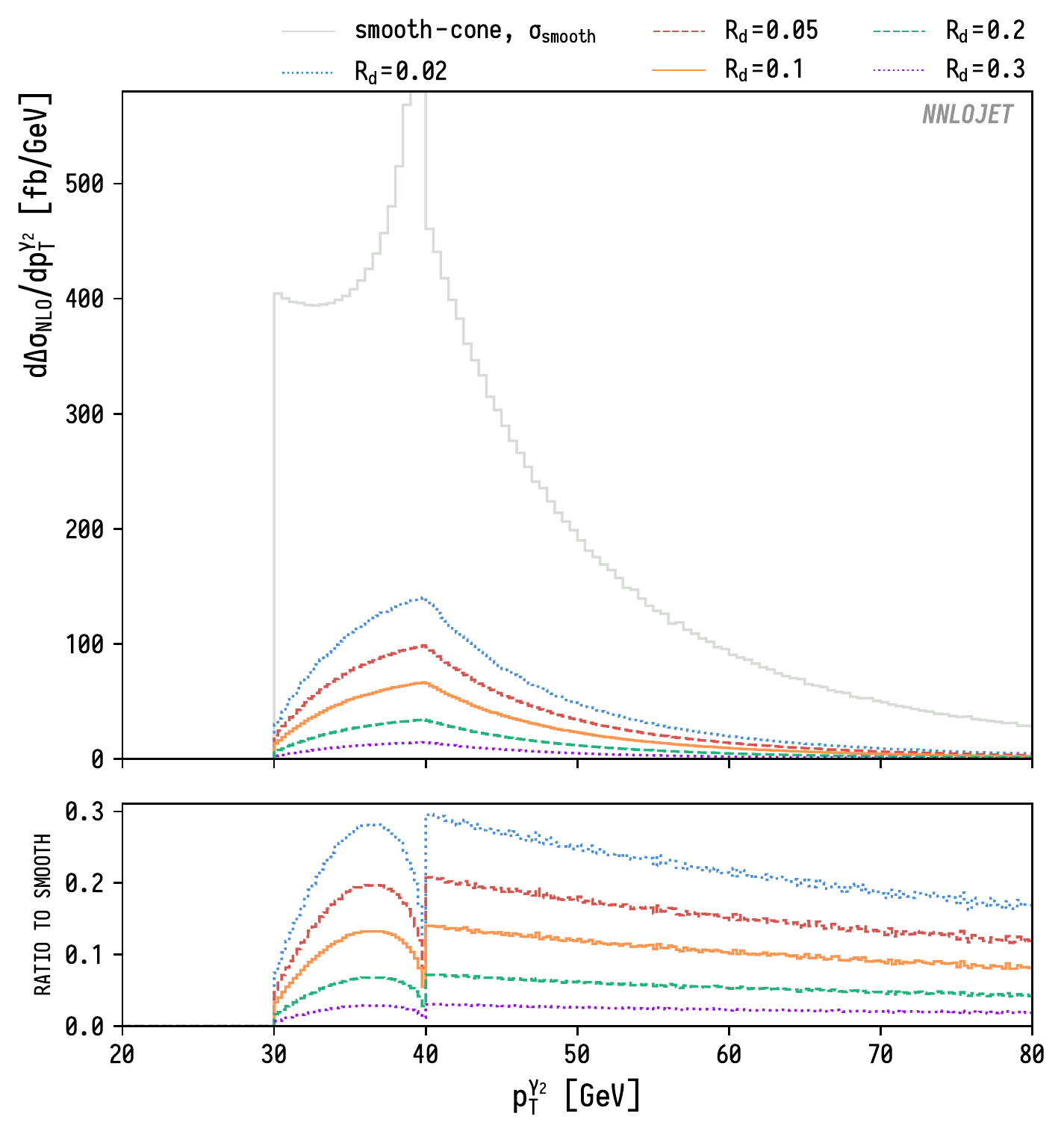}
		\caption{Detailed isolation cone effects at NLO, showing the difference between matched-hybrid and smooth-cone
		isolation $\Delta\sigma$.
		The absolute predictions for smooth-cone isolation are shown for reference.
		At this order, isolation criteria only apply at all in the limited region of
		phase-space defined by $\ptgg \leqslant \Etiso$.  Here, as for the ATLAS 8~TeV data considered throughout,
		$\Etiso=11~\GeV$.
		}
		\label{fig:hybrid_Rd_dists_NLO_rat}
	\end{figure}

	In \cref{fig:isol_hybrid_Rd_dists_NLO,fig:hybrid_Rd_dists_NLO_rat} we show a selection of differential
	cross-sections $\rd\Delta\sigma (R_d) / \rd X$ for a range of values for
	$R_d$ to illustrate the distributional counterparts to \cref{fig:isol_sigma_Rd_NLO}.  As in
	\cref{fig:isol_sigma_Rd_NLO}, the cuts are chosen to match those used in the 8~TeV ATLAS study, whilst the theory
	parameter $R_d$ is varied.

	As can be seen in the first plot of \cref{fig:isol_hybrid_Rd_dists_NLO}, the characteristic kinematic configuration
	of the events additionally allowed by hybrid isolation is very sensitive to the choice of $R_d$.  The peak at $R_d$
	arises because the difference between the smooth-cone and hybrid profile functions is maximised at $R_d$. This
	leads to a localised sensitivity to the $R_d$ parameter in certain distributions.  This exposure of the collinear
	singularity shown in \cref{fig:isol_hybrid_Rd_dists_NLO} with decreasing $R_d$ illustrates the kinematics
	underlying the logarithmic behaviour of \cref{eqn:deltasigma_quark}, and shows a gradual bias within the
	photon-cone towards increasingly collinear events as the inner-cone is reduced in size.  In other distributions
	such as $\rd \Delta\sigma / \rd\Delta y_{\gamma\gamma}$, also shown in \cref{fig:isol_hybrid_Rd_dists_NLO}, the
	logarithmic behaviour manifests itself only as a global normalisation.

	Further distributions in which the effect is localised are shown in \cref{fig:hybrid_Rd_dists_NLO_rat} alongside
	the corresponding smooth-cone distributions.  These illustrate interesting features of the isolated differential
	cross-sections at NLO.  In the first figure, the $\rd \Delta\sigma / \rd \ptgg$ distribution shows a discontinuity
	at $\ptgg = \Etiso$.  The shape of this distribution is sensitive to the parameters of hybrid isolation and the
	offset between asymmetric photon cuts.  Here, the peak occurs at the offset whilst the discontinuity occurs at
	$E_2$, in the notation of \cref{eqn:hybrid_profile_function} (including for non-matched isolation).  If $E_2$ were
	allowed to depend on $\ptg{2}$ this discontinuity would be smoothed over an interval in $\ptgg$, but would reappear
	in another distribution.  This arises directly from the boundary of the fixed-cone criterion in phase-space and
	will be discussed further, including its consequences for higher-orders, in \cref{sec:isol_irsensitivity}.

	The $\rd \Delta\sigma / \rd \ptg{2}$ distribution, and as a direct consequence, the $\rd \Delta\sigma / \rd \Mgg$
	distribution, show discontinuities, in the differential cross-section and its derivative respectively, at the
	boundaries of the Born phase-space.  The latter was analysed in \cite{Catani:2018krb}.  The former arises because
	real soft QCD radiation is kinematically restricted to arise only close to the back-to-back configuration $\ptg{2}
	\lesssim \ptg{1}$, which is permitted by the isolation criteria by design, and cannot cancel as anticipated against
	virtual poles outside the Born kinematics.%
	\footnote{The kinematic prohibition of these soft emissions is the
	underlying mechanism for the unphysical dependence of $\sigma_\NLO$ on the $\ptg{}$ cuts when moving from
	asymmetric to symmetric cuts, first remarked upon in the context of jet production in \cite{Frixione:1997ks}.  This
	can clearly be seen from the lower-right plot in \cref{fig:hybrid_Rd_dists_NLO_rat}.}

	These (unphysical) features arise commonly in both smooth-cone and fixed-cone isolation.  They are a direct
	consequence of the requirement that soft gluon radiation be permitted, to allow the general cancellation of real
	and virtual singularities.  Where the virtual singularities are kinematically prohibited, but real soft
	singularities are not, a miscancellation arises.

	\begin{figure}[htbp]
		\centering
		\includegraphics[width=\textwidth]{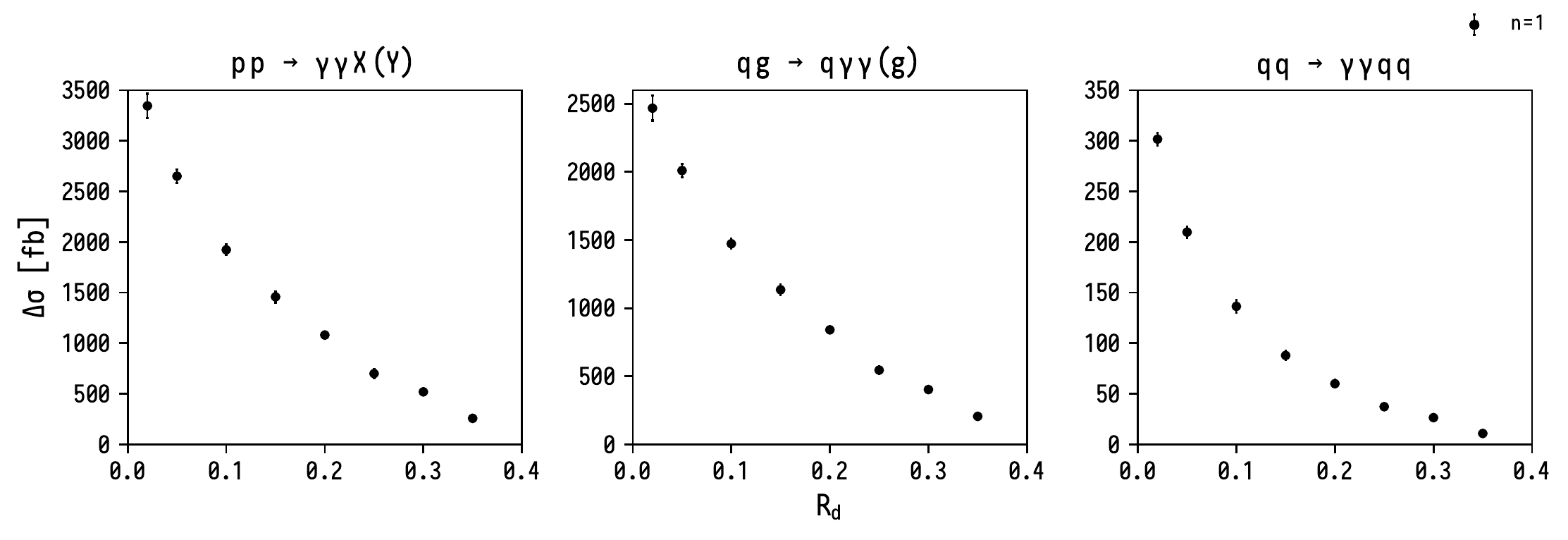}
		\caption{The variation $\Delta\sigma \left(R_d\right) = \sigma_\text{hybrid} - \sigma_\text{smooth}$ at NNLO as
		a function of the inner-cone radius $R_d$, for $R=0.4$.  All other parameters are kept constant.  The two
		channels not shown, $\Pq\Paq$ and $\Pg\Pg$, have comparable shape (but smaller magnitude) to $\Pq\Pg$ and
		$\Pq\Pq$ respectively.}
		\label{fig:isol_sigma_Rd_NNLO}
	\end{figure}

	Since the behaviour of the isolated cross-section at NLO is highly sensitive to the unphysical behaviour in these
	regions, it is \emph{a priori} unclear to what extent the variation of isolation parameters based on NLO behaviour
	will lead to conclusions that hold at higher orders.  Running enough calculations at NNLO with sufficient
	resolution to investigate the $R_d$-dependence of distributions in the regions of non-analyticity shown in
	\cref{fig:hybrid_Rd_dists_NLO_rat} would be prohibitively computationally expensive.  In \cref{sec:isol_comparison}
	we therefore compare smooth-cone isolation to matched-hybrid isolation with fixed $R_d = 0.1$.

	To illustrate the overall dependence of the NNLO cross-section on $R_d$, in \cref{fig:isol_sigma_Rd_NNLO} we show
	the NNLO counterpart to \cref{fig:isol_sigma_Rd_NLO}.  The dependence on $R_d$ is again dominated by the $\Pq \Pg$
	channel.  Overall, the magnitude of the effect is similar to that at NLO despite the contribution from events
	outside the strip of \cref{eqn:NLO_isolkin}, whilst the shape is no longer logarithmic.  Channels in which a parton
	is permitted to enter the photon cone for the first time at NNLO have the same $R_d$-dependence shown in
	\cref{fig:isol_sigma_Rd_NLO}.  This suggests that the procedure used to justify the choice $R_d = 0.1$ above, by
	comparison to the fragmentation calculation, should remain valid at NNLO.  
	
	The $R_d$-dependence illustrated in \cref{fig:isol_sigma_Rd_NLO,fig:isol_sigma_Rd_NNLO} can be used to estimate the 
	uncertainty resulting from the choice of $R_d = 0.1$.  This uncertainty does not arise directly from hybrid 
	isolation, but rather from the inherent uncertainty in the freedom to choose a profile function within generalised 
	smooth-cone isolation.
	The hybrid-isolation profile function parametrises the local regulation of the collinear singularity directly, 
	allowing this uncertainty to emerge from parameter variation, unconstrained by the properties of the profile 
	function far from the collinear point $r=0$.
	Twofold variation in $R_d$ about $R_d = 0.1$ gives an uncertainty band of approximately 1000--1500\fb for both NLO 
	and NNLO, smaller than that induced by scale variation at the corresponding order and so of comparable magnitude
	to the general theory uncertainty we assign to the calculation.
	
	The natural comparison of these isolation uncertainties is with the fragmentation uncertainty that would arise 
	within a fragmentation-inclusive calculation in which the collinear singularity is absorbed via a mass 
	factorisation counterterm. This will eliminate the $R_d$-dependence of the hybrid-isolation predictions, but at the 
	expense of introducing a sensitivity to photon fragmentation functions, which are only loosely constrained by 
	current experimental data. Such a comparison cannot yet be made at NNLO, and must be postponed until such a 
	calculation has been completed. As at NLO, we expect that the NNLO calculation using hybrid isolation should 
	coincide with the fragmentation-inclusive calculation at the same order for some finite positive value of $R_d$, 
	and its associated uncertainty can be assessed by a variation around this value. 
	
	In practice we limit the exposure of the hybrid-isolation cross-section to the $R_d\to 0$ limit through 
	the `perturbative unitarity' heuristic argument outlined above, using the constraints imposed by fragmentation at 
	NLO. Since the $R_d$-dependence at NNLO seems to be dominated by the NLO real-radiation effects, we can reasonably 
	expect this heuristic argument to limit the possible size of deviation between fragmentation-inclusive and 
	hybrid-cone isolated cross-sections just as it does at NLO, and to approximate the uncertainty inherent to the 
	limited knowledge on the fragmentation process.

	\subsection{Infrared sensitivity}
	\label{sec:isol_irsensitivity}

	In general, any parton-level cone-based isolation criterion of the generic form \cref{eqn:isol_smooth-cone} amounts
	to a veto implemented through a measurement function containing factors of the form
	\begin{align}
		\prod_{\gamma} \prod_{i=1}^{n} \; \mathcal{I}_{\gamma i},
	\end{align}
	where the index $i$ ranges over final-state partons, and
	\begin{align}
	\label{eqn:measurement_function}
		\mathcal{I}_{\gamma i} =
		\Theta\left[\Etiso (\gamma) \; \chi\left(\min\left(\Delta R_{\gamma i}, R\right); R\right)
		- \sum_{j=1}^{n} \Et{j} \; \Theta \left[ \min\left(\Delta R_{\gamma i}, R\right) - \Delta R_{\gamma j} \right]
		\right]
	\end{align}
	(using the $\Theta(0) = 1$ convention).  This is zero, and hence vetoes events, in which the accumulated partonic
	energy in the cone exceeds the profile
	function.

	It can readily be seen from this formalism that the Heaviside step function implies a discontinuity in the
	integrand at the bounding surface on which the isolation criteria inequalities are exactly saturated.  This is an
	intrinsic property of veto-based isolation techniques.  At NLO, where there is a single parton that can only enter
	the photon cone of the softer photon, the consequences of this become clearer:
	\begin{align}
	\label{eqn:measurement_function_NLO}
		\mathcal{I}_{\gamma 1} =
		\Theta\left[\Etiso (\gamma) \; \chi\left(\min\left(\Delta R_{\gamma 1}, R\right); R\right)
		- \Et{1} \; \Theta \left[ \min\left(\Delta R_{\gamma 1}, R\right) - \Delta R_{\gamma 1} \right] \right]
	\end{align}
	That is, we expect to have introduced a step-like discontinuity inside the physical region at
	\begin{align}
		\ptgg \equiv \Et{1}
		= \Etiso (\gamma)  \; \chi\left(\Delta R_{\gamma 1}; R\right) \qquad \forall \Delta R_{\gamma 1} \leqslant R
	\end{align}
	where the integrand is zero for $\ptgg > \Etiso (\gamma) \; \chi\left(\Delta R_{\gamma 1}; R\right)$ and
	non-zero below it,
	through the formulation of the isolation criterion.  This is precisely the source of the discontinuity visible in
	\cref{fig:hybrid_Rd_dists_NLO_rat}.  For the complementary region $\Delta R_{\gamma 1} > R $ where the
	parton is outside the cone, there is no	discontinuity in the integrand: the measurement function
	\cref{eqn:measurement_function_NLO} is never zero, and so the
	isolated and unisolated integrands are identical everywhere.  Conversely, examining instead the region defined by
	$\ptgg \geqslant\Etiso(\gamma)$ we see the same step-like discontinuity arising at $\Delta R_{\gamma 1} = R$, the
	boundary of the isolation cone.

	Such discontinuities within the physical region were first described in general in \cite{Catani:1997xc}.  For
	diphoton production they first arise in the NLO-plus-fragmentation calculation and were remarked upon in
	\cite{Binoth:1999qq}, but have not previously been identified in the NNLO direct production calculation.  They
	represent a localised breakdown of perturbation theory in which a step-like discontinuity leads at higher orders to
	infrared Sudakov singularities. These arise from the disruption of the expected cancellation between soft real
	gluons and the corresponding virtual corrections, since isolation vetoes a subset of the former without affecting
	the latter.  Resummation of the generated logarithms is then expected to restore continuity of the distribution,
	resulting in a characteristic `Sudakov shoulder'.  Following the logic outlined in \cite{Catani:1997xc}, the
	step-like isolation behaviour shown in \cref{fig:hybrid_Rd_dists_NLO_rat} leads to a double-logarithmic divergence
	in the region $\ptgg < \Etthresh$,
	\begin{align}
		\Delta^\pm_{\ptgg=\Etthresh} \sim - \ln ^{2}\left[1-\left(\frac{\ptgg}{\Etthresh}\right)^2\right].
	\end{align}

\begin{figure}[tbp]
		\centering
		\includegraphics[width=0.49\textwidth]{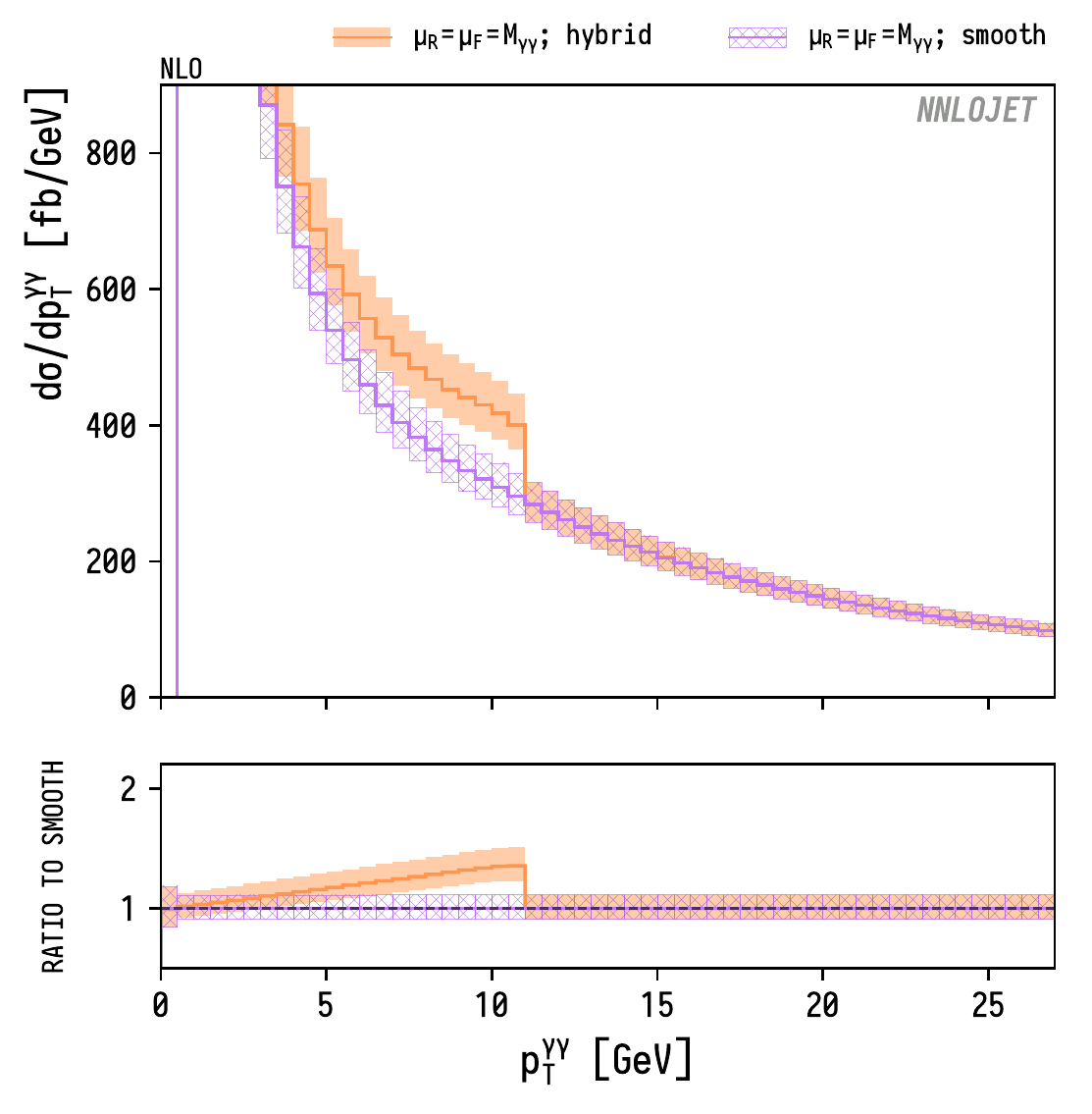}
		\includegraphics[width=0.49\textwidth]{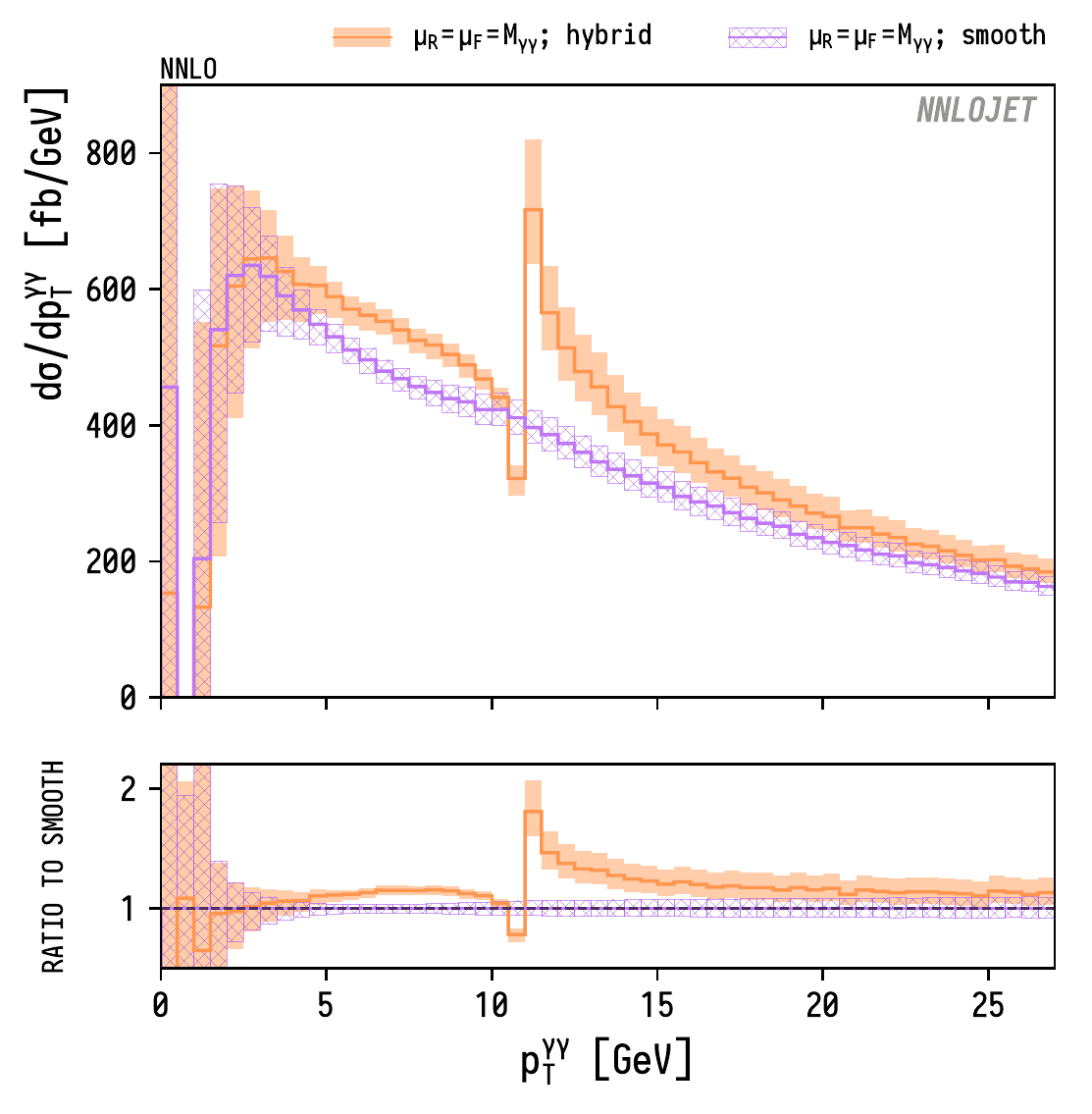}
		\caption{Discontinuity in the $\rd \sigma / \rd \ptgg$ distribution arising from hybrid isolation at NLO with
		$\Etthresh = 11 \GeV$, and the resulting Sudakov singularity at NNLO.}
		\label{fig:isol_infrared_ptgg}
	\end{figure}

	This behaviour does indeed arise in the NNLO $\rd \sigma / \rd \ptgg$ distribution as expected.  It is shown
	alongside the corresponding NLO discontinuity in \cref{fig:isol_infrared_ptgg}, together with the corresponding
	(continuous) smooth-cone distribution.  The distinctive double-singularity shape of the hybrid-isolation
	distribution is as anticipated in \cite{Catani:1997xc}, and represents a clear deviation from the expected
	behaviour of the hybrid-isolation distribution on physical grounds from \cref{eqn:dsigma_inequality}.

	There is an additional Sudakov critical point arising from the boundary of the Born kinematic region at $\ptgg = 0$
	which would also be expected to require resummation to generate reliable predictions.  The practical effect of this
	additional singularity at small $\ptgg$ is therefore to revise upwards the lower boundary of the region of the
	$\ptgg$-distribution at which we might expect NNLO calculations to accurately describe the data.  For current
	experimental binnings, this effect is negligible.  The singularities are integrable, and the positive and negative
	logarithmic contributions typically cancel against each other in a single bin that contains the critical point.
	However, as the target precision of both experimental data and theoretical predictions increases, these effects may
	not remain negligible, especially if a bin-edge coincides with the Sudakov critical point.

	We briefly remark on the second discontinuity implied by \cref{eqn:measurement_function_NLO}, in the $\dRgj$
	distribution.  The NLO isolation function \cref{eqn:measurement_function_NLO} implies a discontinuity in $\Delta
	R_{\gamma 1}$ at the boundary of the isolation cone.   At NLO, where each identified jet comprises a single parton,
	this would lead to a discontinuity in a $\Delta R_{\gamma_2 j_1}$ distribution, were the jet definition set small
	enough to allow partons  to be simultaneously soft enough to be permitted inside the cone by isolation, and hard
	enough to be identified as a jet.  The obvious tension between these two conditions makes this a theoretical,
	rather than a phenomenological concern.  At NNLO, however, the possibility arises for partons soft enough to be
	permitted inside the cone by the isolation criteria to be combined with harder partons outside the cone, resulting
	in a jet with $ \Delta R_{\gamma j}  > R$.  The underlying discontinuity at one order and the resulting Sudakov
	singularities at the next order would then be displaced relative to one another, and would resemble a new
	phenomenon of unclear origin.  These boundary effects can be expected to lead to unphysical results in any
	fixed-order prediction of photon-jet separation.

	At NLO, the nature of the isolation-induced discontinuity shown in \cref{fig:hybrid_Rd_dists_NLO_rat} is specific
	to hybrid- and fixed-cone isolation with $\epsg = 0$.  The surface defined in \cref{eqn:measurement_function_NLO}
	is a surface of constant $\ptgg$, and hence the discontinuity introduced into the integrand remains in the $\rd
	\sigma / \rd \ptgg$ distribution, and at higher orders gives rise to a Sudakov critical point.  More generally, for 
	$\epsg = 0$ a discontinuity in the $\ptgg$-distribution arises from any interval on which $\chi(r; R)$ is constant.

	The discontinuity is fully regulated in smooth-cone isolation in NLO kinematics, since the boundary in $\ptgg$ at
	which the discontinuity would arise is no longer a constant $\Etthresh$, but a monotonic function of $r$, and the
	threshold of permitted events is spread evenly across $\ptgg$ rather than discretely at a boundary.  This masks the
	IR critical point and gives a smooth $\ptgg$ distribution.  However, it instead introduces one into the $\ptgg /
	\chi(r)$ distribution.

	Within hybrid isolation, continuity can be restored to the $\ptgg$-distribution by, for example, introducing a
	small non-zero $\epsg$.  This amounts to a rotation of the boundary surface, and moves the discontinuity from the
	$\ptgg$-distribution into the $(\ptgg - \epsg \ptg2)$ distribution.  The resulting Sudakov singularities in the
	latter distribution then manifest themselves in the $\ptgg$ distribution as an unphysical bump resulting from the
	remainder of the cancellation of positive and negative Sudakov logarithms in each bin.

	These discontinuities, and the resulting singularities, are therefore a necessary consequence of cone-based
	isolation, and can only be moved between distributions, rather than avoided entirely.  The effect of the logarithms
	is not confined to the distribution that is discontinuous at a lower order, but can leak into correlated
	distributions, where it may be harder to identify.

	In general, any observable whose definition is constructed to align with the axis of the step-function will exhibit
	this threshold behaviour.  Where this coincides at a lower order with an observable of physical interest, it is
	likely to lead to infrared sensitivity.  For sufficiently wide histogram bins (including those used for the ATLAS 8
	TeV data), the integrable singularities are masked, whilst binnings that combine both critical points, at $\ptgg=0$
	and $\ptgg = \Etthresh$ into a single bin disguise both Sudakov critical points entirely, as in fig. 13 of
	\cite{Catani:2018krb}.

	Given this, it appears that the phenomenological significance of these singularities is limited, provided that
	deviations from fixed-order predictions in these regions are not misunderstood to have physical significance.  This
	is easier to recognise in distributions such as $\ptgg$ that are directly constrained by photon isolation than it
	might be where the analogous observable is not of direct physical interest.  This is the case, for example, for the
	photon-plus-jet process, where different experimental cuts and attention to different observables change the
	relevance of the expected non-analytic behaviour of $\ptgj$.

	However, for colourless final-states including the diphoton final state, the differential cross section with
	respect to the transverse momentum of the identified final state has particular significance, as it is relied upon
	by alternative subtraction schemes such as $q_\rT$- or $N$-jettiness subtraction.  It is clear from
	\cref{fig:isol_infrared_ptgg} that the $\ptgg$-dependence of the cross-section at small $\ptgg$ is sensitive to the
	details of the isolation used and not universal, which would explain the absence of a plateau in the
	$r_\cut$-dependence plots for diphoton production using $q_\rT$-subtraction with \Matrix in
	\cite{Grazzini:2017mhc}.  These power corrections have been explored analytically in \cite{Ebert:2019zkb}, where it
	was found that for smooth-cone isolation, they grow in magnitude as $\left({Q}/{\Etiso}\right)^{{1}/{n}}$, with a
	proposal for how they could be accounted for.  As a result, the phenomenological significance of these power
	corrections should grow as we move to higher centre-of-mass energies.  It remains to be seen whether they will pose
	a meaningful problem for these alternative subtraction schemes.

	\subsection{Comparison of hybrid and smooth-cone distributions}
	\label{sec:isol_comparison}

	We now outline the key differences of phenomenological significance between hybrid and smooth-cone isolation, as
	applied to a selection of differential cross-sections.  We use a setup corresponding to the ATLAS 8~TeV data
	\cite{Aaboud:2017vol}, which we will return to in \cref{sec:combvar_fourway}, and plot data for those distributions
	where it exists for later reference.	The relevant fiducial cuts are:
	\begin{subequations}
	\label{eqn:ATLAScuts}
	\begin{align}
		\ptg{1} &> 40 \;\GeV \,, 	&	\ptg{2} &> 30 \;\GeV \,,  \\
		\dRgg &> 0.4 \,, & \absyg &\in \left[ 0, 1.37 \right) \cup \left( 1.56, 2.37 \right), \\
		\Etisopart &< 11 \;\GeV  & \text{within cone}\enspace  \dR &\leqslant 0.4.
	\end{align}
	\end{subequations}
	We choose $R_d=0.1$ for hybrid isolation, as outlined in \cref{sec:isol_matchedhybrid}, and smooth-cone isolation parameters
	$n=1$ and $\Etiso = 11~\GeV$ for both $\chi$ and $\chi^\text{hyb}$.  Here and
	throughout we use the NNPDF 3.1 parton distribution functions \cite{Ball:2017nwa}. The QED coupling constant
	$\alpha$ is set at $\alphaem(0) = 1/137$.

	We first explore the effect of moving from smooth-cone to hybrid isolation on differential cross-sections chosen to
	illustrate the underlying features.

	\begin{figure}[htbp]
		\centering
		\begin{subfigure}[t]{\textwidth}
			\includegraphics[width=0.49\textwidth]{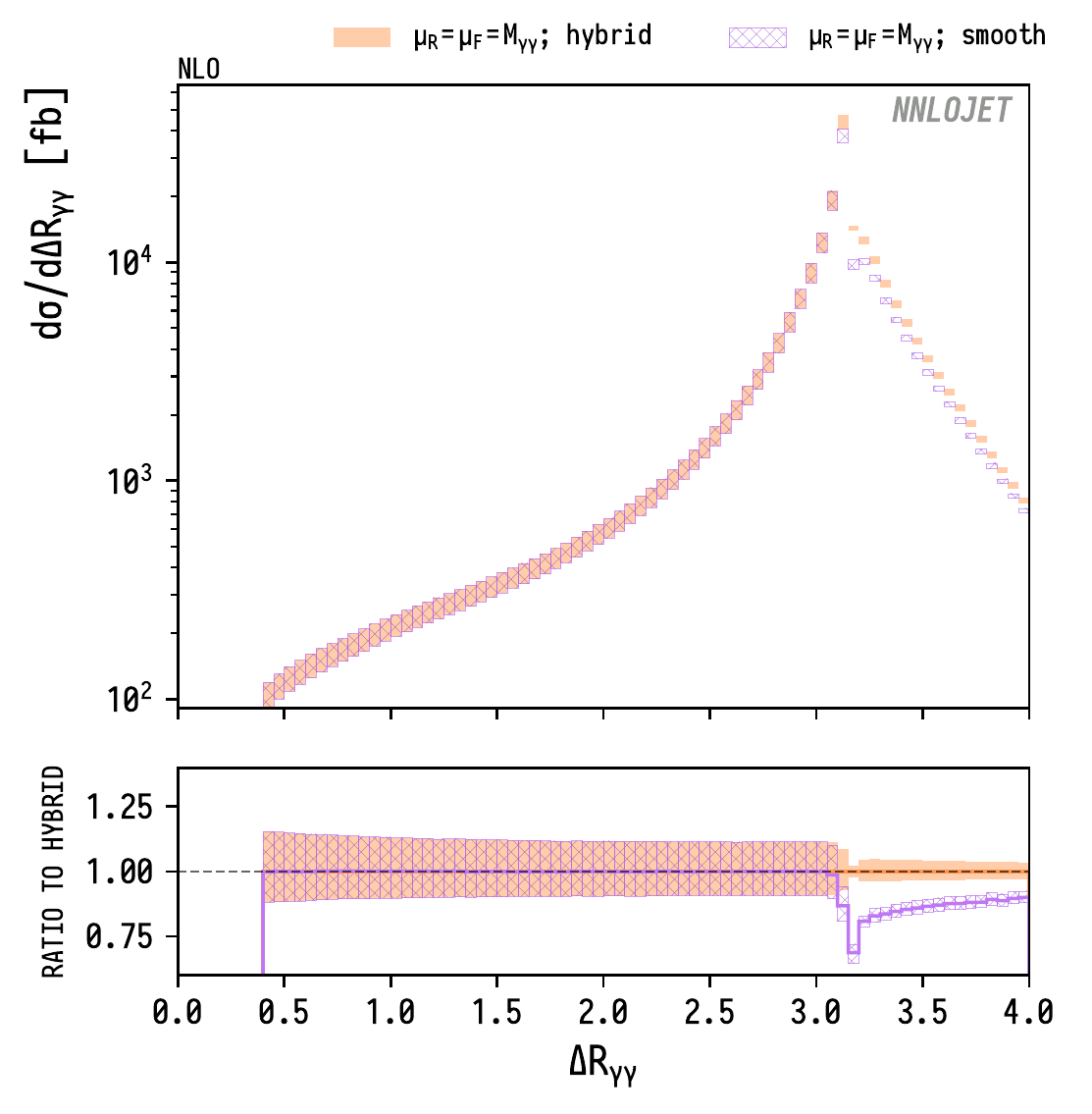}
			\includegraphics[width=0.49\textwidth]{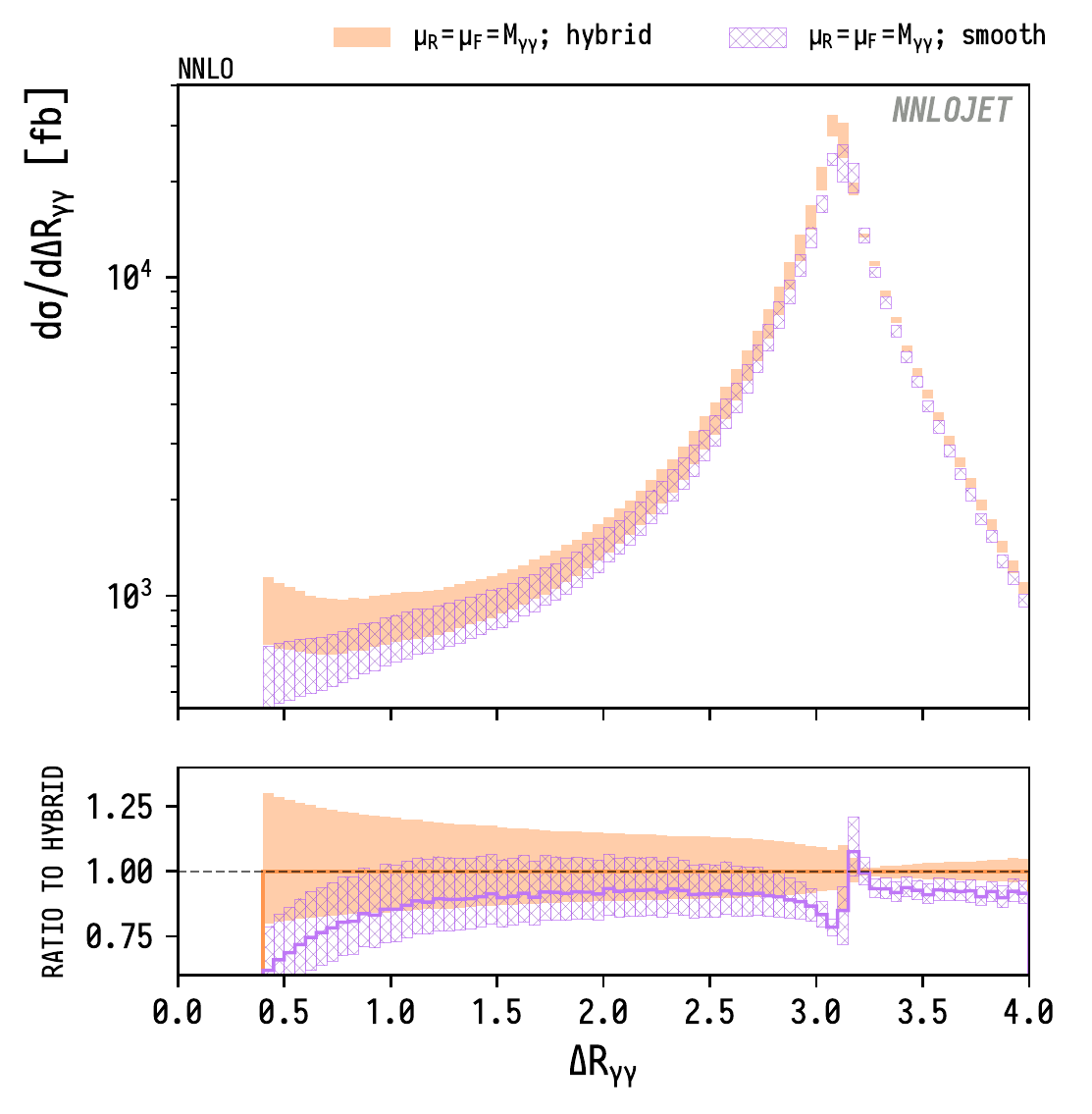}
			\caption{$\dRgg$ at NLO and NNLO using matched-hybrid and smooth-cone isolation.
			}
		\end{subfigure}
		\begin{subfigure}[t]{\textwidth}
			\includegraphics[width=0.49\textwidth]{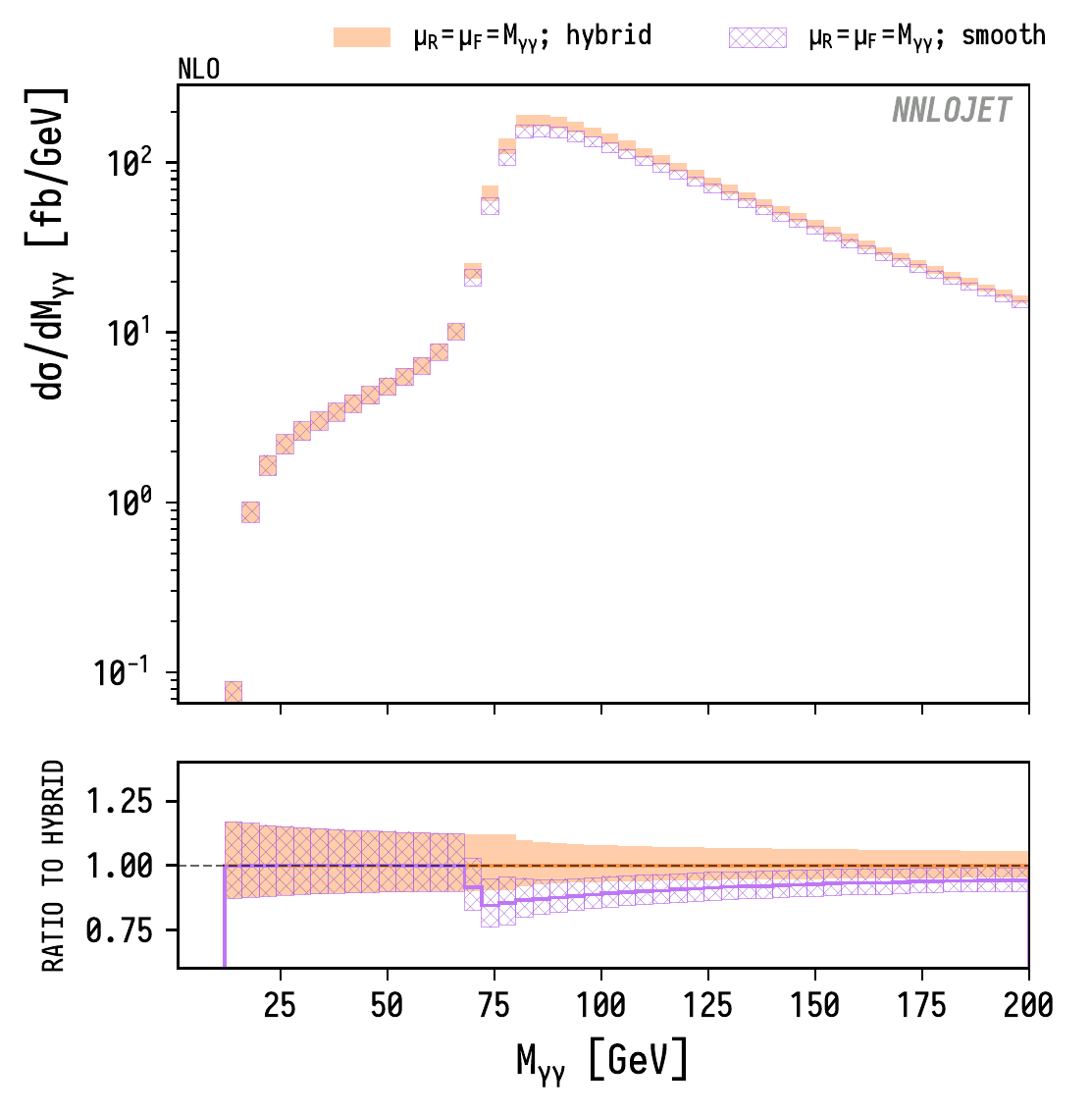}
			\includegraphics[width=0.49\textwidth]{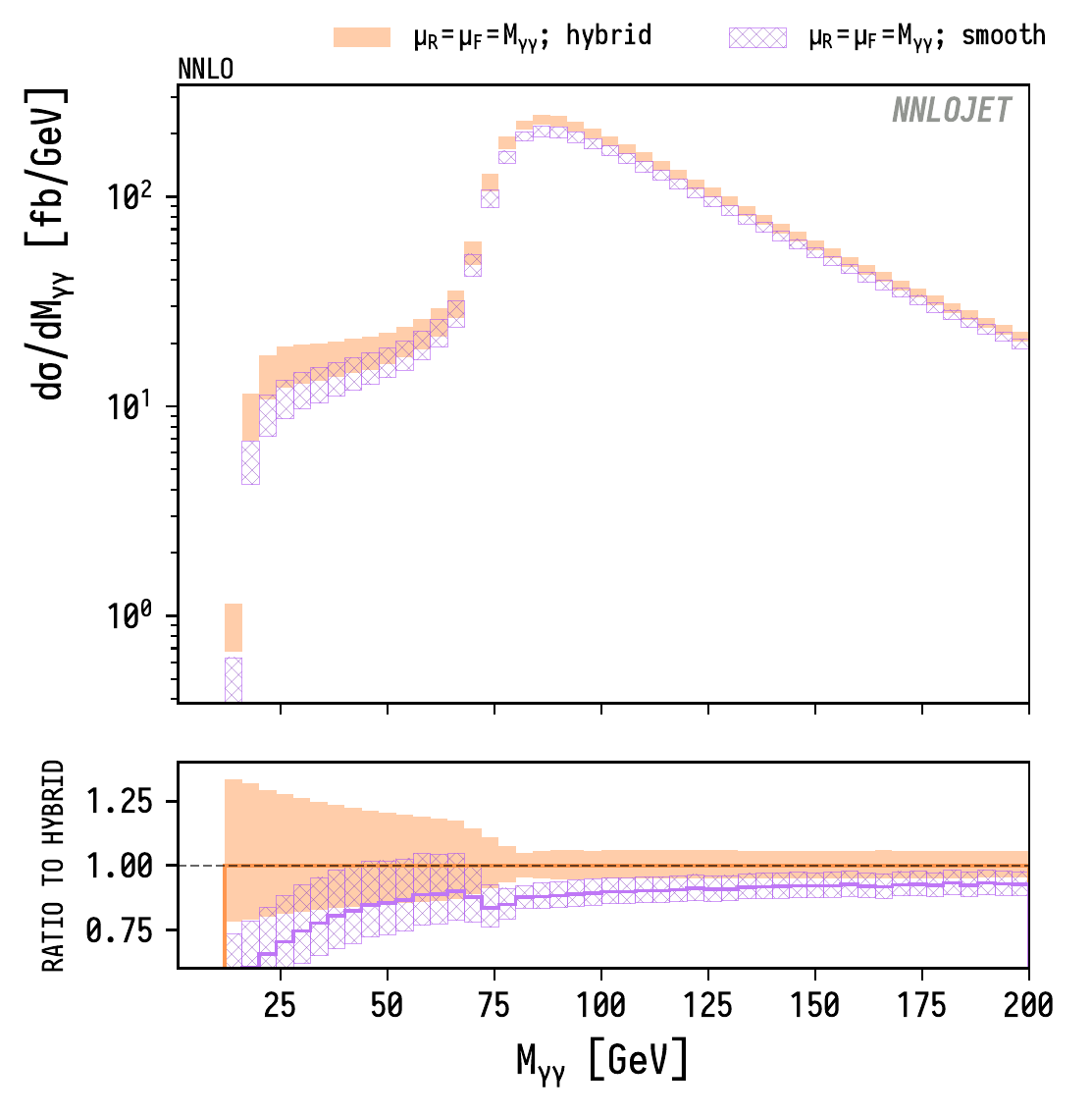}
			\caption{The induced effects at low $\dRgg$ on $\Mgg$.}
		\end{subfigure}
		\caption{$\rd\sigma/\dRgg$ and $\rd\sigma/\Mgg$ at NLO and NNLO using matched-hybrid and smooth-cone
		isolation.  The deviations for small $\Mgg$ and $\dRgg$ are related, as events with small $\Mgg$ can only both
		pass the photon cuts if they have sufficiently small $\dRgg$.  For example, for these cuts, $\Mgg \leqslant
		27~\GeV$ requires $\dRgg \leqslant 0.8$.}
		\label{fig:dR_gg_frix_vs_hybrid}
	\end{figure}

\begin{figure}[htbp]
		\centering
		\includegraphics[width=0.4\textwidth]{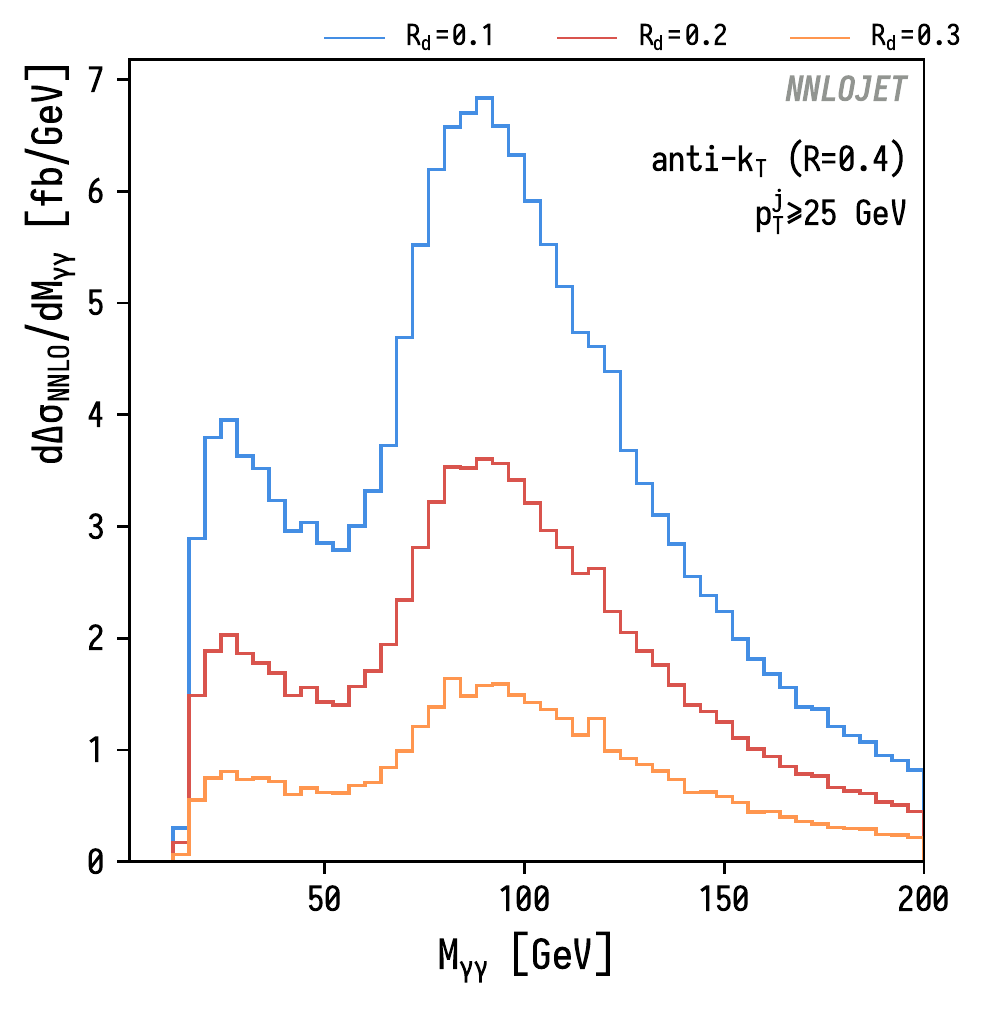}
		\includegraphics[width=0.4\textwidth]{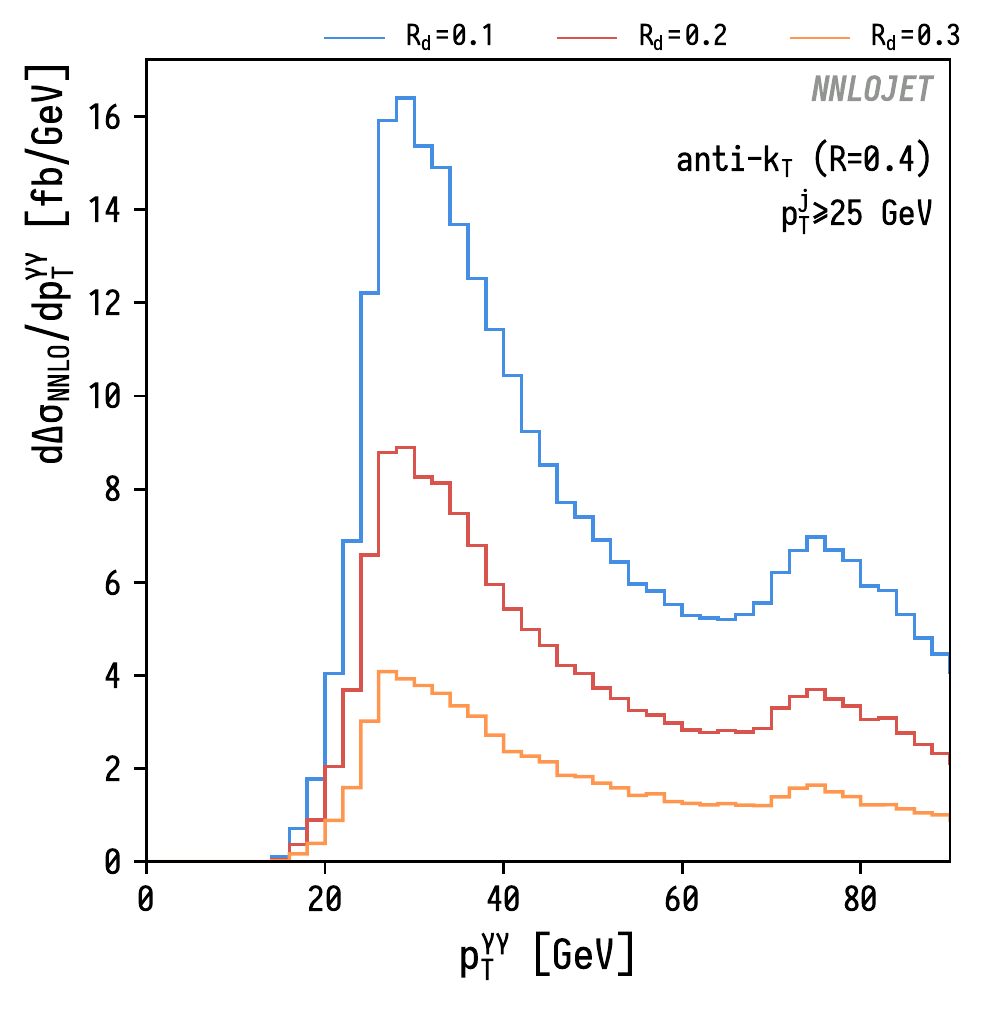}
		\caption{The absolute difference between the hybrid- and smooth-cone isolation differential cross-sections $\rd
		\sigma / \rd \Mgg$ and $\rd \sigma / \rd \ptgg$ for $R_d = 0.1, 0.2, 0.3$, for diphoton production in
		association with an anti-$k_\rT$ jet of $\pt{} \geqslant 25 \;\GeV$, with $R=0.4$.
		}
		\label{fig:Delta_sigma_Mgg}
	\end{figure}

	In \cref{fig:dR_gg_frix_vs_hybrid} we show $\rd \sigma / \rd \dRgg$ and $\rd \sigma / \rd\Mgg$.  The relative
	enhancement is greatest at low $\Mgg$, whilst the absolute enhancement $\rd\Delta\sigma / \rd \Mgg$ follows the
	shape of the underlying distribution, with the difference largest at the Born threshold of $\Mgg = 80~\GeV$.
	Broadly these reflect the two dominant configurations in which soft partonic emissions can enter into isolation
	cones: either the photons are balanced against each other (Born-like), or the diphoton system is relatively
	collimated and balanced against a jet.  Accordingly, configurations with the explicit requirement of an extra jet
	see a further peak in $\rd\Delta\sigma / \rd \Mgg$ at $\Mgg \approx \ptj{\cut}$ corresponding to
	$\dRgg \approx \dRgg^{\cut}$, as shown in \cref{fig:Delta_sigma_Mgg}.  In the $\ptj{\cut} \to 0$ limit, this is 
	effectively truncated by the cuts  on $\dRgg$, which is the configuration corresponding to the small $\dRgg$ 
	effects seen in \cref{fig:dR_gg_frix_vs_hybrid}.  As the jet cut increases this peak will become dominant.

	The requirement of a jet imposes a lower bound on $\ptgg$ and so removes the Sudakov instabilities of the inclusive
	distribution that were discussed in \cref{sec:isol_irsensitivity}.  The two peaks in the two plots correspond to
	the same physics in the opposite order, with the peak at $\ptgg \approx \ptj{\cut}$ corresponding to the
	configuration in which the photons and jet are balanced, and the second peak at $\ptgg \approx 75 \,\GeV$
	corresponding to the threshold at 70 GeV, the smallest value that can be generated within the cuts for every value
	of $\dphigg$.  As discussed in \cite{Binoth:2000zt,Catani:2018krb}, below this threshold the photon cuts imply an 
	implicit minimum for $\dphigg$, restricting the	available phase-space.
	Contributions from this peak  give rise to a distinctive cusp in both the experimental and the NNLO distributions
	which, corresponding to the small-$\dRgg$ region, is especially sensitive to isolation.  Smooth-cone isolation
	suppresses the kinematic peak in this region, which is restored by the less restrictive hybrid isolation profile
	function.

\begin{figure}[htbp]
		\centering
		\includegraphics[width=0.45\textwidth]{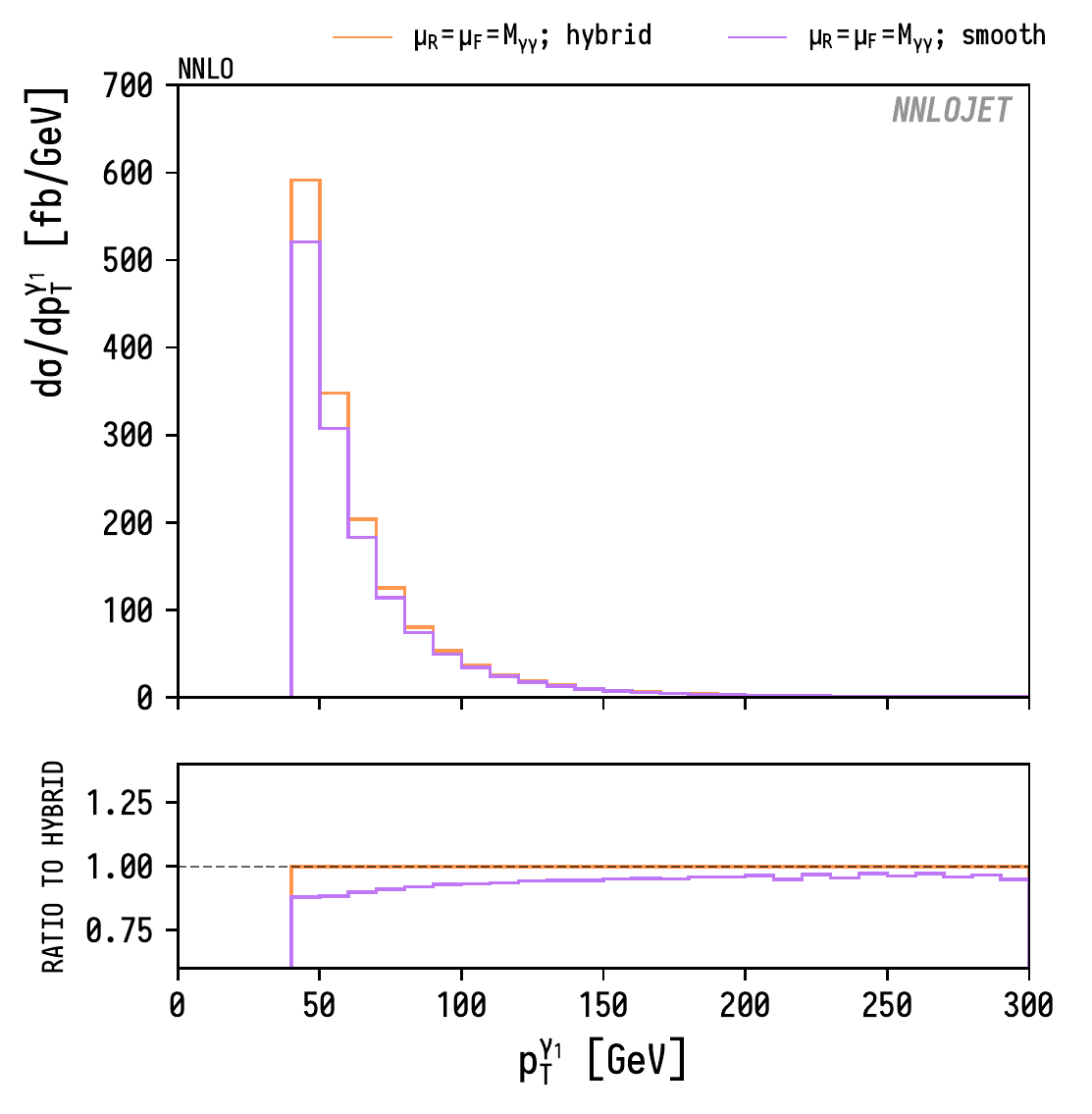}
		\includegraphics[width=0.45\textwidth]{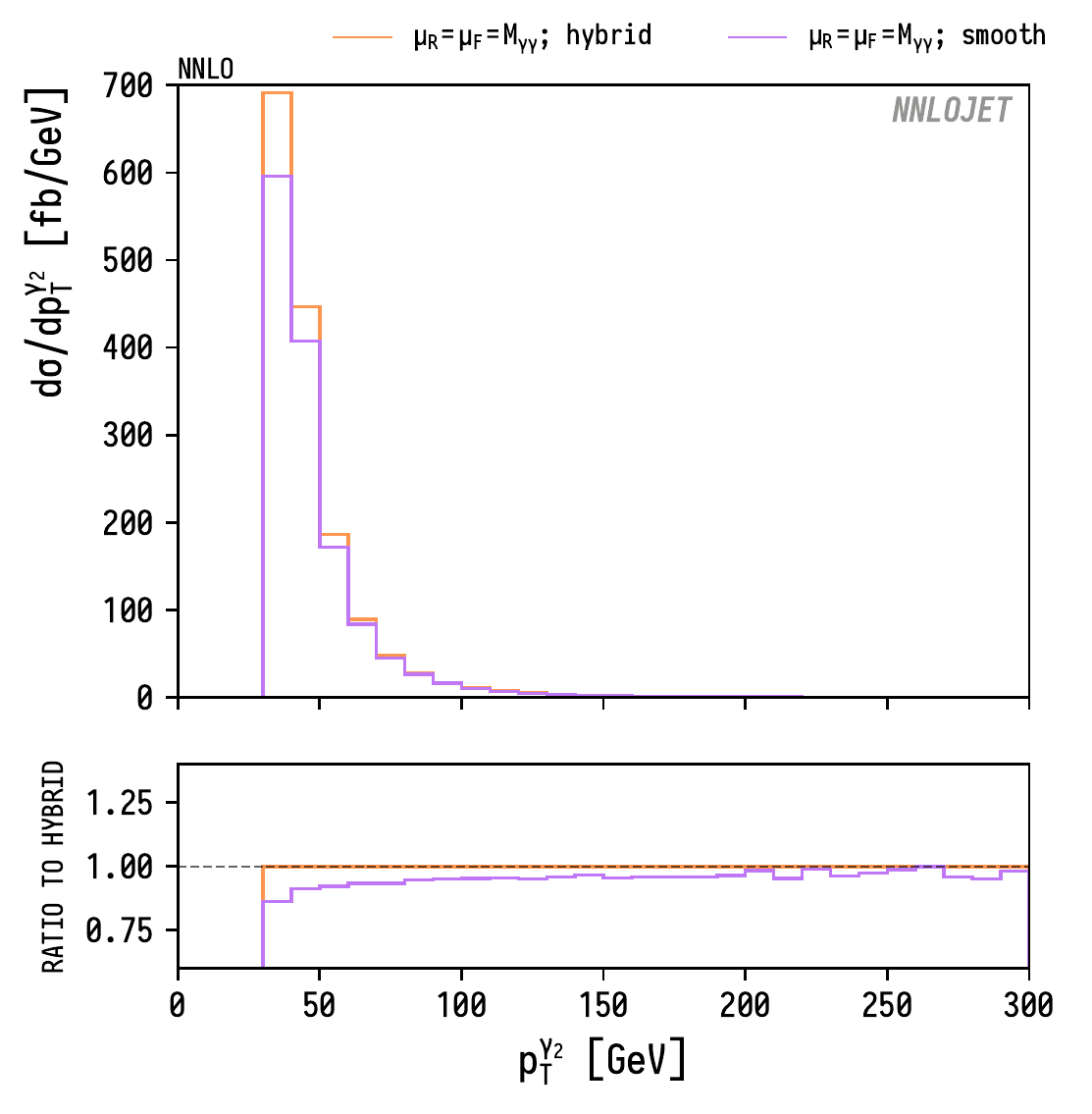}
		\includegraphics[width=0.45\textwidth]{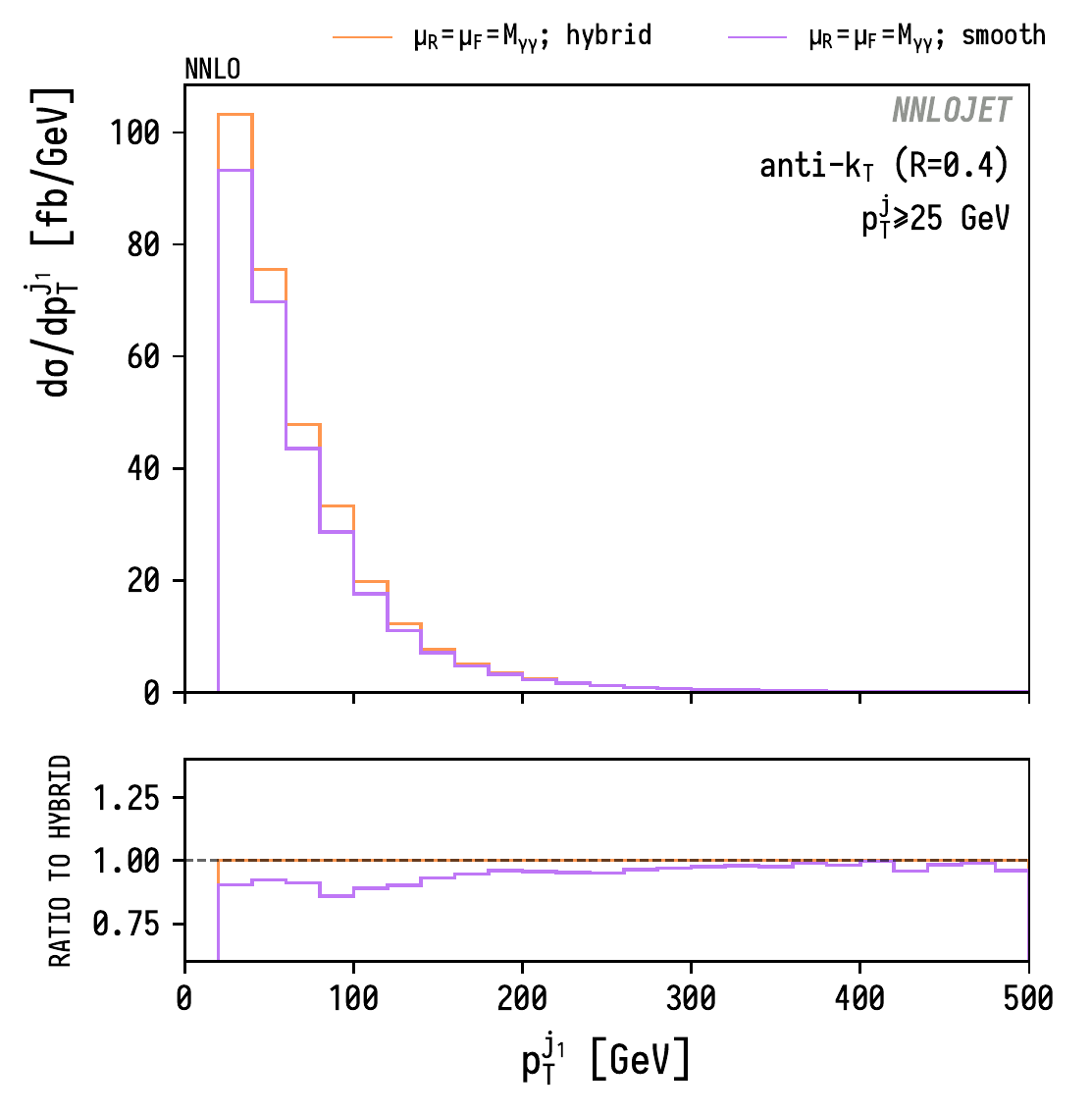}
		\includegraphics[width=0.45\textwidth]{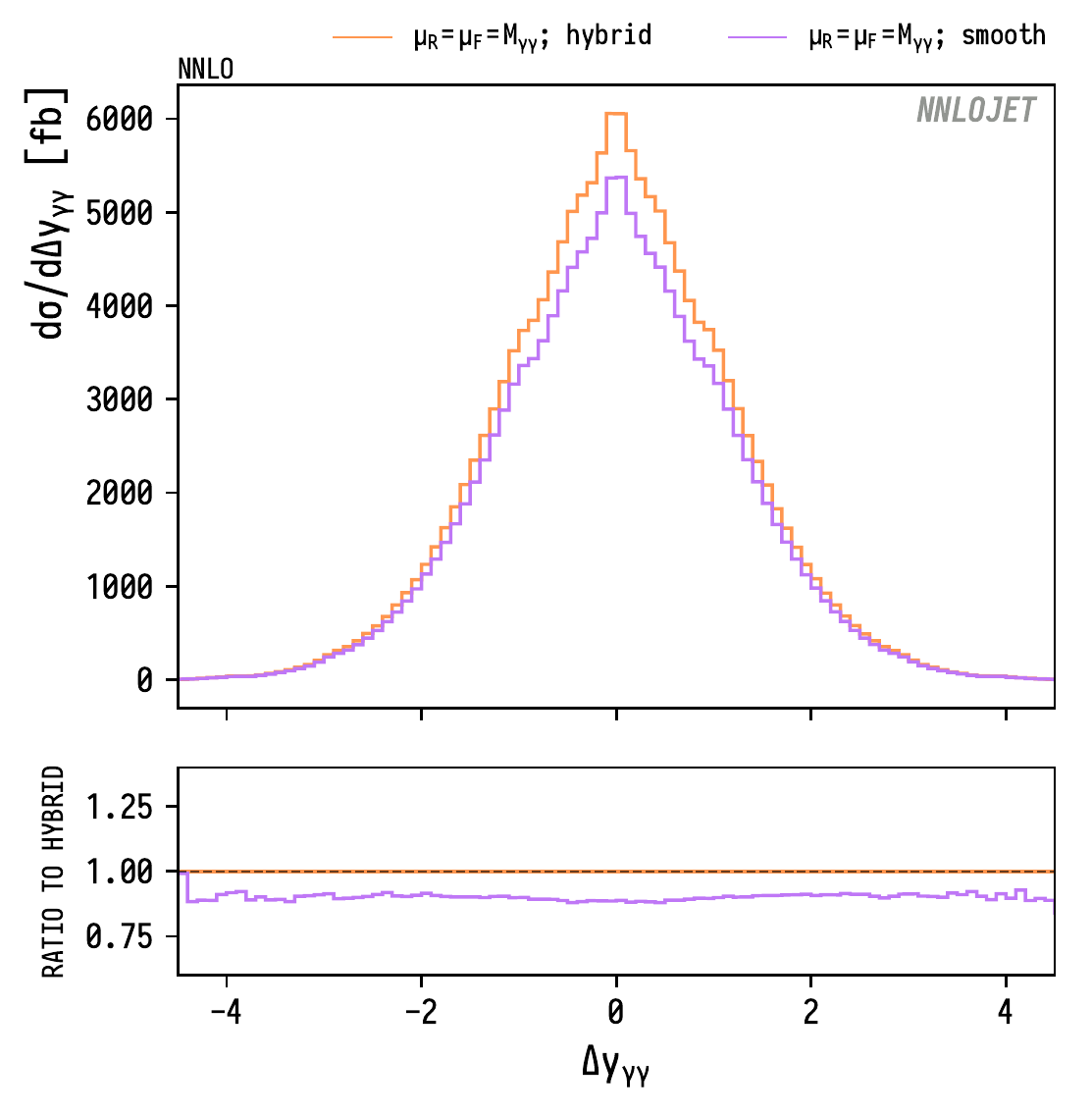}
		\caption{The NNLO distributions $\rd\sigma / \rd \ptg{1}$, $\rd\sigma / \rd \ptg{2}$, $\rd\sigma / \rd \ptj{1}$
		and $\rd\sigma / \rd \dygg$ for hybrid-cone and smooth-cone isolation respectively, and the ratio between the
		smooth-cone and the hybrid distributions.  The defining jet requirement for the third plot is of an 
		anti-$k_\rT$ jet with $\ptj{} \geqslant 25 \GeV$ and $R=0.4$.
		}
		\label{fig:ptg1_ptg2_dygg}
	\end{figure}

	Finally, for completeness, in \cref{fig:ptg1_ptg2_dygg} we consider four further differential cross-sections of
	interest.  The $\ptg1$ and $\ptg2$ distributions are affected most substantially at the boundary of the photon
	cuts, as expected from \cref{fig:hybrid_Rd_dists_NLO_rat}, but are elsewhere mostly unchanged by modifications of
	the cuts.  These regions dominate the cross-section, and explain the large $R_d$-sensitivity of
	\cref{fig:isol_sigma_Rd_NNLO}.  Whether the correction here is purely physical or, particularly for the $\ptg{2}$
	distribution, arises from unphysical behaviour at the boundary of the Born phase-space, is unclear.  As at NLO, for
	the rapidity separation $\dygg$ the additional events permitted by hybrid isolation amount to an overall constant
	factor in  $\rd \Delta \sigma / \rd \dygg$.

	In this section we have compared smooth-cone to hybrid isolation at NNLO for a range of differential cross-sections
	of phenomenological significance.  The effect of interchanging them, which indicates the  uncertainty associated to
	the theoretical implementation of the isolation criteria, is substantial and leads to effects of approximately 10\%
	in uncorrelated distributions, and localised effects of up to 40\% in distributions highly sensitive to the
	specifics of the isolation criteria.  Uncertainties of this magnitude are compatible with the size of the
	scale-uncertainty band, and therefore represent a substantial theory uncertainty that should be accounted for.

	We will return to consider isolation effects in tandem with scale choice in \cref{sec:combvar}.

	\section{Scale choice}
	\label{sec:scale}

	A further uncertainty in the theoretical calculation arises from the choice of functional form $\mu_0$ for the
	renormalisation and factorisation scales.  The conventional choice is $\mu_0 = \Mgg$, the invariant mass of the
	diphoton system, with the magnitude of missing higher-order-uncertainties (MHOUs) estimated through the envelope of
	the variation $\mu_{\rR,\rF} = \xi_{\rR,\rF} \cdot \mu_0$ for $\xi_\rR, \xi_\rF \in \left\{ \frac{1}{2}, 1, 2
	\right\}$.

	Where two \emph{a priori} reasonable choices of $\mu_0$ themselves differ by a factor greater than 2, either
	locally or globally, this procedure fails to span the uncertainty of the calculation even at the known orders.  Any
	estimate of MHOUs is therefore potentially unreliable.

	We begin by briefly reviewing the common scale choices for related processes.  In
	\cref{sec:scale_pertconv,sec:scale_kineffects} we then look at the effects of moving between two choices motivated
	by these, $\mu_0 = \Mgg$ and $\mu_0 = \ptgave$, the arithmetic mean of the photon transverse momenta of the two
	required photons. Finally, in \cref{sec:scale_alt_scales} we generalise to a wider class of possible scale
	choices.

	\subsection{Scale choice for photon processes}
	\label{sec:scale_photon}

	We briefly summarise the scale choices used in the literature for this and related processes.  In
	\cite{Binoth:1999qq}, the first NLO study of diphoton production with fragmentation (\Diphox), the authors used
	$\mu_0 = \frac{11}{20} \ptgave$ for fixed-target data, and $\mu_0=\Mgg$ as the central scale for LHC predictions.
	This scale is also used for NNLO calculations making predictions for or comparisons with data in
	\cite{Catani:2011qz, Campbell:2016yrh, Catani:2018krb} and the experimental papers applying them to measurements at
	the Tevatron \cite{Aaltonen:2012jd,Abazov:2013pua} and the LHC
	\cite{Chatrchyan:2014fsa,Aad:2012tba,Aaboud:2017vol}.  In \cite{Catani:2018krb} the scale $\mu_0 = M_{\rT}^{
	\gamma\gamma} = \sqrt{\Mgg^2 + (\ptgg)^2}$ is additionally considered, finding that the results differ from those
	for $\Mgg$ only in regions of distributions that correspond to the presence of a hard, high-$\pt$ jet.

	For an inclusive single photon and a single photon in association with a jet, $\ptg{}$ is used in the NNLO
	calculations of \cite{Campbell:2016lzl, Chen:2019zmr}.  In the context of PDF fits, it was found in
	\cite{Campbell:2018wfu} that direct photon production data with the former NNLO calculation and scale $\ptg{}$
	could be incorporated into the NNPDF 3.1 global fit without exhibiting tensions with other data.

	For triphoton production, $M_{\gamma\gamma\gamma}$ is used for the MCFM NLO calculation in \cite{Campbell:2014yka},
	and $\frac{1}{4} H_\rT = \frac{3}{4} \ptgave$ and $\frac{1}{2} H_\rT = \frac{3}{2} \ptgave$ are both found to be in
	agreement with data in the NNLO calculation of \cite{Chawdhry:2019bji}.

	Finally, we note that the closest kinematically-related process whose measurements were used in the NNPDF 3.1 fit
	is that of single-inclusive jets, for which the jet $\pt$ was used as the central scale.  A more recent study of the
	scale-choice for single-inclusive jet cross-sections \cite{AbdulKhalek:2020jut} used the central choice
	$\hat{H}_\rT$, the scalar sum of the transverse momenta of all partons in the event.

	This illustrates that the conventional choice for diphoton production of $\mu_0 = \Mgg$ is somewhat atypical among
	related processes.  Its main advantage is for Higgs processes or through the analogy with dilepton final states
	arising from heavy-boson decay.  For such processes the invariant mass of the conditioned-upon two-particle
	final-state particles gives the imputed invariant mass of the virtual boson.  For QCD photon production, however,
	there is no particle to which this invariant mass is expected to correspond, and no QCD vertex with which it can be
	associated.  To explore the significance of this convention we therefore choose to compare $\mu_0 = \Mgg$ against
	alternatives below, focusing on $\mu_0=\ptgave$.

	\subsection{Perturbative convergence}
	\label{sec:scale_pertconv}

\begin{figure}[htbp]
		\begin{subfigure}[t]{\textwidth}
			\centering
			\includegraphics[width=\textwidth]{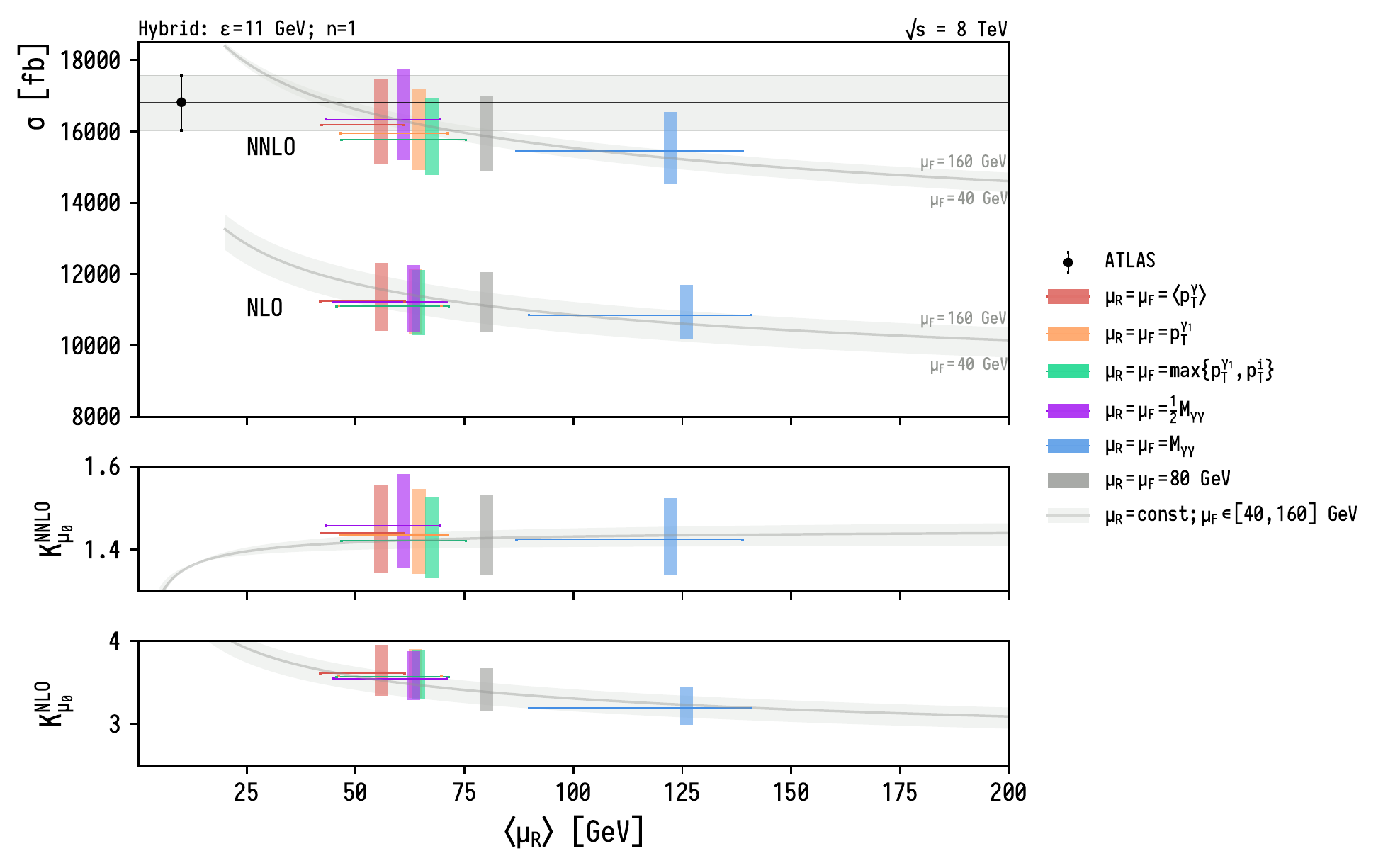}
		\end{subfigure}
		\caption{Scale dependence of the cross-section.
			Each cross-section associated with a dynamical scale $\mu_0$ is plotted against its mean, $ \langle \mu_0
			\rangle $. The $x$-error bars indicate the lower- and upper-quartiles of the scale-variable distribution,
			calculated from the binned data.  The scale bands are the scale uncertainties associated with the usual
			7-point scale variation around the central scale.
			The grey bands give the cross-section for the fixed scales $\mur, \muf$ specified, calculated from the
			renormalisation group equations.}
		\label{fig:xs_vs_scl}
	\end{figure}

	We first consider the perturbative convergence of the cross-section.  In \cref{fig:xs_vs_scl} we show the
	cross-section and $K$-factors at NLO and NNLO for a number of choices of dynamic scale, as well as the scale
	evolution calculated from the renormalisation group equations.

	The NLO $K$-factors are consistently large due to the opening of the $\Pq\Pg$ channel, and vary according to its
	considerable dependence on the scale choice.  The $K$-factor for the $\Pq\Paq$ channel alone is approximately 1.5.
	The cross-sections for dynamic scale choices are largely consistent with the fixed-scale calculation corresponding
	to their mean value, suggesting the reweighting of phase-space by the dynamism of the dynamic central scales has a
	limited effect on the total cross-section.
	At NNLO, the $K$-factor is still considerable (approximately 1.4), due to sizeable NLO corrections in the $\Pq\Pg$
	channel ($K$-factor \textasciitilde1.3), NNLO corrections in the $\Pq\Paq$ channel (\textasciitilde1.2), and
	the opening of the $\Pg\Pg$ and $\Pq\Pq'$ channels, but is stable for all the choices of scales considered.

	Overall, as expected from the running of $\alphas$, dynamic scales which range over smaller values lead to larger
	predictions than those with larger values.  Purely in terms of the distribution of their magnitude, the scales
	$\ptgave$ and $\Mgg$ represent the two extremes between which other reasonable dynamic scales are likely to fall.

	Despite the stability of the NNLO-to-NLO $K$-factor across these choices of scales, it is clear from the gradient of
	the grey band that the scale-dependence remains significant.  The use of a dynamic rather than a fixed scale can be
	seen to bring the scales into closer agreement than would be expected from their central values alone.

	\subsection{Kinematic effects}
	\label{sec:scale_kineffects}

	We now consider the kinematics of the two scales $\Mgg$ and $\ptgave$, focusing on regions of phase-space in which
	we expect the ratio ${\Mgg}/{\ptgave}$ to become large (or small) and potentially lead to discrepancies arising
	from large logarithms of ratios of the scales.  Although we focus on the diphoton context, including the ATLAS
	cuts, the underlying kinematic properties are universal.

	In the Born kinematics, $\ptgave = \ptg{1} = \ptg{2}$ and
	\begin{align}
	\label{eqn:Mgg_LO}
		\Mgg = 2 \ptgave \cosh \left(\frac{1}{2} \dygg \right) \geqslant 2 \ptgave.
	\end{align}
	The fiducial cuts on rapidity separation restrict $\left|\dygg\right| \leqslant 4.74$ and hence in the Born
	kinematics,
	\begin{align}
		2 \ptgave
		\leqslant \Mgg
		\leqslant 10.8 \ptgave.
	\end{align}

	Thus already at leading order, the two scales differ by at least the factor of 2 used in the conventional
	renormalisation and factorisation scale variation.  We can therefore anticipate there to be regions of differential
	distributions in which the scale uncertainty bands around the two choices of $\mu_0$ do not overlap.

	The exponential behaviour of the scale $\Mgg$ at high rapidity separations persists to all orders, with the
	general expression
	\begin{align}
		\label{eqn:Mgg_exponential}
		\Mgg = \sqrt{2 \ptg{1} \ptg{2} \left(\cosh \dygg - \cos \dphigg\right)}.
	\end{align}
	At higher orders, $\Mgg\leqslant\ptgave$ becomes possible.  $\Mgg$ is bounded below only as a result of the photon
	separation cut $\dRgg \geqslant 0.4$, which restricts
	\begin{align}
	\label{eqn:Mgg_lowerbound}
		\Mgg
		\geqslant
		2 \sqrt{\ptg{1}\ptg{2}} \sin \left(\frac{1}{2}\dRgg^\cut\right)
		> 13.76 \;\GeV
	\end{align}
	for the ATLAS cuts described in \cref{eqn:ATLAScuts}.  Without this cut, which is set to be equal to the isolation
	cone radius by experiment specifically to exclude each photon from the isolation cone of the other, $\Mgg$ would in
	principle be permitted within the calculation to get arbitrarily small.  Thus for fixed $\ptg{1}$ and $\ptg{2}$
	(and hence fixed $\ptgave$), $\Mgg$ can vary over a factor of approximately 25:
	\begin{align}
	\label{eqn:Mgg_NNLO}
		0.397
		\leqslant \frac{\Mgg}{\sqrt{\ptg{1}\ptg{2}}}
		\leqslant
		10.8,
	\end{align}
	with the size of this factor entirely dependent on cuts chosen for primarily experimental reasons.  Were the
	photon-separation cut allowed to become smaller (e.g. to $\dRgg\geqslant 0.2$), or the maximum rapidity separation
	allowed to grow (e.g. from 4.74 to 6), this ratio would span two orders of magnitude.

\begin{figure}[tbp]
		\centering
		\includegraphics[width=0.6\textwidth]{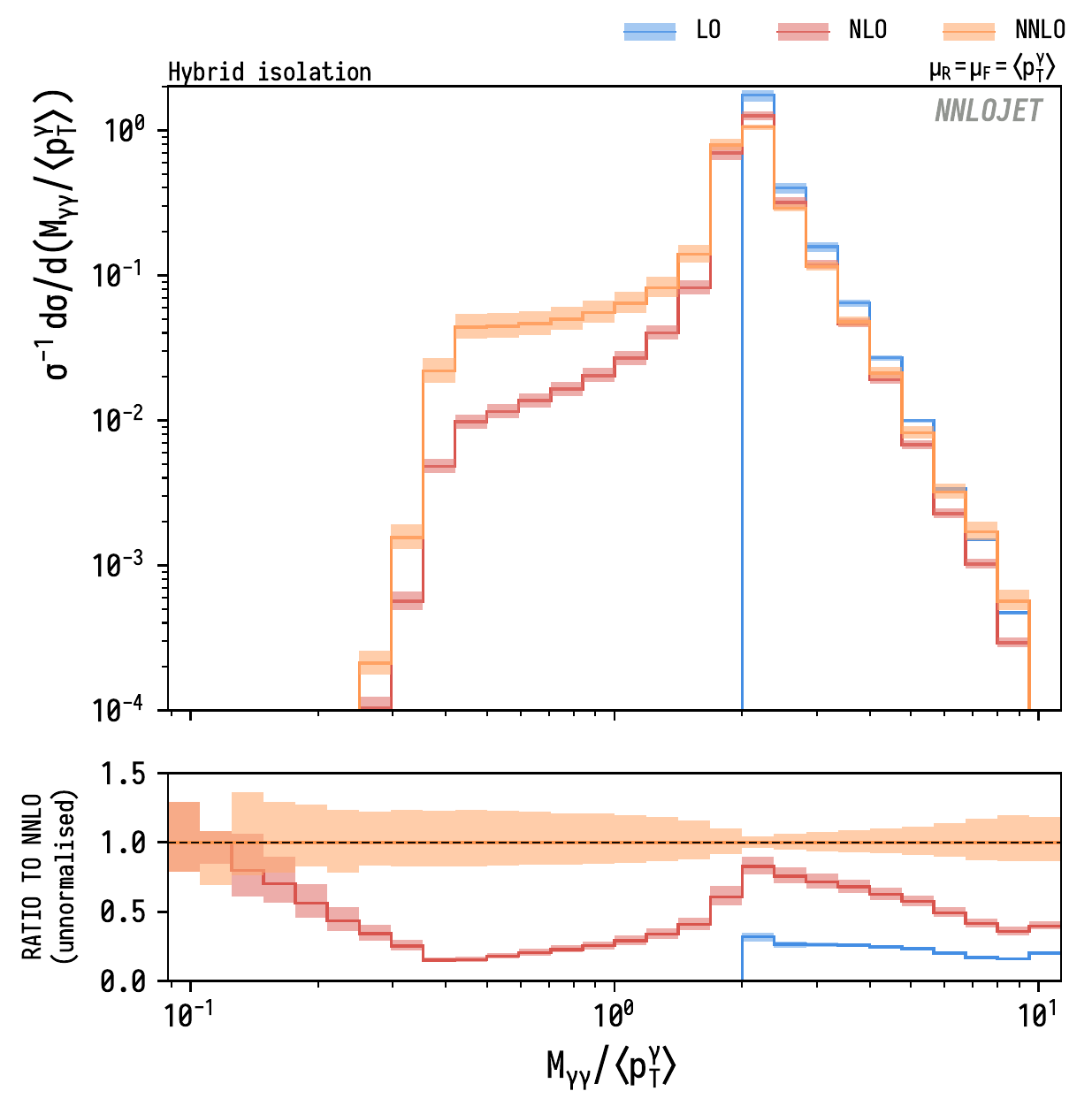}
		\caption{Ratio of scale variables.  As phase-space constraints are lifted by the emission of additional
		partons, large ratios of scales become possible.
		}
		\label{fig:Mggrat}
	\end{figure}

	To illustrate the range of values taken by the ratio ${\Mgg}/{\ptgave}$ we show the corresponding normalised 
	distribution at LO, NLO and NNLO in \cref{fig:Mggrat}.
	We see that the modal value for the ratio is 2, and that
	the regions where the logarithm of the ratio will be large are suppressed in their contribution to the
	cross-section, and predominantly arise from the NNLO contribution as additional partonic radiation allows the
	kinematic configuration to depart further from the Born.

\begin{figure}[htbp]
	\centering
		\includegraphics[width=0.49\textwidth]{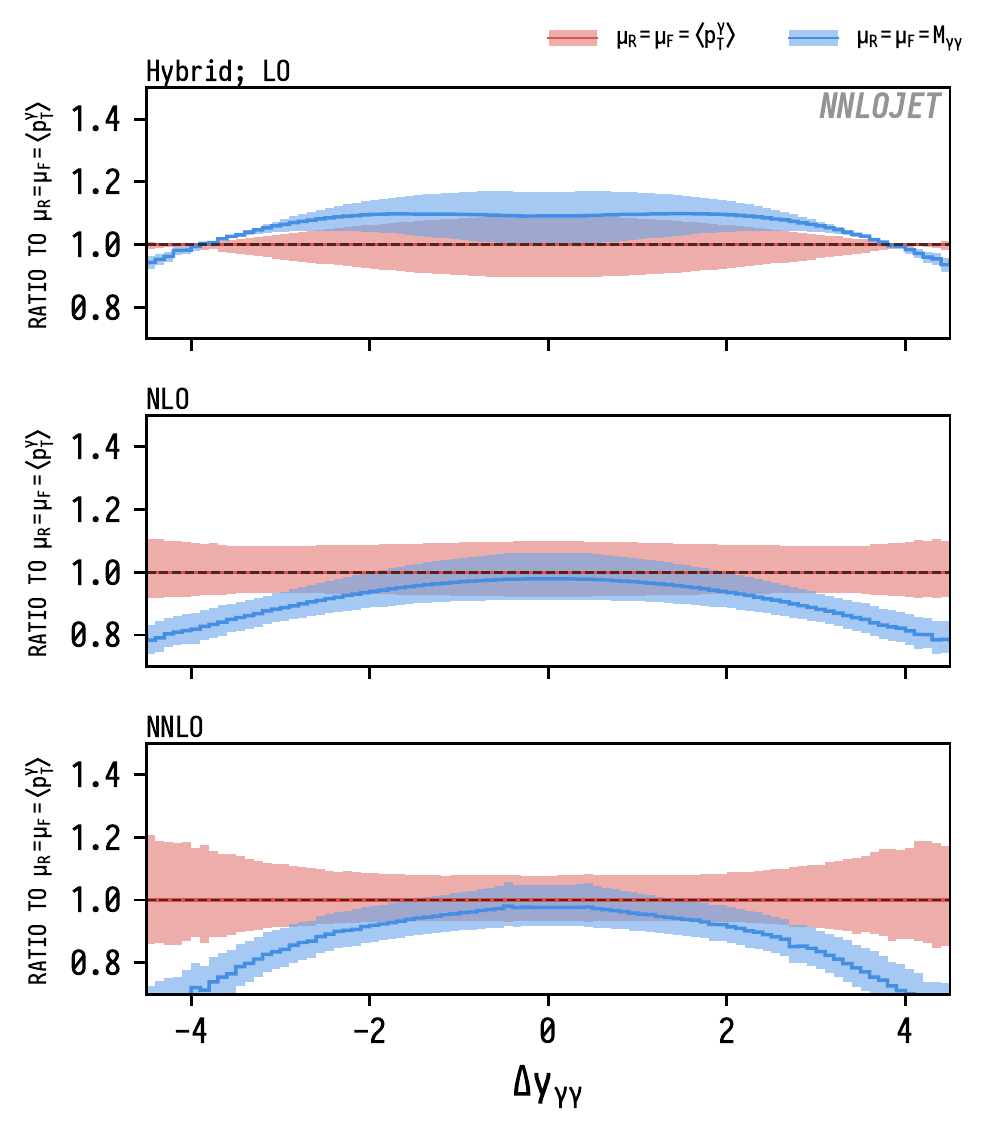}
		\includegraphics[width=0.49\textwidth]{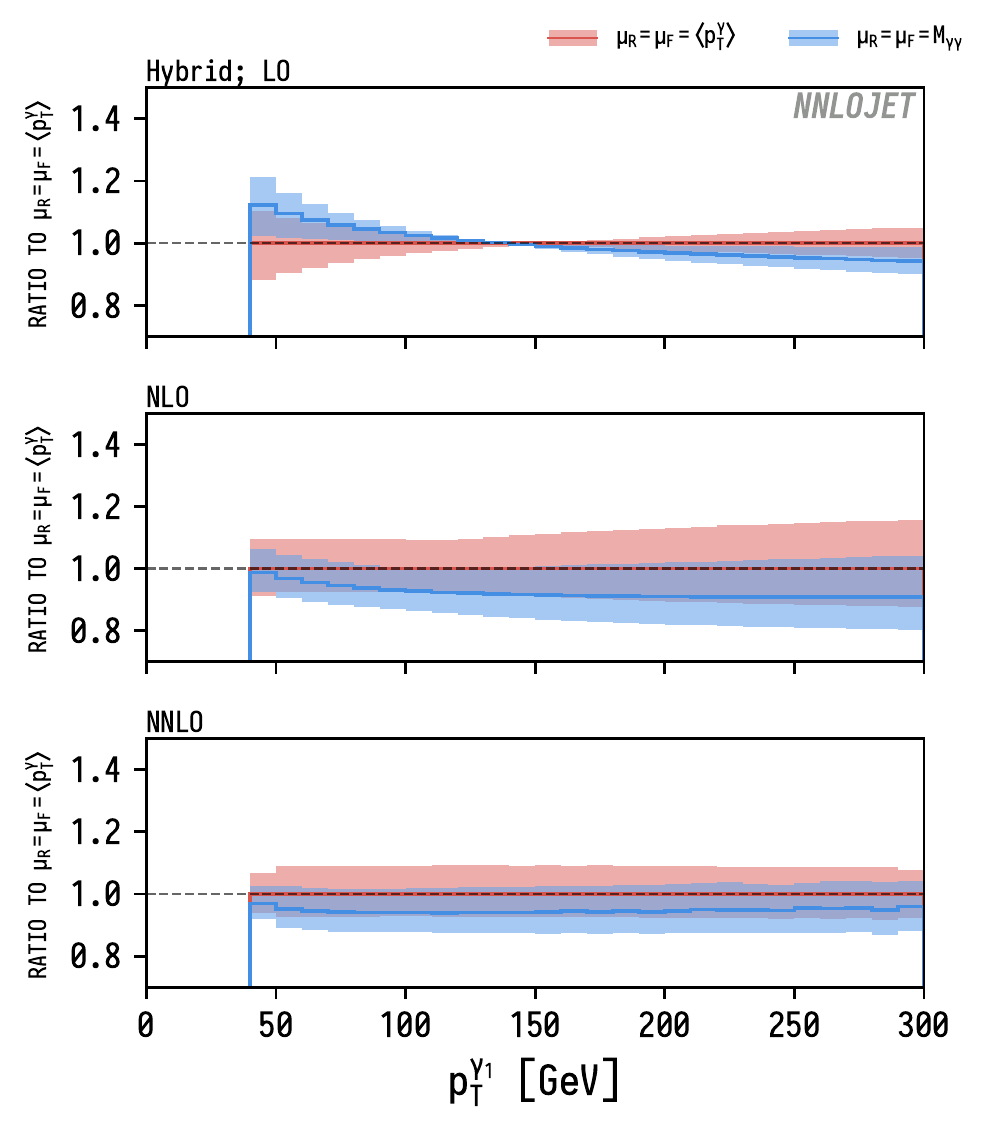}
		\includegraphics[width=0.49\textwidth]{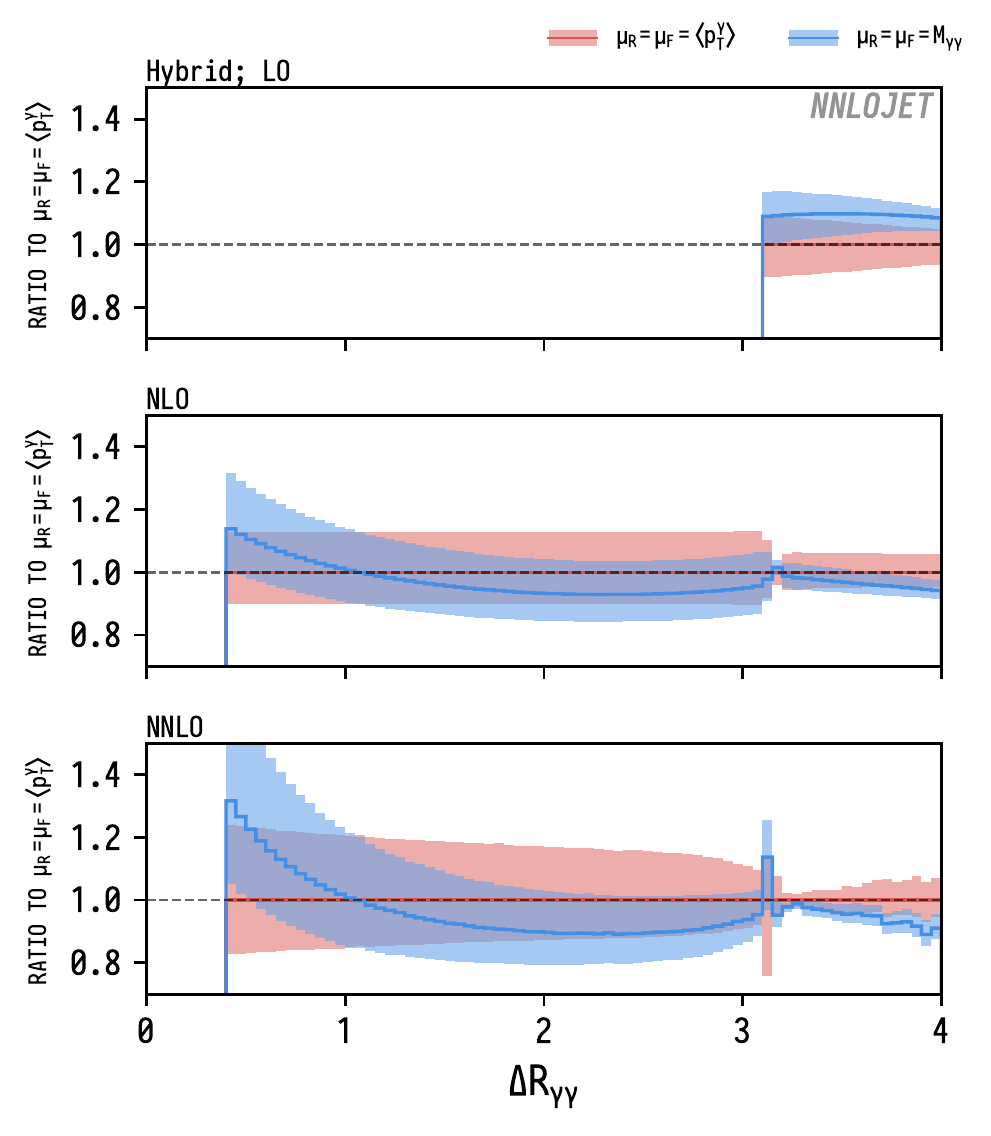}
		\includegraphics[width=0.49\textwidth]{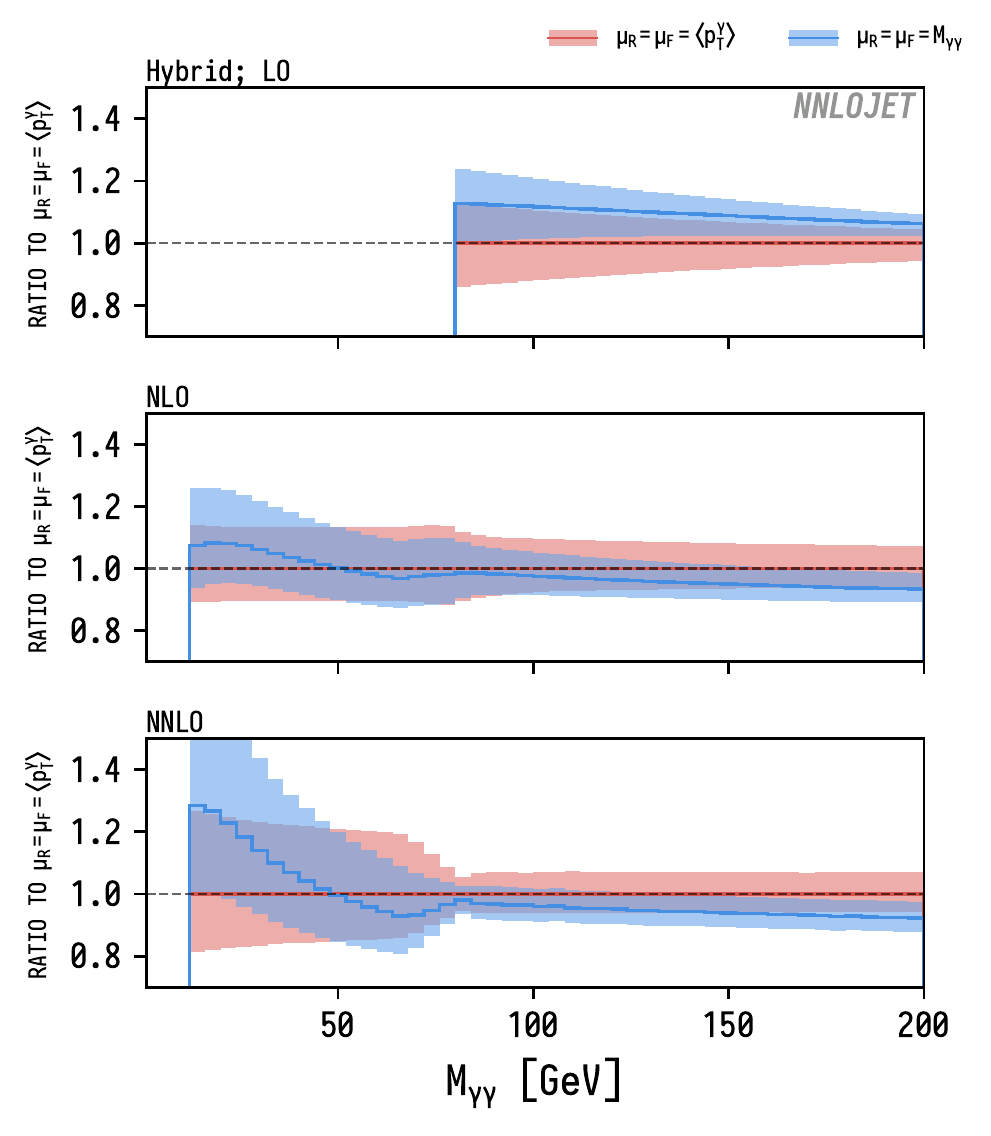}
		\caption{Order-by-order comparison of the difference between the scale choice $\mu_0=\Mgg$ and $\mu_0=\ptgave$.
		Leading order here means $\alphas^0$, so the counterbalancing effects of the PDFs and the running of $\alphas$
		can be deduced.}
	\label{fig:orderscl}
	\end{figure}

	The distortive effect of the scale choice on differential cross-sections depends substantially on the order of the
	strong coupling $\alphas$, through the renormalisation group equations.  This is illustrated in \cref{fig:orderscl}.
	The ${\rd \sigma}/{\rd \dygg}$ distribution exposes the exponential behaviour remarked upon in
	\cref{eqn:Mgg_exponential}.  At leading-order $\alphas^0$, the calculation is independent of $\mur$, and so the 
	dependence
	is only on $\muf$ through the PDFs.  The dependence is mild: the results for the scale choice $\mu_0=\Mgg$ are
	modestly larger than those for $\mu_0=\ptgave$, with the deviation largest for $\dygg=0$ where $\Mgg = 2 \ptgave$
	exactly, due to \cref{eqn:Mgg_LO}, and as can be seen through the coincidence of the scale bands of one scale with
	the central scale of the other.

	Additional powers of the coupling constant $\alphas$ reverse that hierarchy, due to the monotonicity of the running
	of the coupling that ensures
	$\alphas(\mu_1) \geqslant \alphas(\mu_2)$ for $\mu_1 \leqslant \mu_2$.
	Thus in the regions of large rapidity-separation, the $\mu_0=\Mgg$ predictions are suppressed relative to those for
	$\mu_0=\ptgave$ by up to 30\%.

	In the extremes of the distribution, this is driven by the constructive interference of factorisation- and
	renormalisation-scale variation, in the sense that larger $\muf$ and larger $\mur$ both act to suppress the
	result.  The substantial correlation between $\dygg$ and $\Mgg$ in these bins leads to an implicit cut on $\Mgg$ in
	each bin, which leads to artificially small scale-uncertainty bands for the $\mu_0=\Mgg$ result compared to variation
	over an inclusive dynamical scale variable.  This might lead to the conclusion that the $\mu_0=\Mgg$ distributions
	display improved perturbative convergence due to the narrower scale bands, when it is in fact an artefact of
	correlation of the scale with the binned observable, leading to a restricted domain for the scale variation
	procedure.

	The behaviour of the $\rd\sigma / \rd \dRgg $ distribution at low $\dRgg$ shows exactly the inverse behaviour:
	small values of the ratio ${\Mgg}/{\ptgave}$ lead to an enhanced distribution.  As discussed in
	\cref{sec:isol_comparison}, the low-$\Mgg$ distribution corresponds exactly to small values of $\dRgg$, as a result
	of the cuts on photon transverse momenta. This accounts for the common behaviour between the bottom two plots.  For
	an event in the lowest $\Mgg$-bin, the NNLO contributions to the cross-section with the scale $\mu_0=\Mgg$ are
	weighted relative to the $\mu_0=\ptgave$ contribution with a factor proportional to the ratio of $\alphas^2$
	evaluated at the two scales, which is imperfectly compensated by the corresponding dependence in the real-virtual
	matrix elements.
	This gives rise to the extreme \char`\~ 30\% deviations between the scale choices in this region; the
	factorisation-scale dependence is negligible.  Since the lower bound on $\Mgg$ is set by the experimental $\dRgg$
	and $\ptg{}$ cuts rather than any theory considerations, smaller values of these cuts would lead to still greater
	distortions between the scale choices.  Note that this is in contrast to the problem of scale choices for the dijet
	process, in which scale choices $M_{jj}$ and $\langle\ptj{}\rangle$ differ substantially at NLO but less so at NNLO
	\cite{AbdulKhalek:2020jut}.

	\subsection{Alternative scale functional forms}
	\label{sec:scale_alt_scales}

	We remark on the elements of the above discussion which carry over to scale choices with functional forms other
	than $\mu_0= \Mgg$ and $\mu_0=\ptgave$.  Popular candidates commonly found in studies of other processes typically
	involve a weighted average, mixing four-momentum-invariant-type observables with transverse-plane observables,
	schematically of the form
	\begin{align}
		\mu_0 = \left(\alpha \Mgg^r + \beta f\left(\left\{\vecptof{i}\right\}\right)^r \right)^{\frac{1}{r}}
	\end{align}
	where common choices for $f$ include $\ptgg$, the transverse momentum of the diphoton system, or the total
	transverse
	momentum of all partons, all jets, or both photons.  A variety of functional forms of this type were considered in
	\cite{Badger:2013ava} for the production of a photon pair in association with  up to three identified jets.

	Functional forms containing $\Mgg$, i.e.\ with $\alpha\neq 0$, are dominated by the exponential function of
	rapidity separation in the (sufficiently) large rapidity-separation region discussed above, and so behave like
	$\Mgg$ there.  The results in this region therefore lie within the envelope bounded by the scale variation $\mu \in
	\left\{\frac{1}{2}\Mgg, 2\Mgg\right\}$.  Of particular importance is the choice
	\begin{align}
	\label{eqn:MTgg}
		\mu_0 = \MTgg &= \sqrt{\Mgg^2 + \left(\ptgg\right)^2}
		\nonumber \\
		&= \sqrt{ \left(\ptg{1}\right)^2 + \left(\ptg{2}\right)^2 + 2 \ptg{1} \ptg{2} \cosh \dygg }
	\end{align}
	which was considered in \cite{Catani:2018krb} and shows identical behaviour in this limit.  In addition, the scales
	$H'_\rT, \hat{H}'_\rT, \sqrt{\Sigma^2}$ and $\sqrt{\hat{\Sigma}^2}$ that were investigated for diphoton production
	in association with up to three jets in \cite{Badger:2013ava} all have similar behaviour, as a consequence of their
	dependence on $\Mgg$.

	From a physical perspective, the behaviour in this limit represents the scale ambiguity between transverse-plane
	and four-momentum observables.  For central final-states, both classes of observable are of the same order of
	magnitude and induce a similar ordering of events by scale.  For events with large rapidity separation, the
	projection onto the transverse plane dramatically changes the apparent energy scale of the event.  In the extremes
	of rapidity separation we enter the two-large-scales regime, in which resummation or other approaches may become
	relevant to correct for large logarithms of the form $\ln\left(\hat{s}/{\Et{2}}\right)$. It is possible that
	compensating behaviour partially accounting for these logarithms would arise in the parton distribution functions
	if one or the other type of scale was used consistently in fits.

	The second region discussed above, of small $\dRgg$, arises as a direct consequence of the specific form of the
	angular factor $\left(\cosh\dygg - \cos\dphigg\right)$ in $\Mgg$, which reduces to $\dRgg$ in this limit.  As a
	result, modifying $\Mgg$ by any offset function $f$ with a non-zero limit as $\dRgg \to 0$ rectifies the
	problematic behaviour.  This is the case for $\MTgg$ in \cref{eqn:MTgg} above, and all other scales considered with
	non-zero $\beta$.  Whilst candidates for $f$ with similar asymptotic behaviour to $\Mgg$ do exist (e.g. $f =
	\ptg{1}\ptg{2} \dRgg$), they do not arise naturally from a consideration of the scale of the process.  From a
	physical perspective, problematic behaviour in this region can be explained as the failure of the scale $\Mgg$ to
	capture the natural scale of the underlying process, in which the collimated diphoton pair recoils against a hard
	jet.  The scale variable vanishes as the two photons become collinear, restricted only by the experimental cut,
	even as the event maps onto a photon-plus-jet event of characteristic scale $\ptg{}\sim\ptj{}$.  This leads to
	exaggerated contributions from $\alphas$ which are not compensated by the real-virtual matrix elements.

	We can understand this substantial exposure as follows.  The diphoton final-state is a two-particle final-state, so
	the Born-level kinematics are highly restricted; it is colourless, so only the $\Pq \Paq$-channel is fully NNLO,
	and there is no resonant propagator, so the cross-section is not dominated by a single modal value of the
	final-state invariant mass.  It might therefore be expected that other final-states are unlikely to yield similar
	sensitivities.  Nevertheless, with the same cuts, the same ratios of scales would arise for, e.g., the $Z
	\to 2\ell$ process, and it may be worth investigating their impact further.

	\section{Combined effect of isolation and scale variation}
	\label{sec:combvar}

	Finally we illustrate the combined effect of the simultaneous variation of scale and isolation choice on the
	distributions.  We have previously seen in \cref{fig:dR_gg_frix_vs_hybrid} that the region of phase-space most
	affected by the difference between smooth-cone and hybrid isolation is that in which $\dRgg$ is small, and that the
	same region is highly sensitive to the scale choice, growing starkly with the running coupling relative to a
	prediction using a scale independent of $\dRgg$.

\begin{figure}[tbp]
		\centering
		\includegraphics[width=0.49\textwidth]{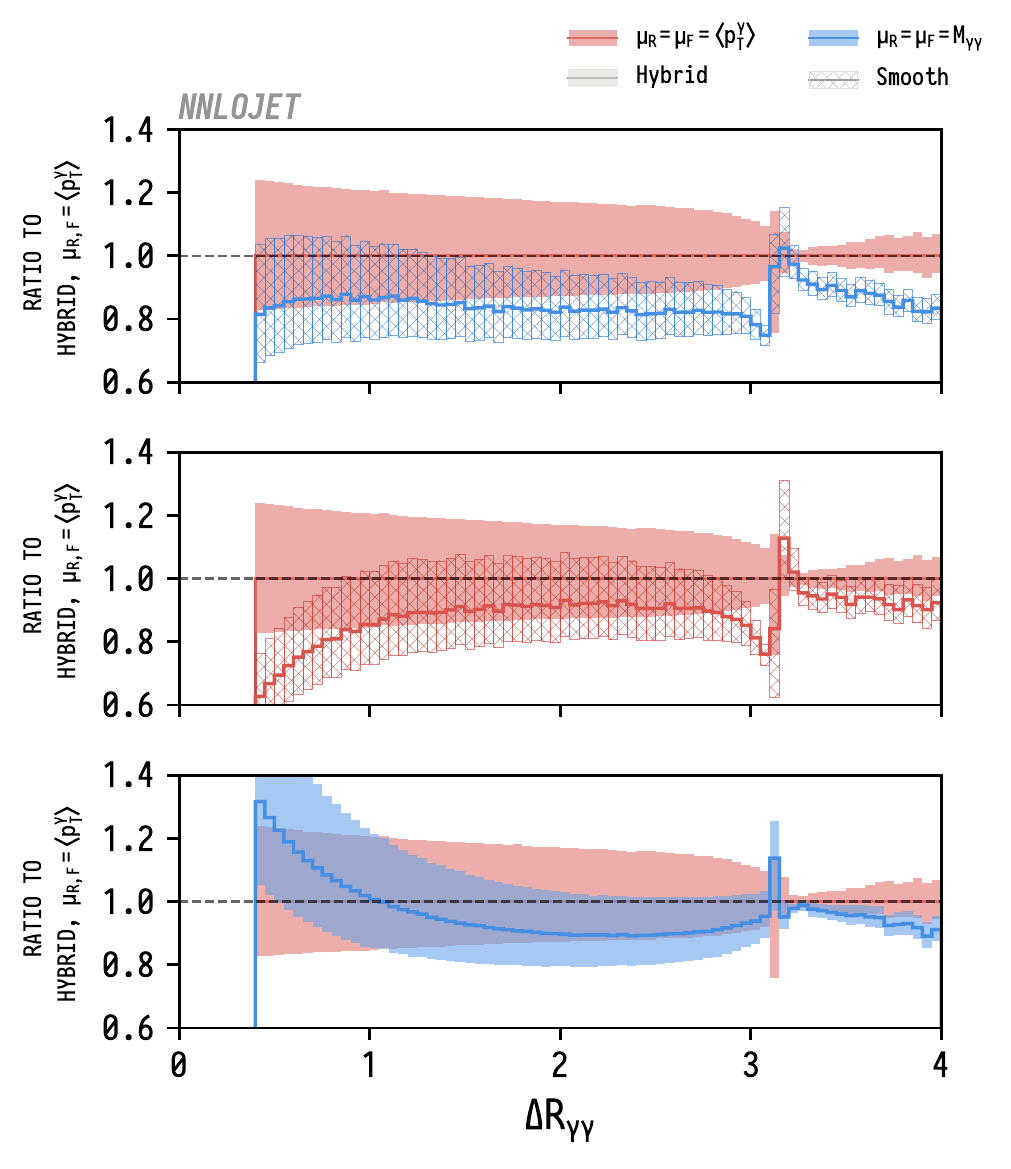}
		\includegraphics[width=0.49\textwidth]{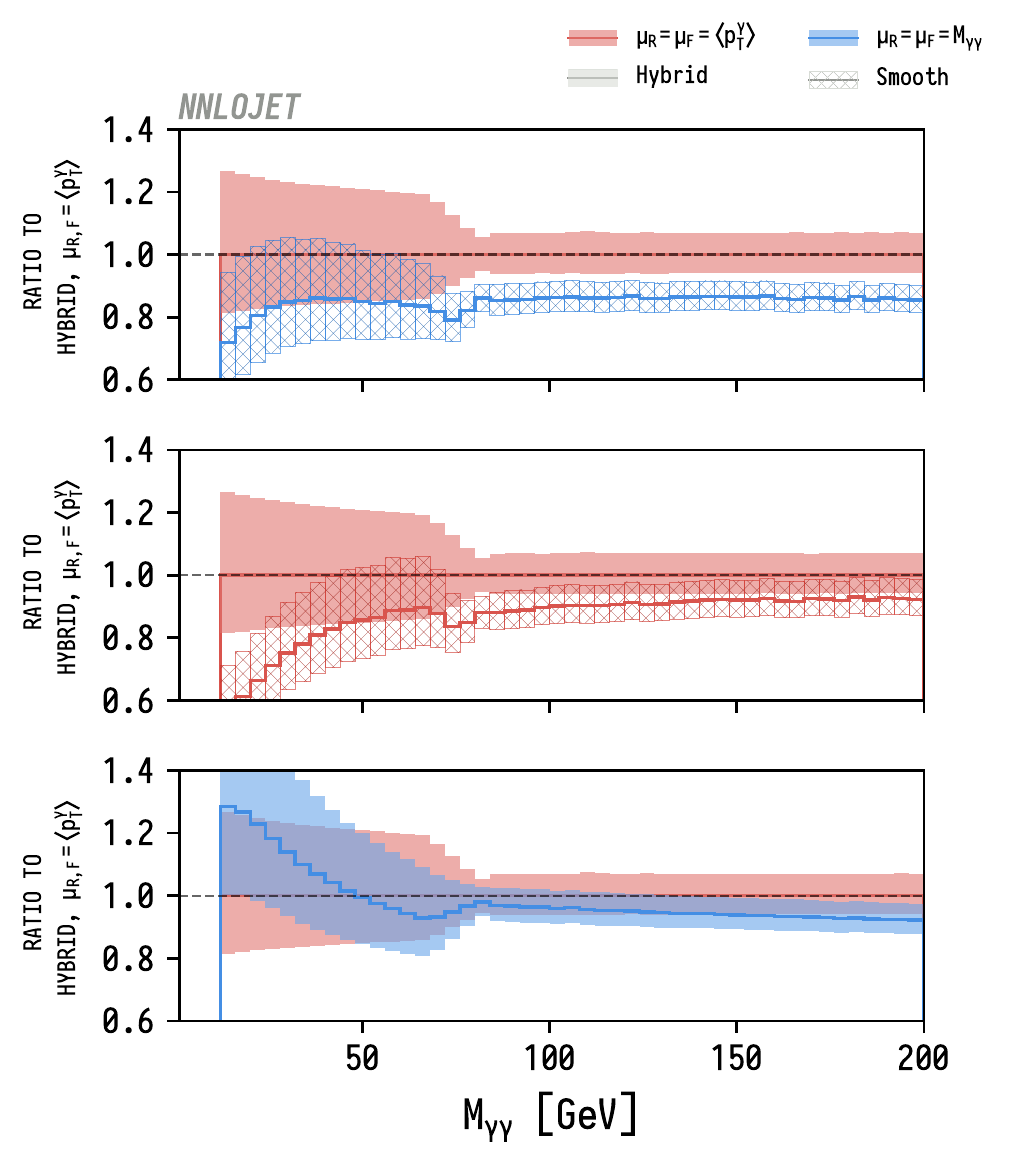}
		\caption{Combined ratio plots for four-way scales and isolation comparison, at NNLO
			(ratio to $\muF=\muR=\ptgave$ with hybrid isolation).
		}
		\label{fig:isolscl_nodata}
	\end{figure}

	We therefore examine the relative size of these competing effects in \cref{fig:isolscl_nodata}.  In the top panel,
	suppression of the cross-section for smooth-cone isolation as $\dRgg \to 0$ competes with the enhancement from the
	scale $\Mgg$ to leave the ratio almost flat.  As a result, for this specific combination of isolation procedure and
	scale choice, the competing effects of each choice shown in the lower two panels are disguised, leaving
	distributions that differ by an overall normalisation.

	Away from this region, which is the region not populated by the Born kinematics, the ratio is stable.

	\subsection{Comparison to ATLAS data: four-way comparison}
	\label{sec:combvar_fourway}

	In this section we compare the four combinations of choices for isolation and scale to ATLAS 8~TeV
	data~\cite{Aaboud:2017vol}, with the cuts of \cref{eqn:ATLAScuts}.
	As elsewhere, for both smooth-cone and (matched) hybrid isolation we use a cone of radius $0.4$ and a threshold
	$\Etthresh = 11~\GeV$, whilst for matched-hybrid isolation we use inner-cone radius $R_d = 0.1$.

\begin{figure}[tbp]
		\centering
		\includegraphics[width=0.49\textwidth]{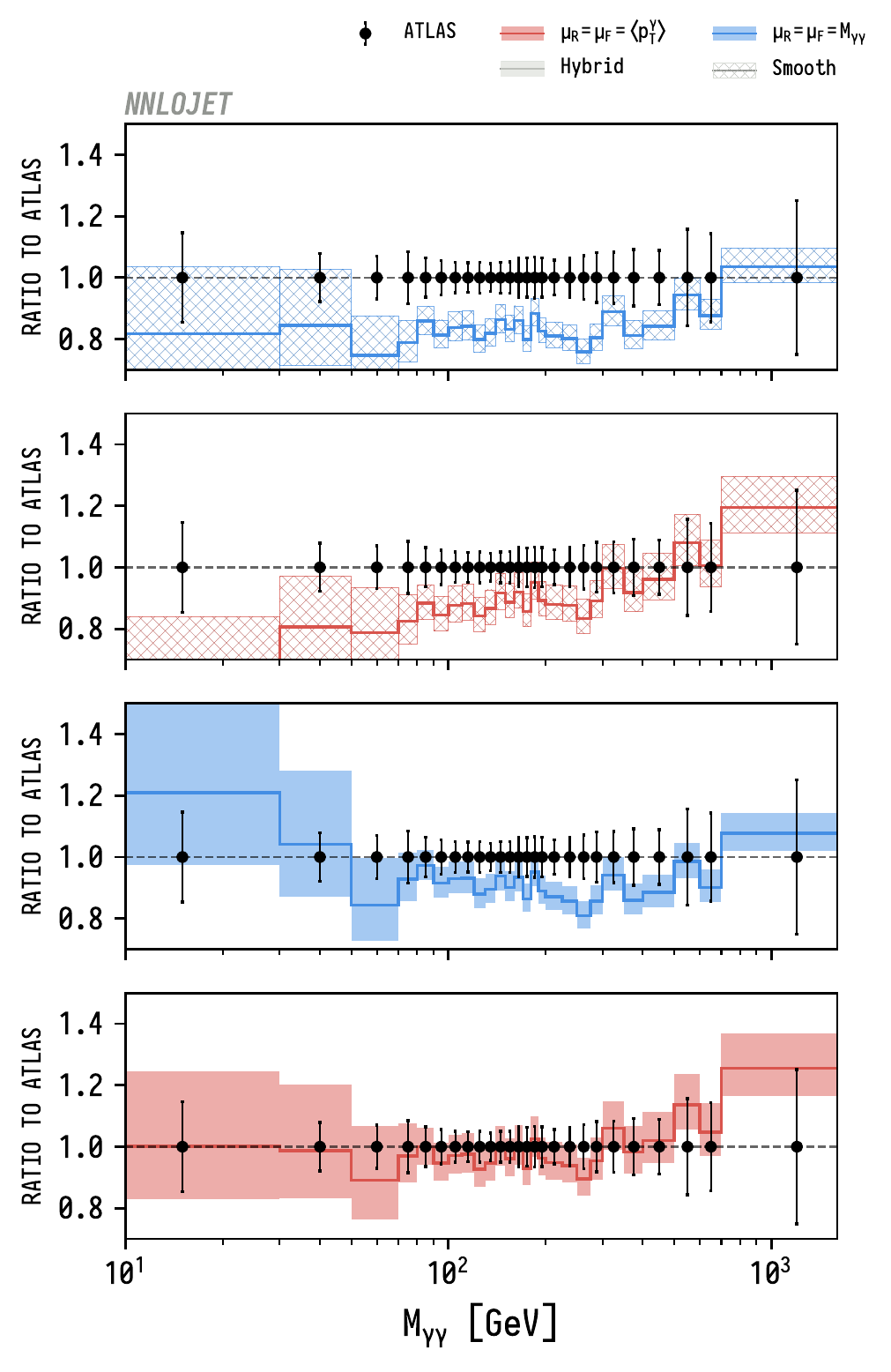}
		\includegraphics[width=0.49\textwidth]{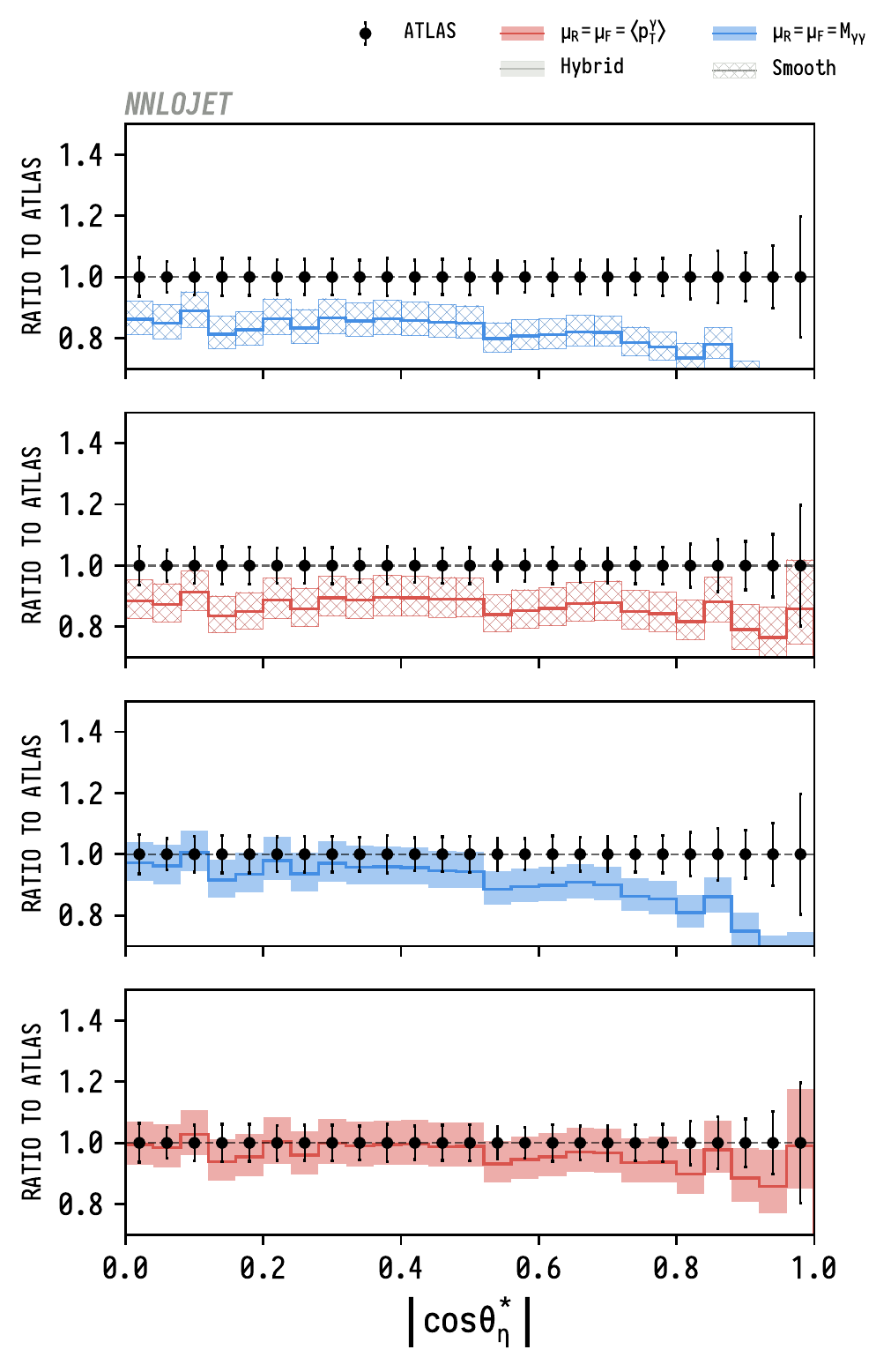}
		\caption{Combined ratio plots for the four-way scales and isolation comparison, at NNLO (ratio to data).}
		\label{fig:isolscl_data}
	\end{figure}

	We begin in \cref{fig:isolscl_data} with the two fully-NNLO distributions $\rd \sigma / \rd \Mgg$ and $\rd \sigma /
	\rd \abscosthetaetastar$.  The features highlighted above can now be seen to dramatically improve
	the overall agreement of the prediction with the data.

	We consider first the $\Mgg$ distribution.  The first panel shows that the overall prediction for the conventional
	scale choice and isolation procedure, $\mu_0=\Mgg$ with smooth-cone isolation, consistently underestimates the data
	by about 20\%, except in the largest $\Mgg$ bins.  Agreement within the scale uncertainty band of the NNLO
	prediction occurs only at the extremes of the distribution, in the lowest and highest $\Mgg$ bins.

	The second panel shows that, in the low-$\Mgg$ region, the agreement observed in the first panel is a direct
	consequence of the low-$\Mgg$ enhancement for $\mu_0=\Mgg$ outlined previously.  Without it, the suppression
	resulting from smooth-cone isolation prevents agreement in this region.  Conversely, the third panel shows that
	without the additional suppressive behaviour of smooth-cone isolation on the low-$\Mgg$ prediction, it grows
	substantially relative to the data, which does not follow the same low-$\Mgg$ behaviour.

	Comparing the first and third panels, we see that with $\mu_0=\Mgg$, moving from smooth-cone to hybrid isolation
	leads to a prediction in better agreement with the data, though still not consistently within the scale
	uncertainties of the theory calculation.  We also see that with the scale choice $\Mgg$, and without the
	suppression due to smooth-cone isolation, the low-$\Mgg$ behaviour arising from the scale choice is untamed, and
	leads to a growing deviation between theory and data as $\Mgg$ decreases.

	The last panel shows that without either the enhancement due to $\mu_0=\Mgg$ for small $\Mgg$, or the suppression
	in the same region due to smooth-cone isolation for small $\dRgg$, we see agreement in this region between the
	theory prediction and the data.  The combined effects on the overall normalisation of more permissive isolation and
	of the alternative scale choice $\mu_0=\ptgave$ correct the 20\% suppression throughout the distribution, resulting
	in theory predictions and experimental measurements largely agreeing within the scale uncertainty bands throughout
	the distribution, except in the highest $\Mgg$ bin where we might expect missing electroweak contributions  to
	become significant.

\begin{figure}[htbp]
		\centering
		\includegraphics[width=0.4\textwidth]{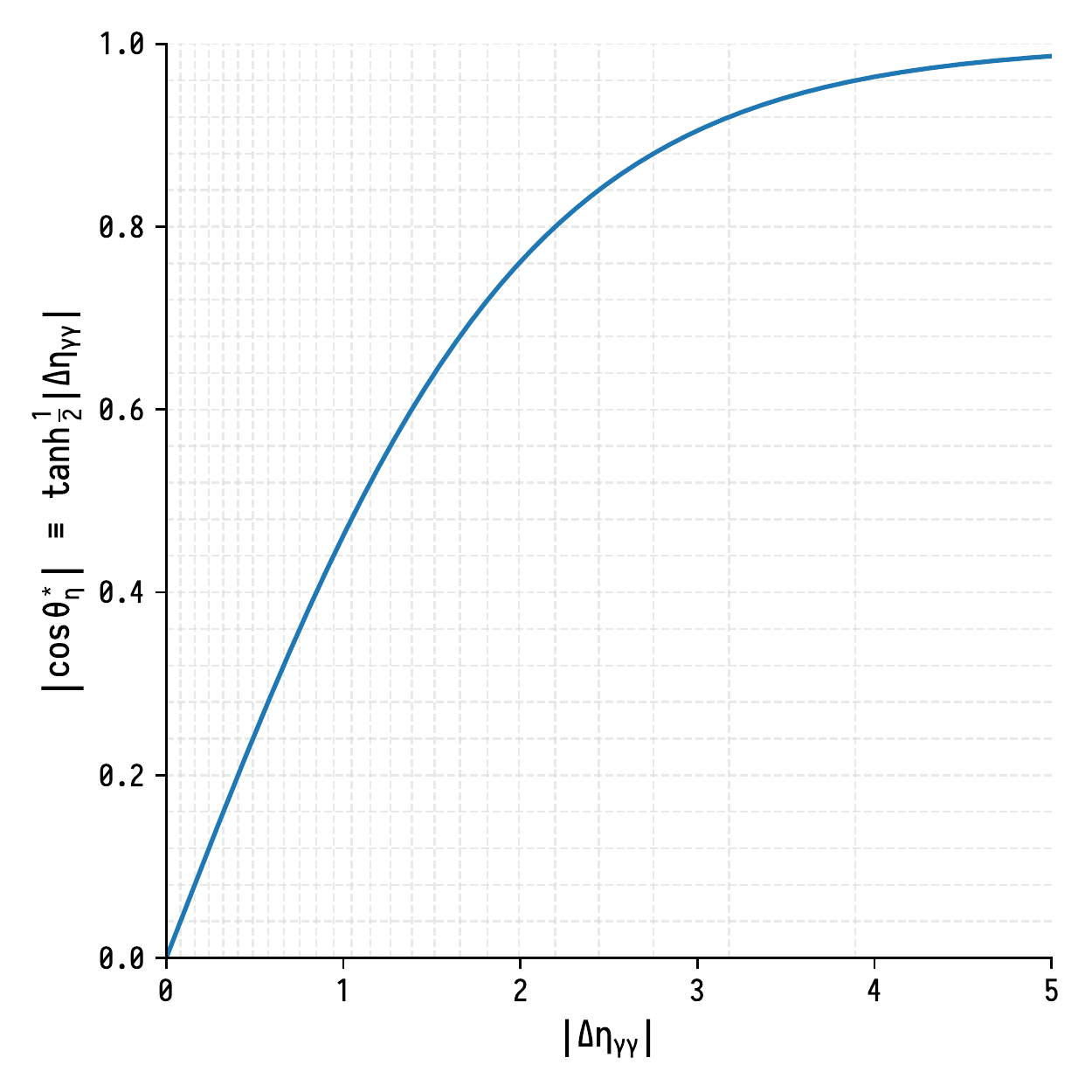}

		\caption{Relationship between $\abscosthetaetastar$ and $\Delta \eta_{\gamma\gamma}$.  The dashed grey lines
		indicate the relationship between ATLAS bins for $\abscosthetaetastar$ and the corresponding intervals for
		$\Delta \eta_{\gamma\gamma}$.  At ATLAS, the experimental cuts $\left| y^\gamma \right| < 2.37$ restrict the
		rapidity separation to $\left|\Delta \eta_{\gamma\gamma}\right| < 4.74$.  This cut only affects the result the
		in final bin, which otherwise extends to infinite rapidity separations.
		}
		\label{fig:costhetaetastar}
	\end{figure}

	We now turn to the $\abscosthetaetastar$ distribution, defined by
	\begin{align}
		\abscosthetaetastar = \tanh{\left(\frac{1}{2}\absdetagg\right)}
	\end{align}
	which is plotted for reference in \cref{fig:costhetaetastar}.  In the first panel in
	\cref{fig:isolscl_data} we see that the prediction with the scale choice $\mu_0=\Mgg$ and smooth-cone isolation
	substantially undershoots the data, by 15\% at small rapidity-separations and 40\% at high rapidity-separations.
	This is absent for the scale choice $\mu_0=\ptgave$ in panels 2 and 4, and is therefore an artefact arising directly
	from the scale $\Mgg$ and its approximately-exponential growth with rapidity separation as discussed in
	\cref{sec:scale_kineffects}.  Any other scale that is independent of $\dygg$ (or, in the notation of
	\cref{sec:scale_alt_scales}, with $\alpha=0$) would be expected to show a similarly flat ratio to the data.
	Clearly, for fixed-order predictions made with $\mu_0=\Mgg$ to exhibit such a ratio, the PDFs would need to grow to
	counterbalance the suppression of the cross-section.  It is not clear that this would be possible in such a way as
	to allow simultaneous agreement with data with both categories of scales.

	As expected, between panels 1 and 3, and 2 and 4, the change in isolation between smooth-cone and hybrid-isolation
	yields an flat upwards normalisation, resulting in very good agreement across the rapidity range for the
	combination $\mu_0=\ptgave$ and hybrid isolation.

	\subsection{Comparison to ATLAS data: two-way comparison}
	\label{sec:combvar_twoway}

\begin{figure}
		\centering
		\includegraphics[width=0.48\textwidth]{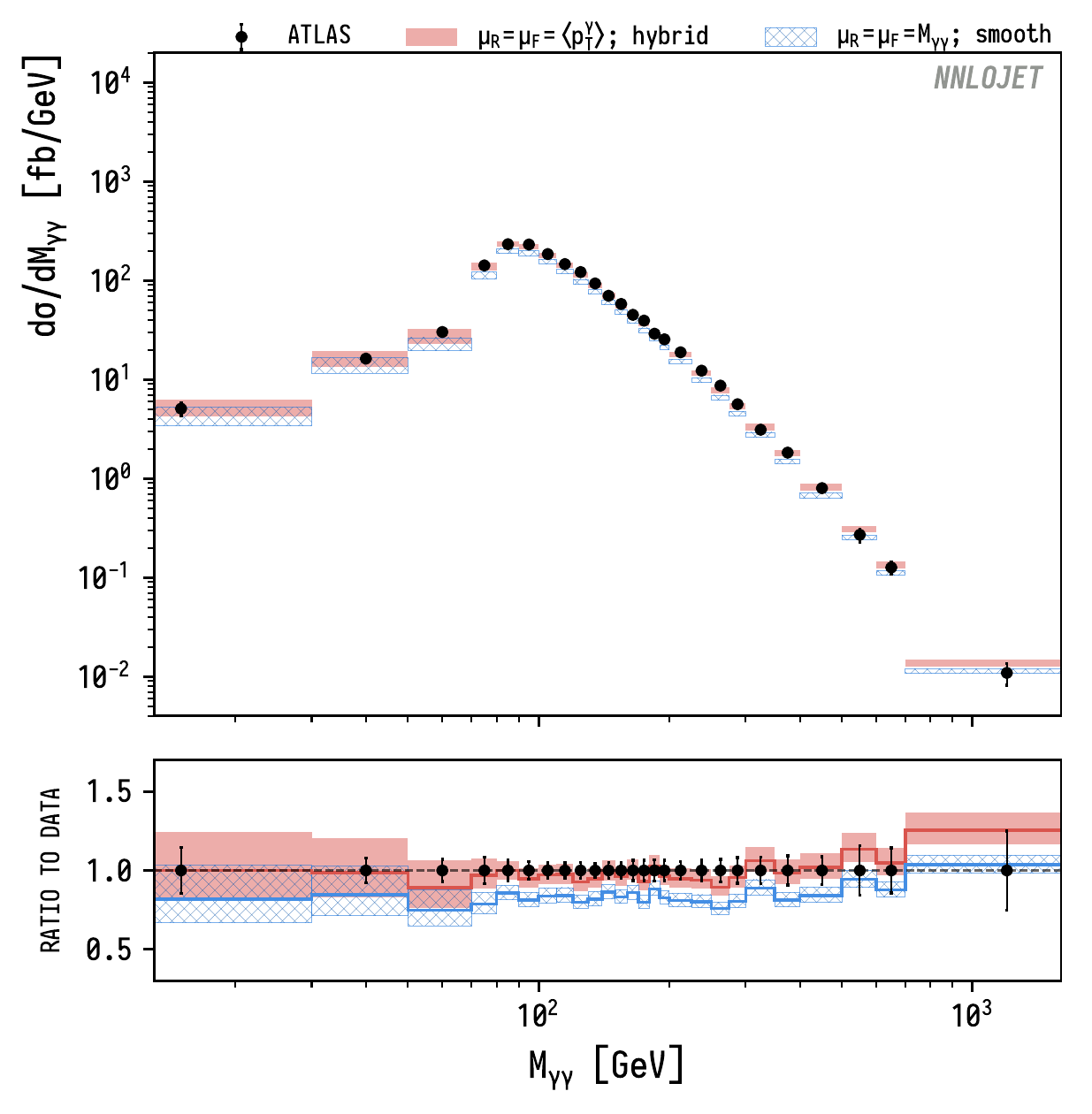}
		\includegraphics[width=0.48\textwidth]{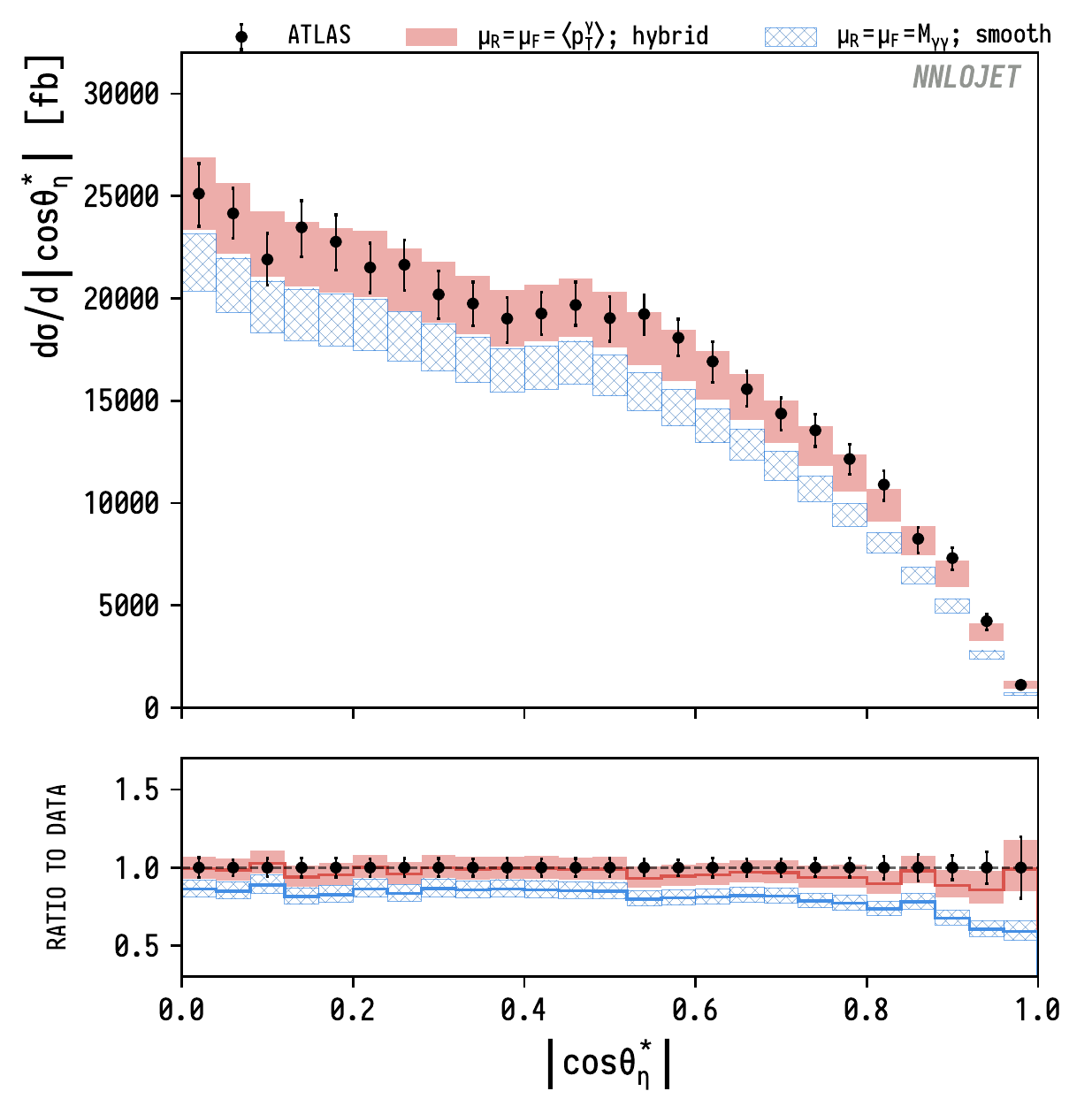}
		\includegraphics[width=0.48\textwidth]{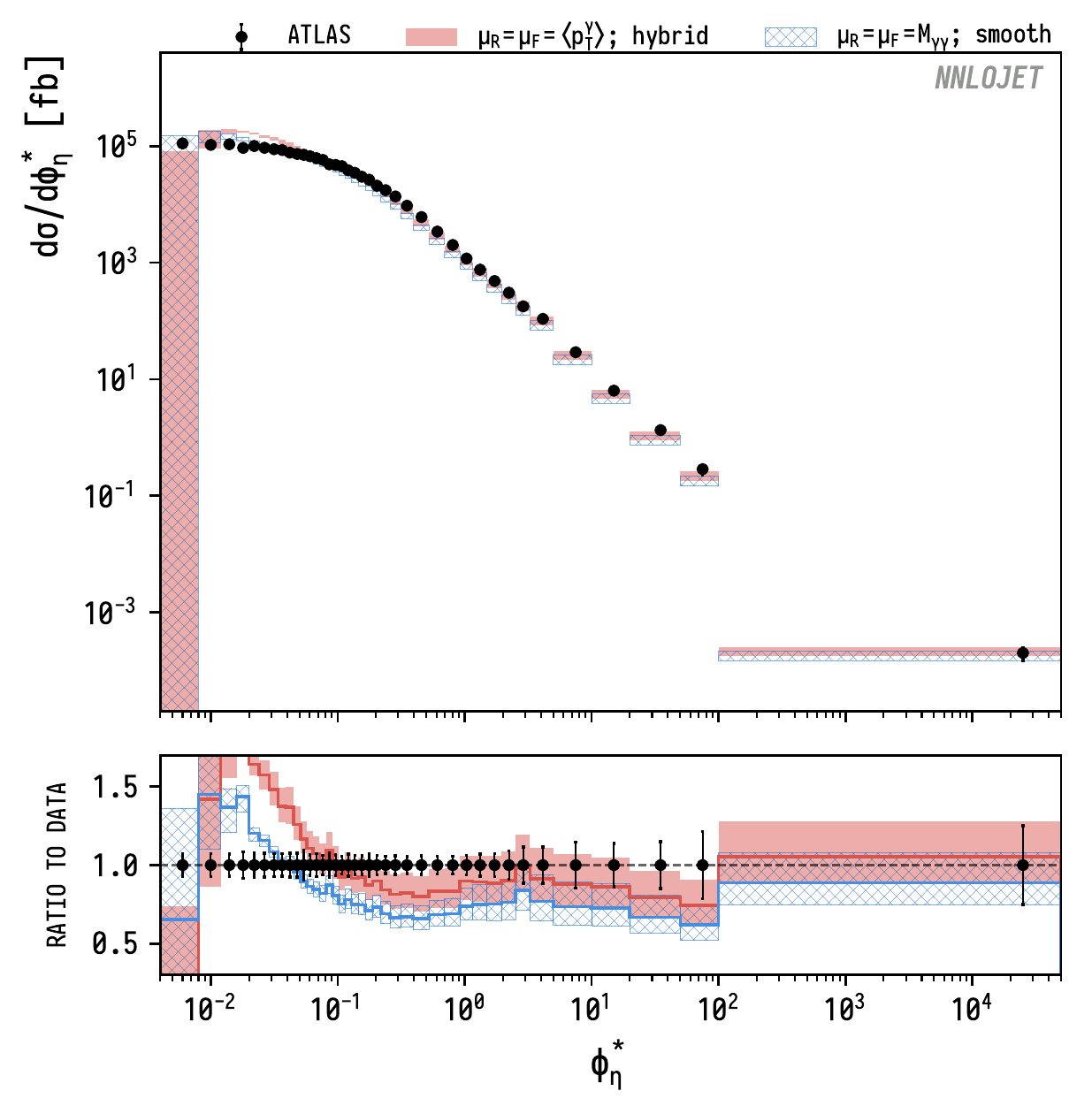}
		\includegraphics[width=0.48\textwidth]{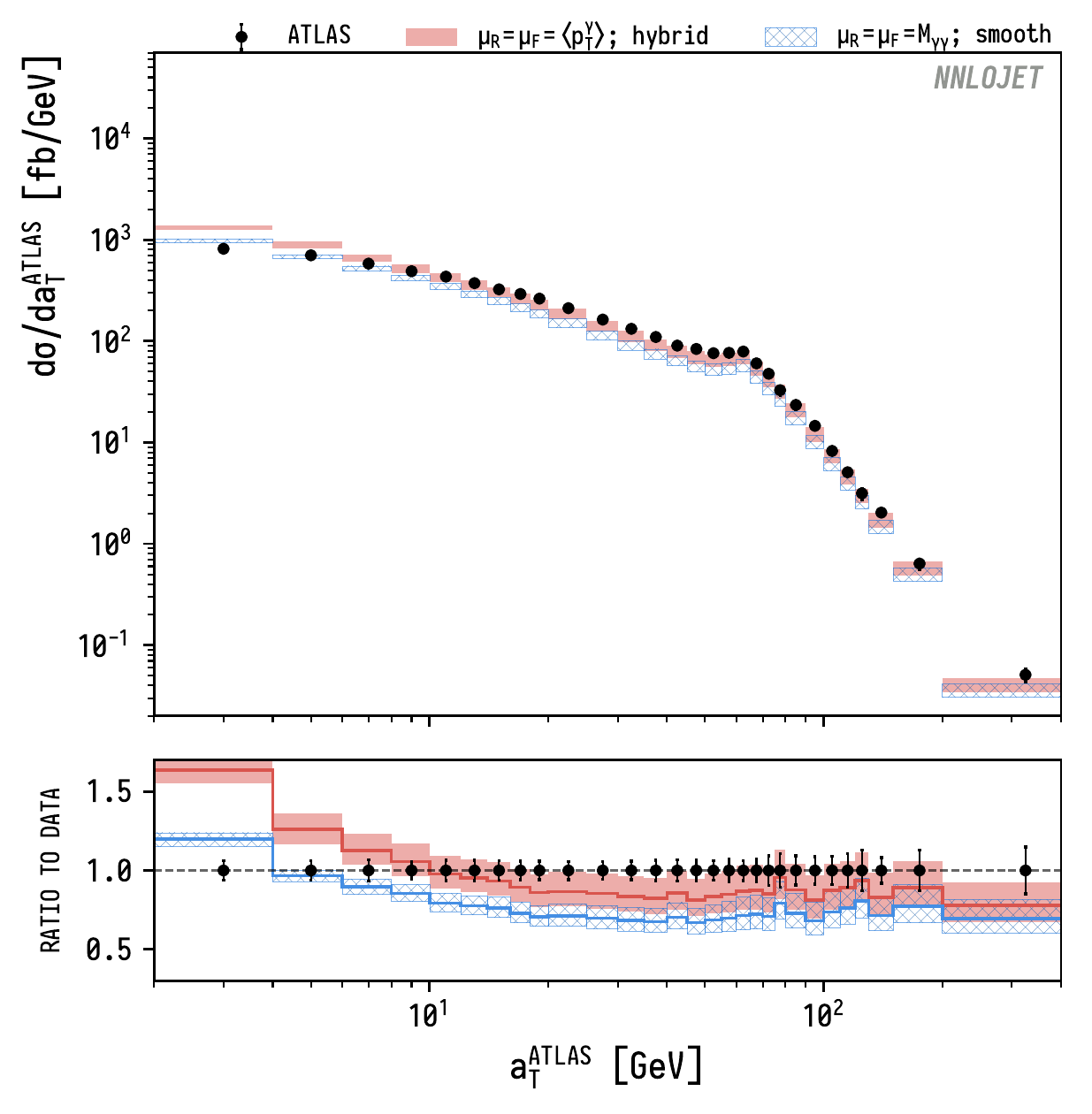}
		\includegraphics[width=0.48\textwidth]{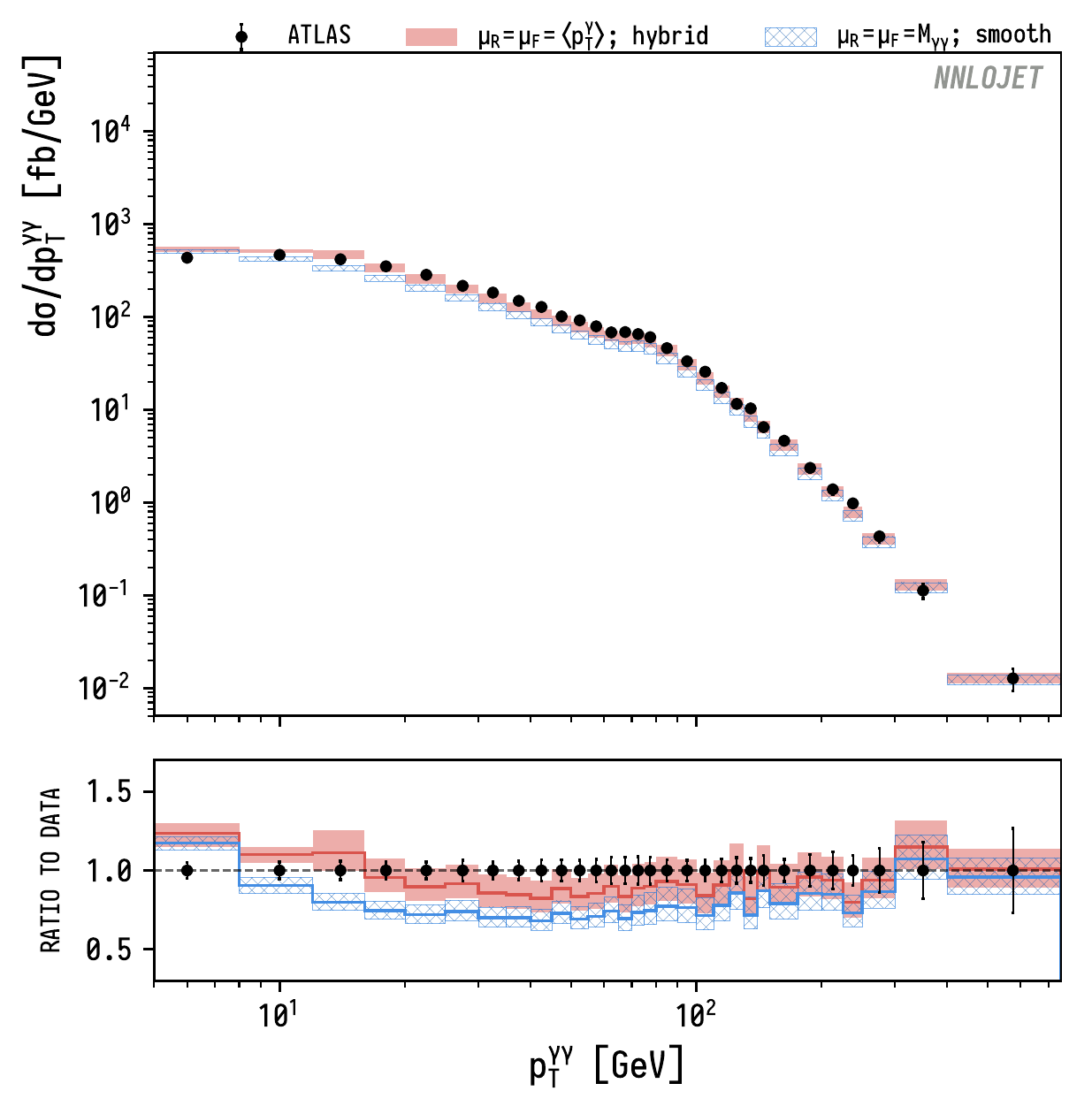}
		\includegraphics[width=0.48\textwidth]{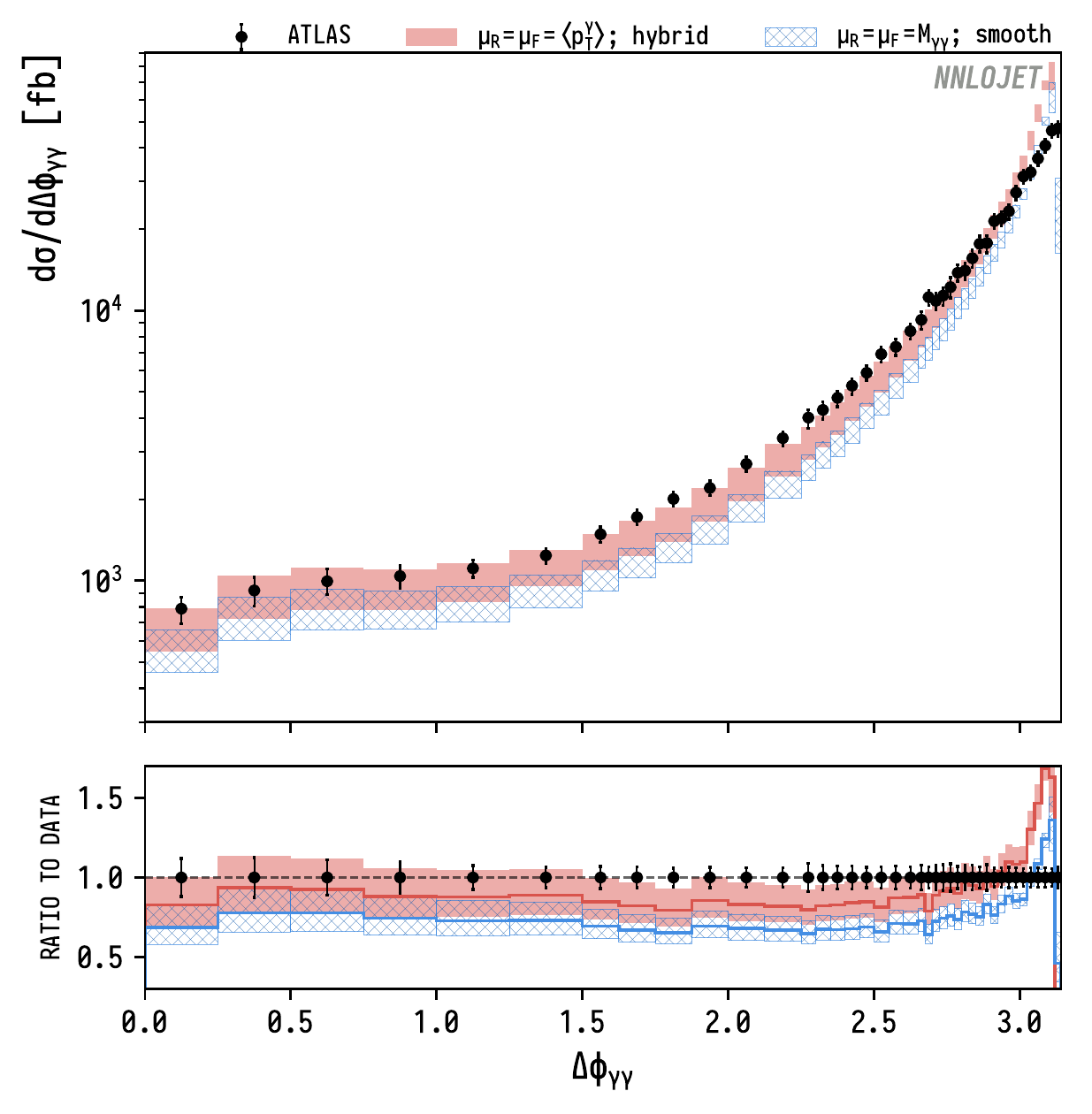}

		\caption{Re-evaluation of Figure 5 from \cite{Aaboud:2017vol} showing the effects of the modified scale choice
		and isolation criteria on the prediction.}
		\label{fig:ATLASvspred}
	\end{figure}

	We have up to now separately investigated the effect of altering scale and isolation independently.  Here we
	examine the combined effect on the agreement with ATLAS data of the simultaneous transition between the
	combinations corresponding to panels 1 and 4 of the plots in \cref{fig:isolscl_data}, namely
	\begin{itemize}
		\item[{(a)}] $\mu_0 = \Mgg$ with smooth-cone isolation, and
		\item[{(b)}] $\mu_0 = \ptgave$ with hybrid isolation.
	\end{itemize}
	These are plotted for the six observables which ATLAS measured in \cref{fig:ATLASvspred}, with axis limits and
	layout set to enable easy comparison with the corresponding figure (fig.\ 5) in the ATLAS experimental paper
	\cite{Aaboud:2017vol}.

	Across all six distributions, combination~(b) gives better agreement with data almost everywhere.  The regions
	where agreement is notably worse are those in the neighbourhood of the Sudakov singularities described in
	\cref{sec:isol_irsensitivity}, and hence where poor agreement is expected in the absence of resummation.  In these
	effectively-NLO distributions we continue to see an incomplete description of the data.  We can infer from the
	\Sherpa results of \cite{Aaboud:2017vol} that the missing radiative corrections that would feature in an NNLO
	diphoton-plus-jets calculation are required to adequately describe the data in these distributions.

\begin{figure}
		\centering
		\includegraphics[width=0.48\textwidth]{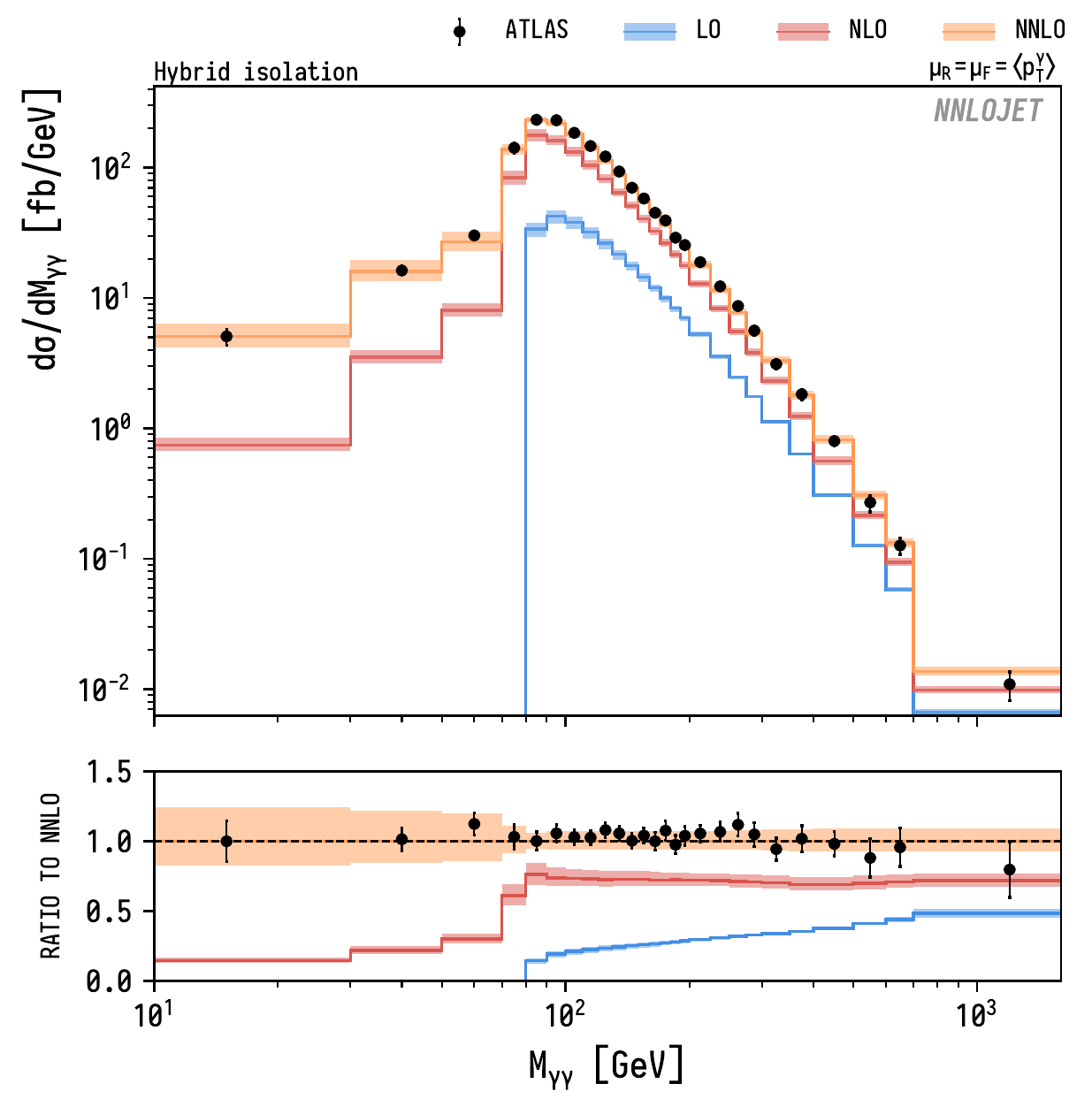}
		\includegraphics[width=0.48\textwidth]{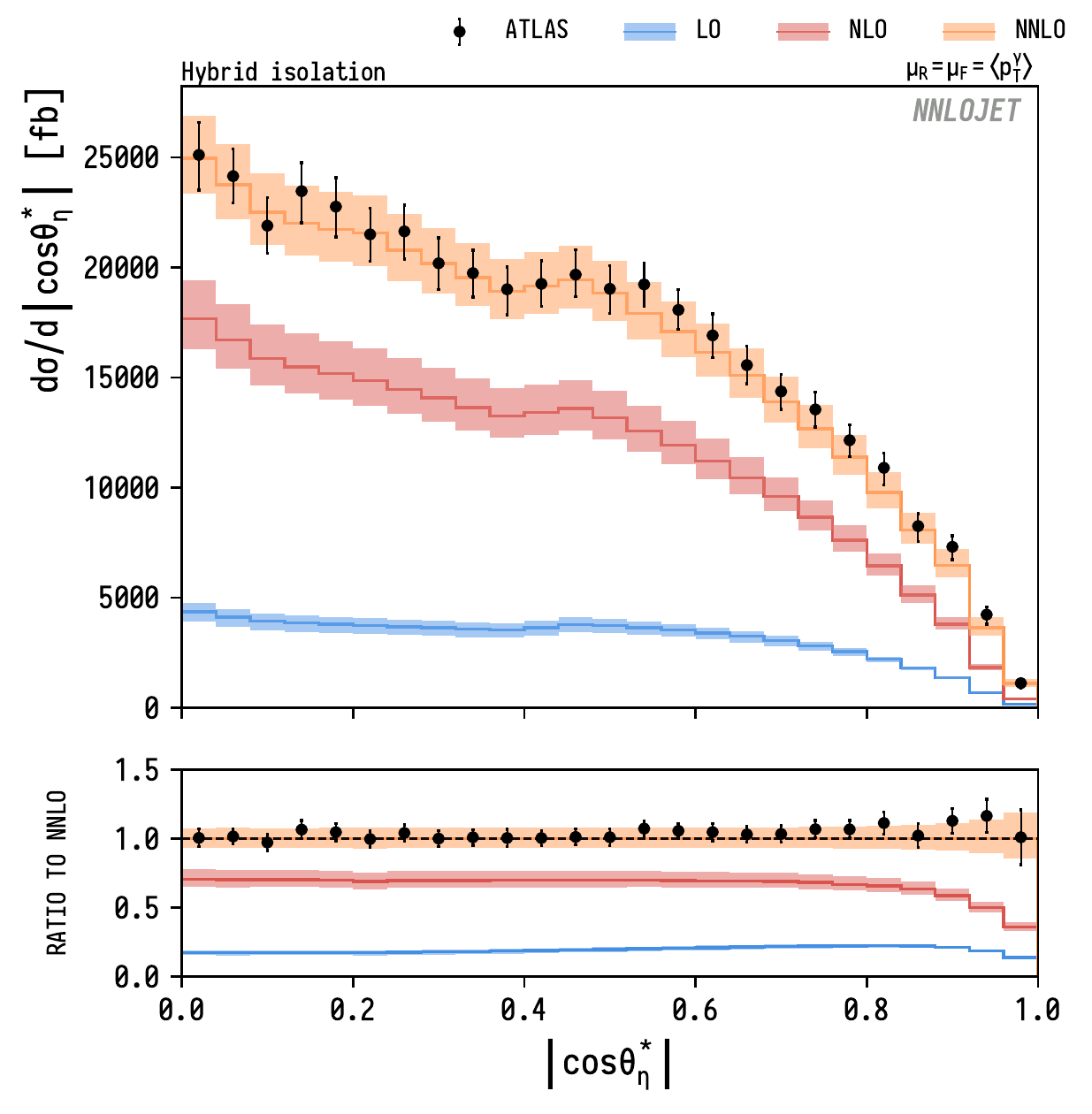}
		\includegraphics[width=0.48\textwidth]{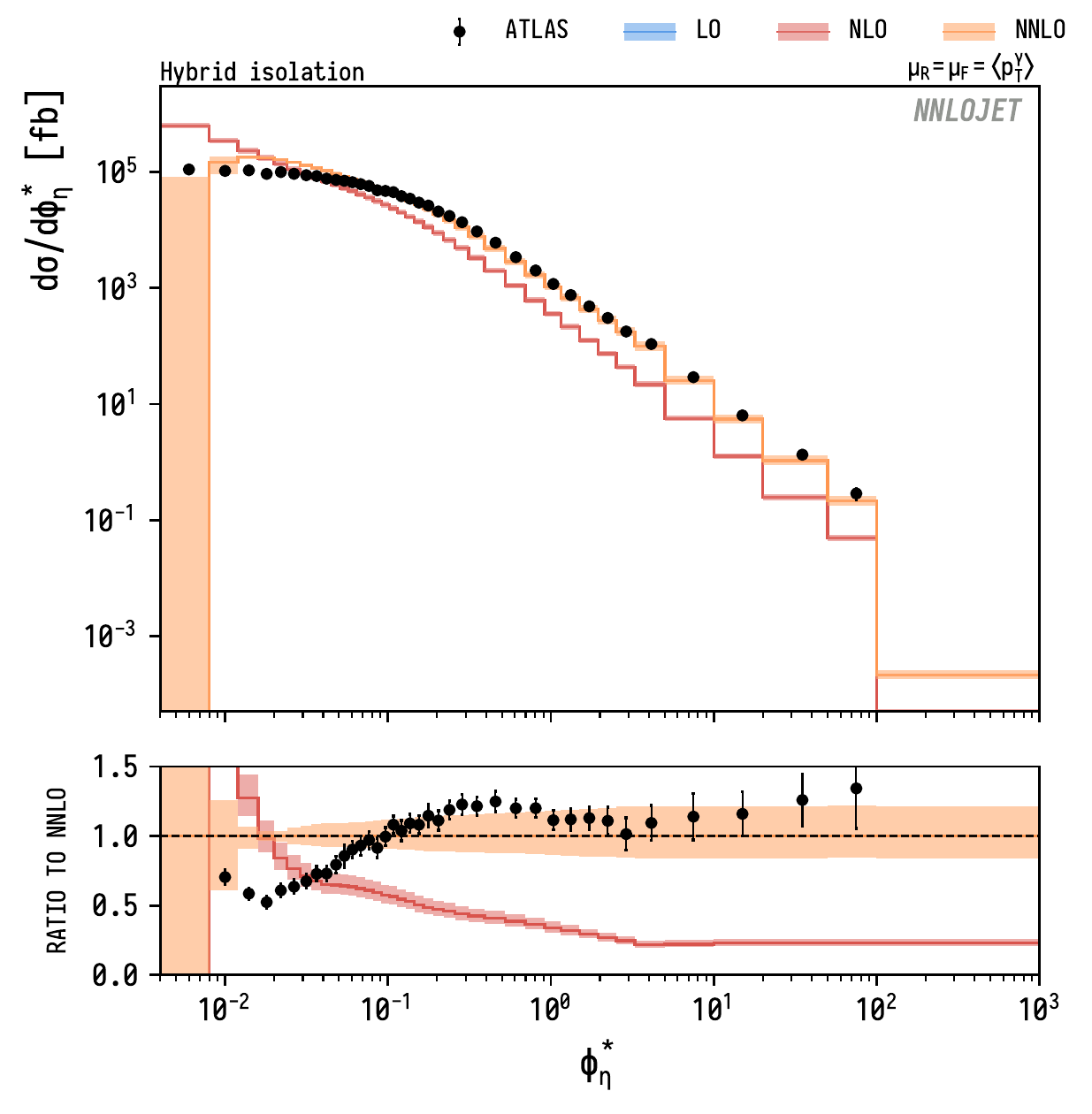}
		\includegraphics[width=0.48\textwidth]{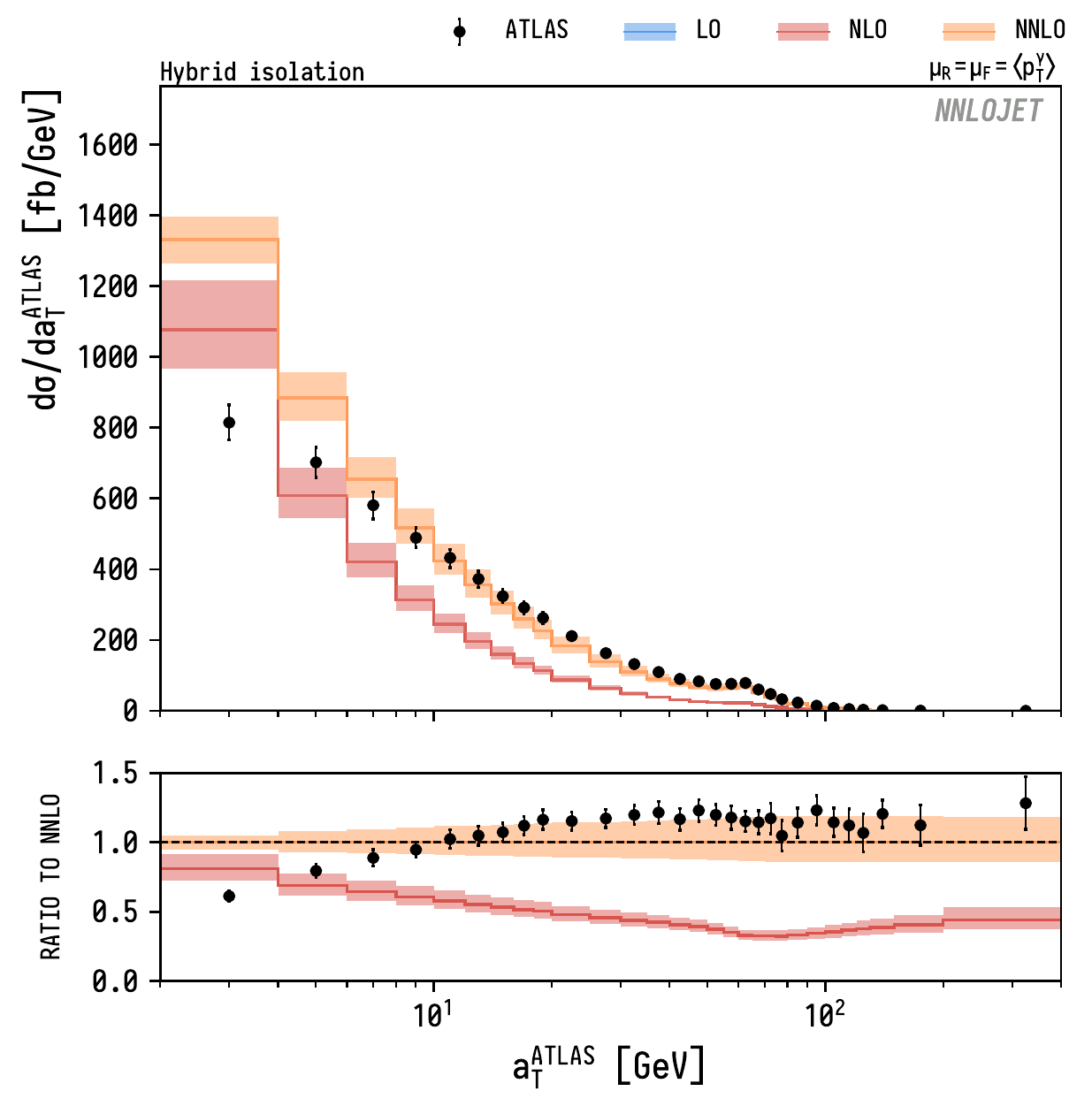}
		\includegraphics[width=0.48\textwidth]{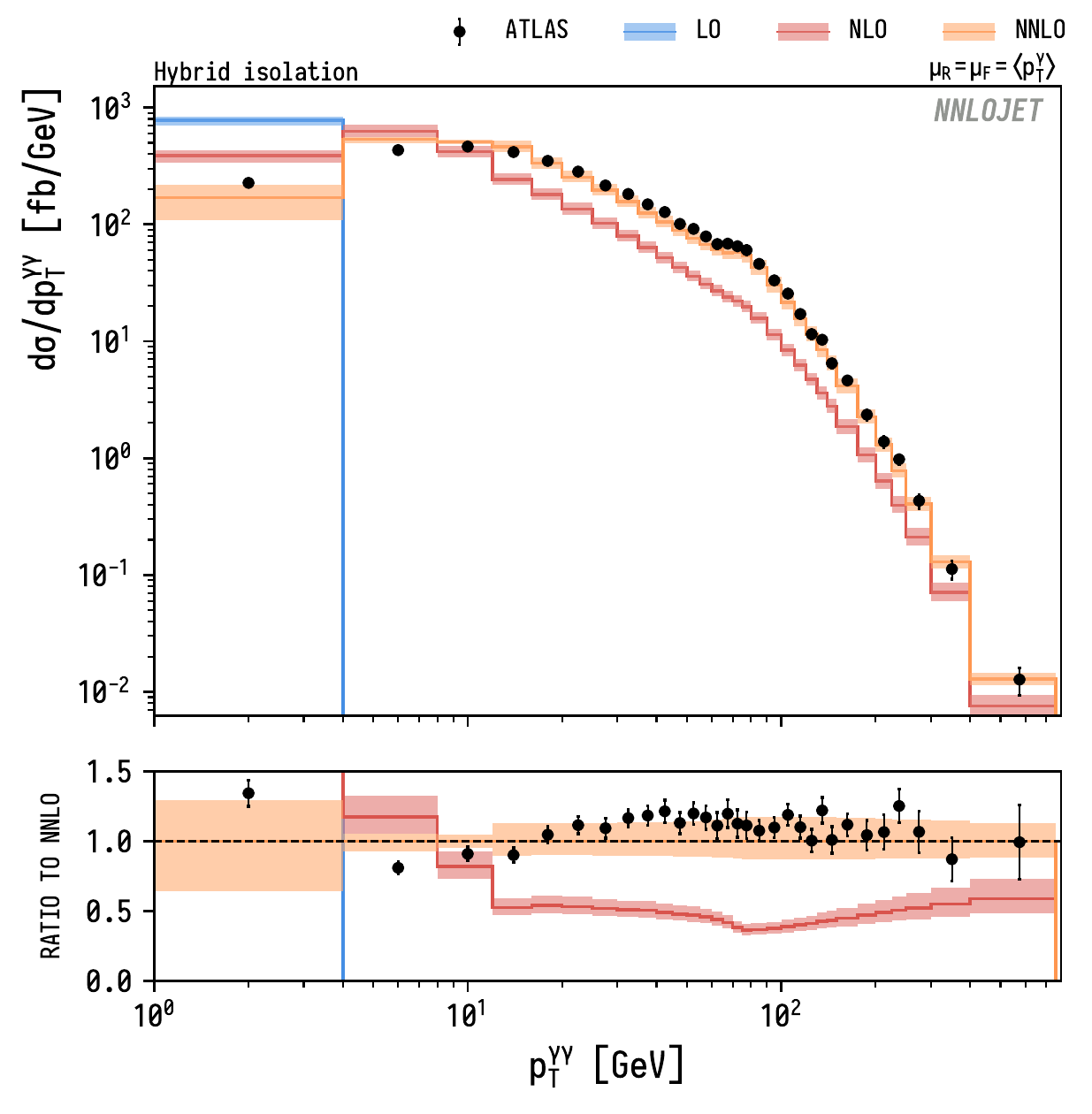}
		\includegraphics[width=0.48\textwidth]{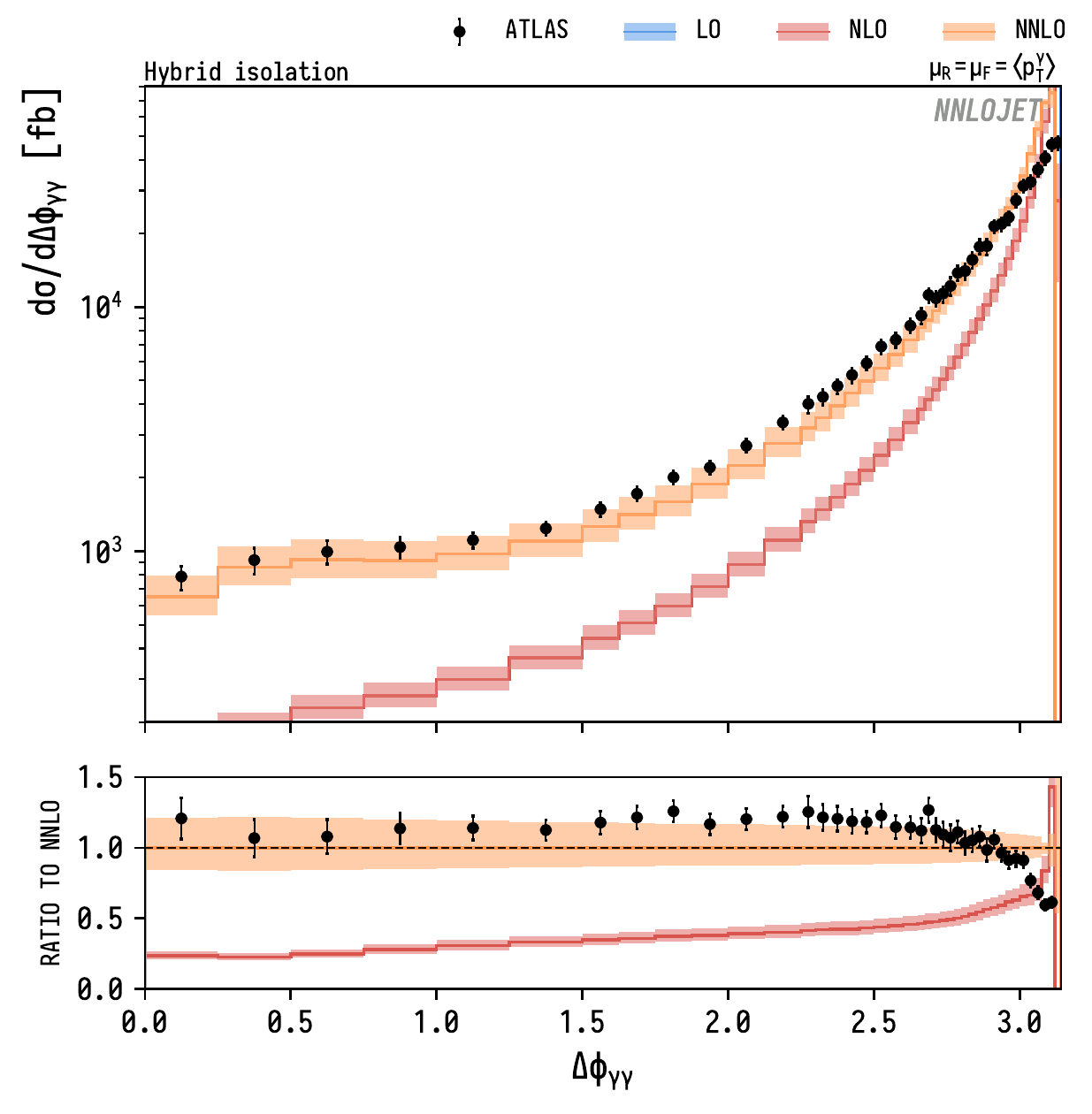}
		\caption{Illustration of the perturbative convergence of the fixed-order predictions, for the six measured
		ATLAS distributions.}
		\label{fig:pertconv}
	\end{figure}

	For completeness, in \cref{fig:pertconv} we show the order-by-order breakdown of the NNLO calculation for choice~(b)
	of scale and isolation criterion, showing the relative magnitude of the NNLO corrections with these parameters.

	\section{Conclusions}
	\label{sec:conc}

	Photon isolation is a substantial source of uncertainty in precision calculations, whose subtleties have not yet
	been fully explored.
	These uncertainties will only become more significant for phenomenology as the target precision of
	experiment and theory narrows.  Whilst they can be mitigated through the careful choice of alternative isolation
	parameters, such as smaller isolation cones, it is important to understand the full effect of approximations made
	in the theoretical modelling of experimental isolation.

	We have shown that better approximating the fiducial region defined by the experimental isolation criterion, using
	so-called hybrid isolation, leads to substantially improved agreement with data than the presently-favoured
	smooth-cone isolation.  Comparing the two, smooth-cone isolation results in a suppression of the cross-section that
	is consistently of the order of 10\%, and in regions of some distributions up to 50\%.

	This has re-exposed the issue of the infrared sensitivity of isolated photon differential cross-sections to
	fixed-cone isolation cuts.  Although this was first discussed in the context of fragmentation cross-sections at NLO
	\cite{Binoth:1999qq}, it has been absent from more recent discussions of fixed-cone and hybrid isolation.  We have
	found that, as might be expected from the step-function formulation of all cone-based isolation criteria,
	discontinuities and resulting Sudakov singularities arise in all of them, but are most significant for the $\rd
	\sigma / \rd \ptgg$ distribution for fixed-cone isolation with a constant threshold.  These pathological regions
	are not currently of direct phenomenological significance, but may become so in future.  Indirectly, they are
	likely to have implications for phenomenology by hindering cut-based subtraction procedures for higher-order
	perturbative calculations.

	We have likewise studied the uncertainty resulting from the choice of functional form for the dynamic
	renormalisation and factorisation scales.  As for the profile-function-induced uncertainty, we have found that the
	envelope of predictions spanned by different reasonable choices of functional form is not adequately described by
	the usual scale variation procedure of varying a central scale up and down by combinations of factors of 2 in each
	direction.  We have identified the regions of phase-space in which two reasonable choices are most likely to give
	very different results, and verified that the sensitivity in these regions is indeed substantial.

	We have identified competing effects arising from the conventional choices of the two theoretical functions
	described above, each of which disguises the effect of the other on the result.  We have shown that these choices,
	of smooth-cone isolation and $\mu_0=\Mgg$, are interdependent, in that unphysical behaviour introduced by the choice
	of smooth-cone isolation is only absent from the result with the scale choice $\mu_0 = \Mgg$, and vice versa.

	Comparing the above findings to ATLAS 8~TeV data, we conclude that these two effects account both for the deviation
	of the central NNLO predictions from the experimental data, and for the underestimation of the theoretical
	uncertainty that places the experimental result outside of the theoretical uncertainty bands.  Without properly
	accounting for this uncertainty, the natural conclusion is that the experimental measurements disagree with the
	theoretical predictions at NNLO.  In fact, reasonable choices for scale setting and isolation procedures give
	excellent agreement.  Further measurements of other genuine NNLO distributions will be required to test whether
	this agreement persists at higher centre-of-mass energies and in other NNLO distributions.

	\acknowledgments
	The authors thank Xuan Chen, Juan Cruz-Martinez, James Currie, Rhorry Gauld, Aude Gehrmann-De Ridder,
	Marius H\"ofer, Imre Majer, Jonathan Mo, Thomas Morgan, Jan Niehues, Jo\~ao Pires and Duncan Walker for useful 
	discussions and 
	their many
	contributions to the \nnlojet code.
	This research was supported in part by the UK Science and Technology Facilities Council and by the
	Swiss National Science Foundation (SNF) under contract 200020-175595.

	\newpage

	\bibliographystyle{JHEP}
	\bibliography{refs}
\end{document}